\documentclass[apj,numberedappendix,twocolappendix,appendixfloats]{emulateapj}

\usepackage{rotating}
\usepackage{epsfig}

\usepackage[fleqn]{amsmath}

\usepackage{color}

\newcommand{\lt}{\ifmmode\,<\,\else \,$<$\,\fi}
\newcommand{\kms}{\ifmmode\,{\rm km}\,{\rm s}^{-1}\else km$\,$s$^{-1}$\fi}

\newcommand{\magarc}{\ifmmode {{{{\rm mag}~{\rm arcsec}}^{-2}}}
             \else {{{mag}$~${arcsec}$^{-2}$}}
             \fi}
\newcommand{\hunit}{km~s$^{-1}$~Mpc$^{-1}$}
\newcommand{\Nunit}{cm$^{-2}$\,(\kms)$^{-1}$}

\newcommand{\Lya}{Ly$\alpha$}

\newcommand{\Mstar}{\mathrm{M_{\star}}}
\newcommand{\Msun}{\mathrm{M_{\odot}}}

\newcommand{\sfrunit}{\mathrm{M}_{\sun}~\mathrm{yr}^{-1}}

\newcommand{\AFe}{[$\alpha$/Fe]}

\newcommand{\Hi}{H~{\sc i}}
\newcommand{\Hii}{H~{\sc ii}}

\newcommand{\Ciia}{C~{\sc ii}}
\newcommand{\Civa}{C~{\sc iv}}
\newcommand{\Nia}{N~{\sc i}}
\newcommand{\Oia}{O~{\sc i}}

\newcommand{\Mgii}{Mg~{\sc ii}}
\newcommand{\Mgia}{Mg~{\sc i}}

\newcommand{\Aliia}{Al~{\sc ii}}
\newcommand{\Aliiia}{Al~{\sc iii}}
\newcommand{\Siia}{Si~{\sc i}}
\newcommand{\Siiia}{Si~{\sc ii}}
\newcommand{\Siiiia}{Si~{\sc iii}}
\newcommand{\Siiva}{Si~{\sc iv}}
\newcommand{\Suiia}{S~{\sc ii}}  
\newcommand{\Piia}{P~{\sc ii}}
\newcommand{\Criia}{Cr~{\sc ii}}
\newcommand{\Mniia}{Mn~{\sc ii}}

\newcommand{\Feiia}{Fe~{\sc ii}}
\newcommand{\Niiia}{Ni~{\sc ii}}
\newcommand{\Zniia}{Zn~{\sc ii}}
\newcommand{\Oii}{[O~{\sc ii}]}
\newcommand{\Oiii}{[O~{\sc iii}]}
\newcommand{\Ciii}{C~{\sc iii}]}

\shorttitle{ISM composition and kinematics at $z=2$--3}
\shortauthors{Jones, Stark, \& Ellis}

\begin{document}

\title{Dust in the Wind: Composition and Kinematics of Galaxy Outflows at the Peak Epoch of Star Formation}


\author{Tucker Jones\altaffilmark{1}, Daniel P. Stark\altaffilmark{2}, Richard S. Ellis\altaffilmark{3}
}

\altaffiltext{1} {Department of Physics, University of California, Davis, 1 Shields Avenue, Davis, CA 95616, USA; tdjones@ucdavis.edu}
\altaffiltext{2} {Steward Observatory, University of Arizona, 933 N Cherry Ave, Tucson, AZ 85719, USA}
\altaffiltext{3} {Department of Physics and Astronomy, University College London, Gower Street, London WC1E 6BT, UK}


\begin{abstract}

Galactic-scale outflows regulate the stellar mass growth and chemical enrichment of galaxies, yet key outflow properties such as the chemical composition and mass loss rate remain largely unknown. We address these properties with Keck/ESI echellete spectra of nine gravitationally lensed $z\simeq2$--3 star forming galaxies, probing a range of absorption transitions. Interstellar absorption in our sample is dominated by outflowing material with typical velocities $\sim-150$ \kms. Approximately 80\% of the total column density is associated with a net outflow. Mass loss rates in the low ionization phase are comparable to or in excess of the star formation rate, with total outflow rates likely higher when accounting for ionized gas. Of order half of the heavy element yield from star formation is ejected in the low ionization phase, confirming that outflows play a critical role in regulating galaxy chemical evolution. 
Covering fractions vary and are in general non-uniform, with most galaxies having incomplete covering by the low ions across all velocities. Low ion abundance patterns show remarkably little scatter, revealing a distinct ``chemical fingerprint'' of outflows. Gas phase Si/Fe abundances are significantly super-solar ([Si/Fe]~$\gtrsim0.4$) indicating a combination of $\alpha$-enhancement and dust depletion. Derived properties are comparable to the most kinematically broad, metal-rich, and depleted { intergalactic absorption systems} at similar redshifts, suggesting that these extreme systems are associated with galactic outflows at impact parameters conservatively within a few tens of kpc. 
We discuss implications of the abundance patterns in $z\simeq2$--3 galaxies and the role of outflows at this epoch.

\end{abstract}


\section{Introduction}\label{sec:intro}

Energetic feedback from star formation and supermassive black holes is widely regarded as a fundamentally important component of galaxy formation. Powerful feedback-driven outflows of interstellar media are invoked to explain such properties as the galaxy stellar mass function \citep[e.g.,][]{bower2012}, mass-metallicity relation \citep[e.g.,][]{dekel1986,tremonti2004}, and chemical enrichment of the circumgalactic and intergalactic medium \citep[CGM and IGM, e.g.,][]{cowie1995,tumlinson2011}. Observations of the CGM surrounding massive galaxies find that the majority of all metals ever produced in stars have been ejected from their origin galaxies \citep{peeples2014}, presumably driven by such feedback. At low masses, stellar metallicity distributions of dwarf spheroidal galaxies show that $>90-99$\% of their metals have been ejected \citep{kirby2011}. 

Outflows driven by star formation and active galactic nuclei (AGN) are expected to be most influential at redshifts $z\simeq1$--3 when these energy sources were most active \citep[e.g.,][]{madau2014}. Indeed, early spectroscopy of Lyman break galaxies revealed powerful galactic-scale outflows of interstellar media \citep[ISM;][]{pettini1998} which are now known to be ubiquitous among the star forming population at high redshifts \citep[e.g.,][]{shapley2003,steidel2010,jones2012}. However, despite widespread evidence for dramatic effects of large-scale outflows, their fundamental properties remain largely undetermined. Outflow mass loss rates, recycling rates (via so-called galactic fountains), and chemical compositions are critically important parameters yet they vary widely among different galaxy formation models and cosmological simulations \citep[e.g.,][]{dave2011,gibson2013}. While there is general agreement about the broad effects of feedback-driven outflows on galaxies and the IGM, precise knowledge of their physical properties is needed for cosmological galaxy formation models to gain testable predictive power.

Observational probes of outflow properties largely come from two complementary approaches. 
Spectra of background sources provide information on the distribution of gas seen in absorption as a function of projected impact parameter. Generally only one background sightline is available for a given galaxy, but statistical ensembles have revealed the average radial absorption profiles for well-defined galaxy populations \citep{steidel2010,menard2010,tumlinson2011,rudie2012,turner2014}. These data show that the majority of gas absorption is found relatively close to galaxies (within a few tens of kpc), while the metal-enriched CGM extends to $\gtrsim100$ kpc. This is supported by spatial mapping of outflows in a few rare cases at $z\gtrsim1$ \citep{finley2017,martin2013}; more examples may soon become available thanks to sensitive integral field spectrographs such as MUSE and KCWI. 
The other common probe is a ``down the barrel'' view of outflowing material detected as blueshifted absorption in galaxy spectra. This provides good kinematic data but little or no information on the radial profile relative to the host galaxy. 

A major limitation of current data is that spectroscopic measurements sampling the bulk of material -- namely, down the barrel or at very close impact parameters -- are almost exclusively limited to the strongest absorption features. These transitions are typically saturated such that their equivalent widths encapsulate the gas kinematics and geometric covering fraction \citep[e.g.,][]{steidel2010,jones2012,jones2013,leethochawalit2016} but not the column density. In essence, there is little information about the actual mass flux or chemical composition which are critical for understanding {\em how} outflows regulate galaxy growth and the CGM+IGM. The simple reason for this shortcoming is that typical galaxies are too faint for the detailed study of optically thin absorption lines. 
To date this has been overcome only with gravitationally lensed galaxies whose apparent luminosities are magnified by factors $\gtrsim10\times$. This was first accomplished by \cite{pettini2002} who measured column densities and kinematics for a range of ionic species in the spectacular lensed galaxy cB58 at $z=2.7$, revealing outflows with chemical enrichment to a few tenths of the solar value and mass loss rates likely exceeding the star formation rate (SFR). While only two additional $z>2$ lensed galaxies have been studied in such detail \citep{quider2009,dessauges-zavadsky2010}, enlarged samples of bright gravitationally lensed galaxies \citep{stark2013,rigby2018} now provide the opportunity for high quality spectroscopy of a representative population at high redshifts. 

This paper is concerned with characterizing the typical physical properties of galaxy ISM and outflows during the peak periods of their star formation at $z\simeq2$--3. Our goal is to obtain similar measurements to those in the pioneering study of \cite{pettini2002} for a statistically interesting sample representative of the star forming population, for which thousands of spectra exist yet only the strongest ISM features have been characterized thus far. Our study leverages the Cambridge And Sloan Survey Of Wide ARcs in the skY \citep[CASSOWARY, which we abbreviate as CSWA;][]{belokurov2007,belokurov2009}. The CASSOWARY survey has identified more than 100 candidate strong gravitational lens systems in Sloan Digital Sky Survey imaging, with typical lensed galaxy AB magnitudes of 20--21 at optical wavelengths. We have conducted followup campaigns with various telescopes to spectroscopically confirm the nature of these systems \citep{stark2013}. The resulting catalog represents a collection of galaxies which are characteristic of the population at their redshifts, yet are among the brightest examples on the sky thanks to strong lensing magnification. As such these are ideal sources for deep followup spectroscopy. We have previously utilized this sample to measure kinematic properties and to spatially map nebular metallicities using integral field spectroscopy, demonstrating the power of lensing systems to better understand the complex galaxy formation process \citep{jones2013a,leethochawalit2016a}. We now present a study of the ISM and outflow properties for 9 galaxies at $z=1.4$--2.9, in total quadrupling the available sample of quality absorption line spectroscopy at these redshifts.

The paper is structured as follows. The spectroscopic observations are described in Section~\ref{sec:data}. Measurements of gas covering fractions and column densities are presented in Section~\ref{sec:fcov}. We present our analyses of the gas kinematics in Section~\ref{sec:kinematics}, and chemical properties including intrinsic abundances, dust depletion, and ionization state in Section~\ref{sec:abundances}. We discuss implications of our findings for the galaxy population and its evolution in Section~\ref{sec:discussion}, and conclude with a summary of the main results in Section~\ref{sec:summary}. 
Throughout this work we adopt vacuum wavelengths from the NIST Atomic Spectra Database \citep{kramida2016} and oscillator strengths $f$ from the compilations of \cite{morton2000} and \cite{morton2003}, with the exception of \Niiia$\lambda$1317 where we use $f=0.07786$ \citep{morton1991}. Solar abundances refer to the photospheric values given by \cite{asplund2009}, and literature measurements are adjusted to this solar scale as needed. Where necessary we adopt a flat $\Lambda$CDM cosmology with $H_0=70$ \hunit, $\Omega_{\Lambda}=0.7$, and $\Omega_M=0.3$.

\section{Observations and data reduction}\label{sec:data}

All objects were observed with the Echelle Spectrograph and Imager \citep[ESI;][]{sheinis2002} on the Keck II telescope. Data were obtained on two observing runs in November 2012 and March 2013. A summary of the targets and exposure times is given in Table~\ref{tab:arcs}. Spectra were taken in echellete mode with a 0\farcs75 slit, resulting in a spectral resolution $R=6300$ measured from sky lines (velocity resolution = 48 \kms\ FWHM). Data were reduced using the IDL-based ESIRedux code written by J. X. Prochaska.
We first apply a bias subtraction and flat-field correction to all frames using calibration data taken on the same night. Sky subtraction is performed using polynomial and spline fits to the corrected 2D spectra. We extract the 1D spectrum of each object using a boxcar aperture, based on the observed spatial extent of each target along the slit. In some cases we observe multiple lensed images of the same target and extract their traces separately. Individual exposures are combined by first normalizing all extracted spectra to the mean continuum level, and then taking the weighted mean with cosmic ray rejection. Where applicable, spectra of multiple images are then combined. We normalize the spectra to a running median of the continuum level over regions free of strong spectral features. 
{
The continuum level is determined to a precision of $\sim$2\% (1-$\sigma$) for the typical signal-to-noise of 10 per resolution element. This does not substantially contribute to uncertainty in column density measurements, which are instead dominated by the limited signal-to-noise of individual absorption lines. We further confirm that the continuum normalization does not introduce large systematic errors, as we find consistent column densities from different transitions of the same ion (Section~\ref{sec:multiple_transitions}). 
Visual inspection shows that the continuum level behaves well near intrinsic absorption and emission features in the galaxy spectra, which are masked from the continuum estimate. However, the continuum is less reliable near strong telluric features, including sky emission lines. Features which are visibly affected by telluric absorption or sky line residuals are therefore excluded from analysis.
}

\begin{figure*}
\includegraphics[width=\textwidth]{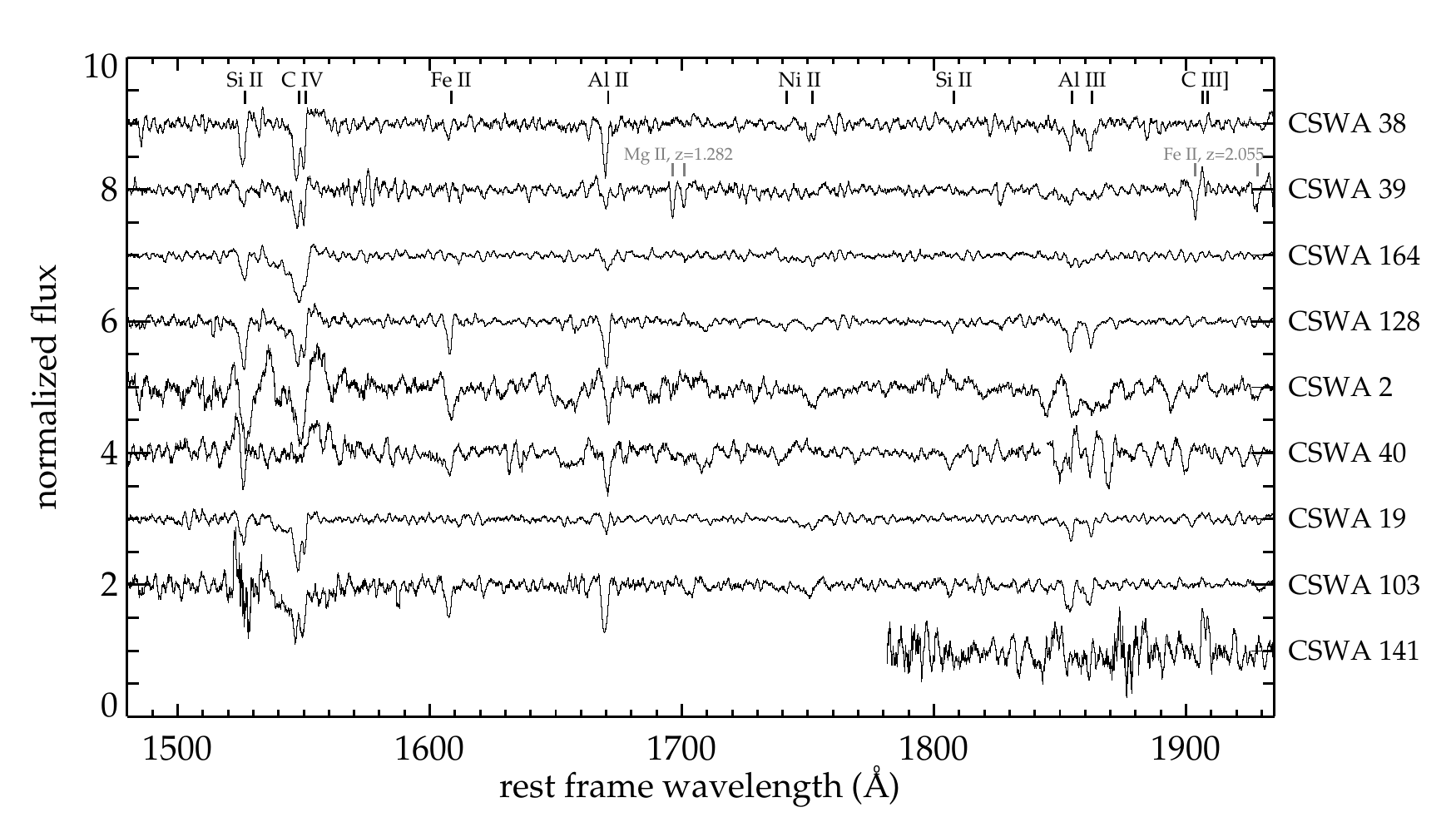}
\caption{
\label{fig:spectra}
Spectra of the nine galaxies in the ESI sample in the rest-frame wavelength range 1500--1900 \AA, normalized and offset for clarity, ordered from highest (top) to lowest redshift. In the case of CSWA 141 we show only the spectrum redward of 4300 \AA\ in the observed frame. The spectra are smoothed to $\sim$250 \kms\ FWHM resolution for display purposes. 
Several prominent features are labeled at top. We also label strong lines from two intervening absorption systems identified in the spectrum of CSWA 39. 
Significant dispersion can be seen in the equivalent width of features such as \Siiia$\lambda$1526 and \Aliia$\lambda$1670, which are typically optically thick and therefore trace the covering fraction and velocity profile of low-ionization gas. Weaker lines such as \Siiia$\lambda$1808 and \Niiia$\lambda$1741,$\lambda$1751 are additionally sensitive to the gas column density. 
}
\end{figure*}

Spectra of each object are shown in Figure~\ref{fig:spectra}. 
In most cases the stellar continuum is detected at $\gtrsim10\sigma$ per resolution element in good sky regions (Table~\ref{tab:arcs}). In three cases we smooth the data to improve the precision of absorption line measurements. This results in final FWHM spectral resolution of 85 \kms\ for CSWA 141, 102 \kms\ for CSWA 40, and 157 \kms\ for CSWA 2, improving the signal-to-noise ratio to $\sim$6--8 in each case.

\subsection{Systemic redshifts}

Systemic redshifts are measured for each source from spectral features which originate from either stellar photospheric absorption or nebular emission in \Hii\ regions. The features used to derive redshifts are listed in Table~\ref{tab:zsys} and we give details for individual sources in Appendix~\ref{sec:zsys}. We also list the redshifts derived for \Feiia* and \Siiia* fine structure emission lines which typically peak near the systemic velocity, although these features arise predominantly in outflowing gas \citep[e.g.,][]{jones2012,prochaska2011} and thus are not expected to trace the stellar kinematics. The root mean square scatter between fine structure emission line centroids and the adopted systemic redshift is 31 \kms\, with mean and median offset $< 10$ \kms. While the fine structure redshifts have larger scatter than true systemic features, they are nonetheless more accurate than other methods used to estimate redshifts when no other systemic features are available \citep[e.g., \Lya\ emission or interstellar absorption lines;][]{steidel2010}.

Redshift uncertainties in Table~\ref{tab:zsys} correspond to a formal 1$\sigma$ uncertainty in the centroid of Gaussian fits to the features. In general these fits have reduced $\chi^2$ values of order unity indicating that the estimated measurement errors are accurate. However, the dispersion in redshift measured from different spectral features is somewhat larger than the statistical uncertainty. This could be caused by different physical origins of the various features, spurious fits of noisy features, underestimation of the true uncertainty, or a systematic error in the wavelength calibration of $\sigma(\lambda) / \lambda \simeq 5 \times 10^{-5}$ in different echelle orders. In any case the adopted redshifts are accurate to $\lesssim 20$ \kms\ in velocity across the observed ESI spectra.

\subsection{Demographic properties of the sample}\label{sec:demographics}

Galaxies in our sample are representative of the intermediate-mass star forming population studied by other large surveys at $z\simeq2-3$ \citep[e.g.,][]{wuyts2016,steidel2016,sanders2015}. At present we have obtained robust stellar masses and SFRs for 7 galaxies in our sample with near-IR imaging (all except CSWA 38 and 164; Mainali et al. in prep). The sample median properties from SED fitting with an assumed constant star formation history, metallicity of 0.2 solar, and a Chabrier IMF are $\log \Mstar/\Msun = 9.8$, SFR~$=13 \,\sfrunit$, and sSFR~$=2$ Gyr$^{-1}$.

\subsection{Intervening absorption systems}

\begin{figure}
\includegraphics[width=\columnwidth]{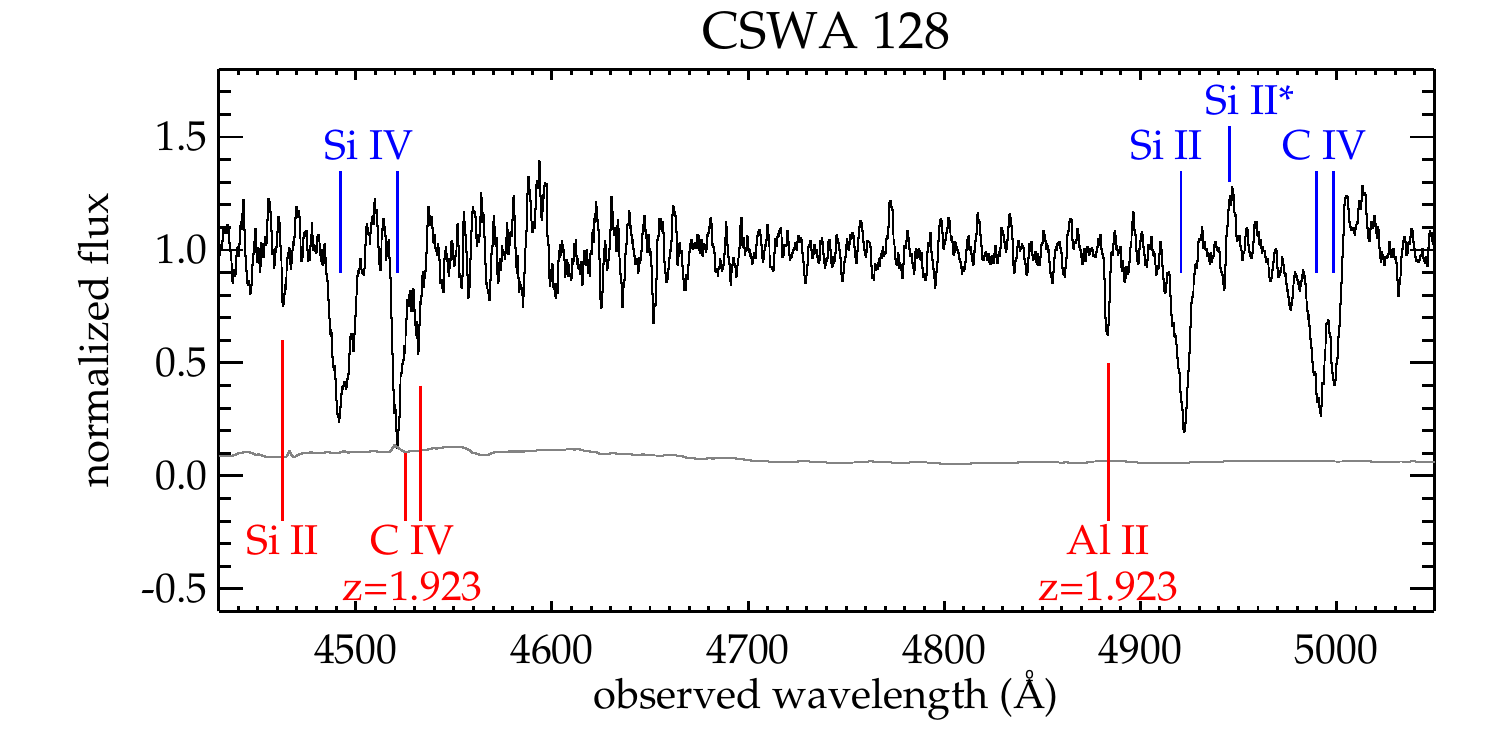}
\includegraphics[width=\columnwidth]{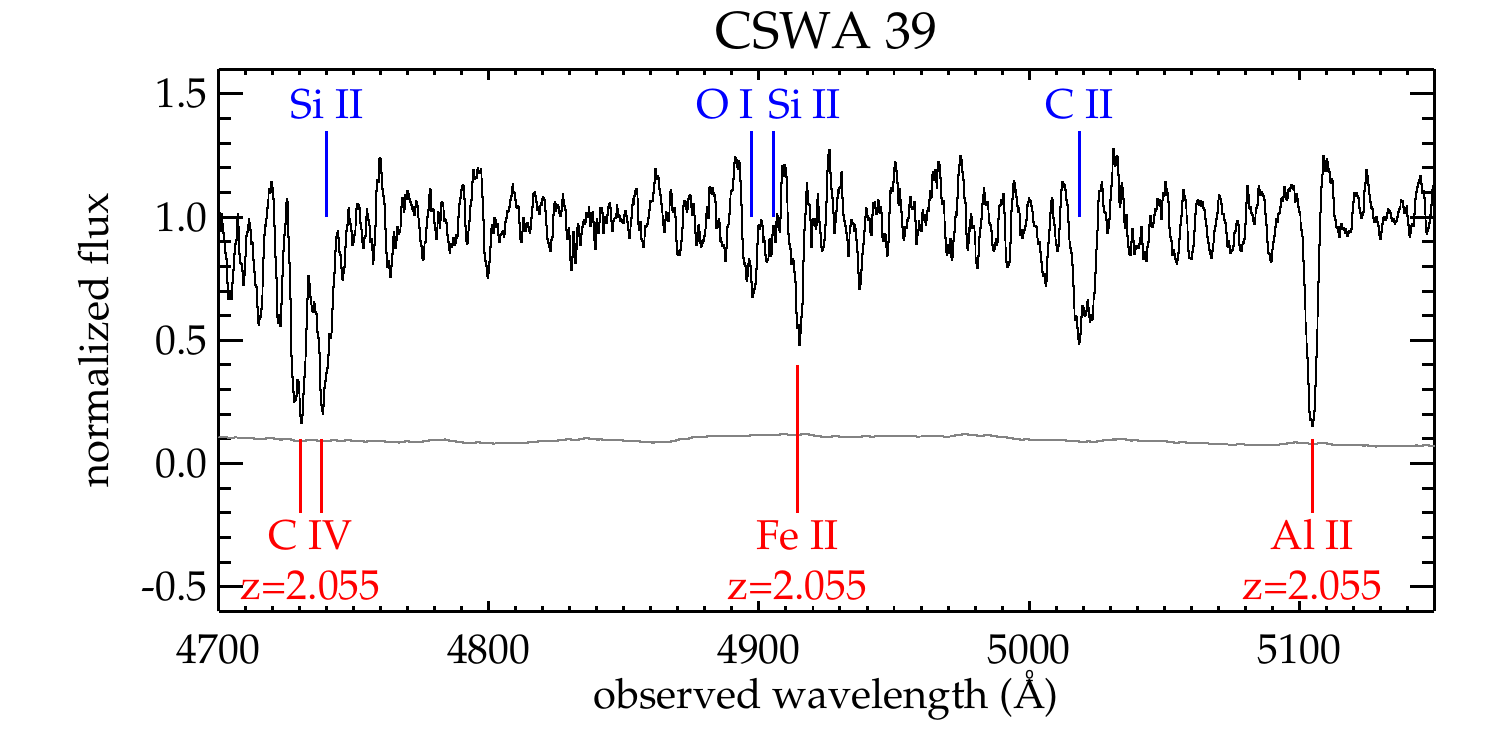}
\includegraphics[width=\columnwidth]{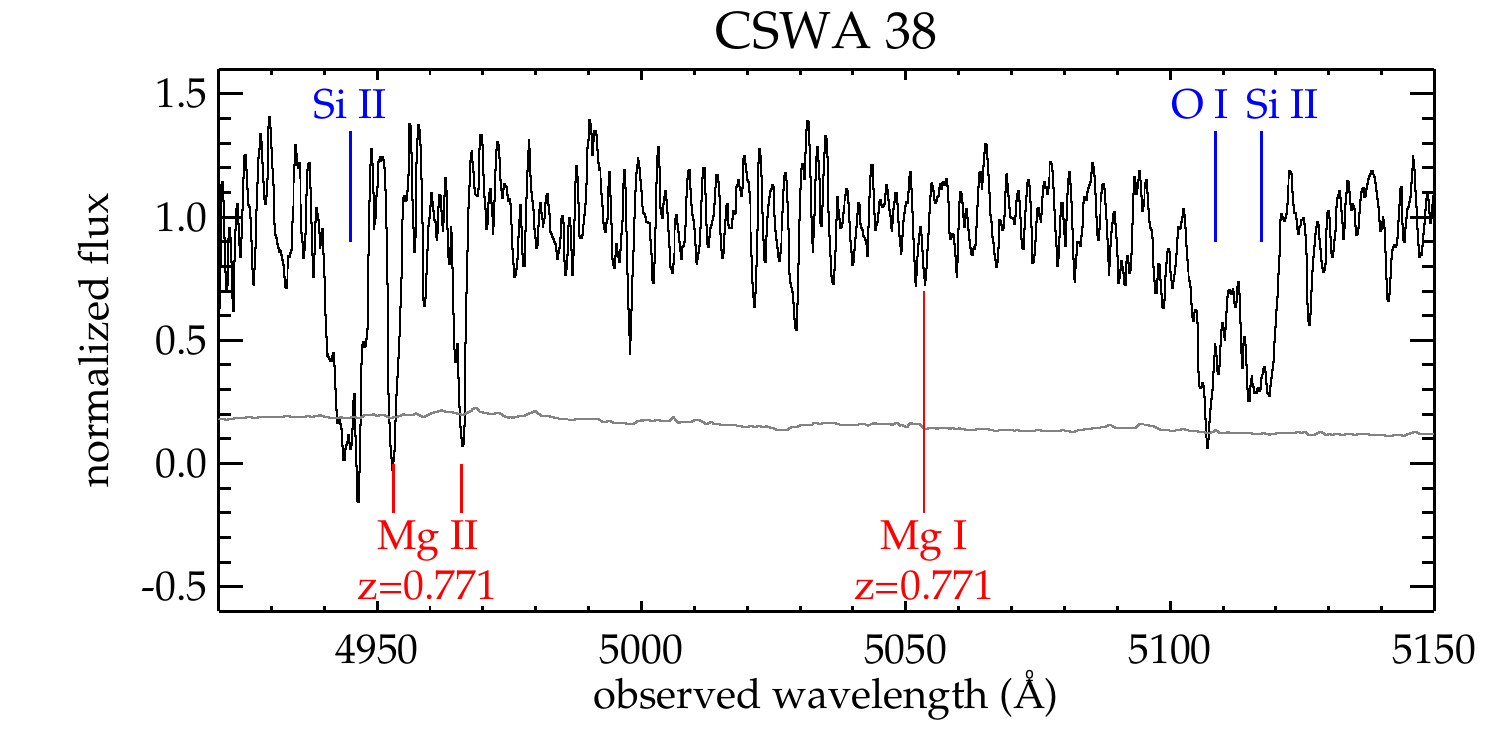}
\caption{
\label{fig:intervening}
Intervening absorption systems in the ESI spectroscopic sample.
Absorption lines intrinsic to the galaxy's interstellar and circumgalactic medium are labeled above in blue, with intervening absorption lines labeled below in red. Error spectra are shown in grey. 
Absorption from foreground systems can be problematic by blending with intrinsic features in the galaxy spectrum, such as \Siiva$\lambda$1402 in CSWA 128 (blended with intervening \Civa\ at $z=1.923$) and \Siiia$\lambda$1260 in CSWA 39 (blended with intervening \Civa\ at $z=2.055$). These blends are excluded from analysis.
}
\end{figure}

In addition to the ISM of our target galaxies, the spectra are sensitive to absorption arising from intervening material at lower redshift along the line-of-sight. Figure~\ref{fig:intervening} shows three such systems at $z\simeq1-2$ detected in the ESI spectra. A detailed census of their properties is beyond the scope of this paper. For our purposes we are interested in whether any absorption lines used in this work are affected by intervening systems. \Siiva$\lambda$1402 in CSWA 128 and \Siiia$\lambda$1260 in CSWA 39 are blended with strong features arising from lower redshift systems (Figure~\ref{fig:intervening}), and we therefore do not use these in subsequent analysis. We find no other cases where intervening absorbers significantly affect the results.

\section{Gas covering fractions and column densities}\label{sec:fcov}

The goals of this work demand measurements of column density from a variety of ions. Given the extended nature of the continuum source, emergent absorption line profiles may arise from a blend of components at different velocities and spatial positions. We are most interested in ions which are physically associated such that their intrinsic abundance ratios can be determined. We thus focus on the ``low ions'' which are predominantly associated with a common \Hi\ gas phase, arise in the same absorption components \citep[e.g.,][]{wolfe2000,werk2013}, and for which multiple elements are probed. Wherever possible we also seek to determine the contribution of different ionization states to total mass and elemental abundances.

We make the simplifying assumptions that at a given velocity, the medium in which absorption from a given ion arises is characterized by 
(1) uniform chemical composition, 
(2) uniform column density, 
and (3) finite spatial extent, such that a fraction $f_c$ of the stellar continuum is covered by the absorbing medium. 
Following \cite{jones2013} we then calculate ion column densities and covering fractions from measured absorption line profiles as follows. 
The residual intensity $I$ of an absorption line is given by
\begin{equation}\label{eq:fcov}
\frac{I}{I_0} = 1 - f_c (1 - e^{-\tau})
\end{equation}
where $I_0$ is the continuum level. The optical depth $\tau$ is in turn related to column density by
\begin{equation}\label{eq:N}
\tau = f \lambda \frac{\pi e^2}{m_e c} \times N = \frac{f \lambda}{3.768 \times 10^{14}} \times N
\end{equation}
where $f$ is the ion oscillator strength, $\lambda$ is the rest-frame wavelength of the transition expressed in \AA, and $N$ is the ion column density in cm$^{-2}$\,(\kms)$^{-1}$. Combining equations~\ref{eq:fcov} and \ref{eq:N} yields an expression for $I$ as a function of $f_c$ and $N$. In the following analysis we will treat all variables as functions of velocity, i.e. $N(v)$ and $f_c(v)$.

\begin{figure}
\includegraphics[width=\columnwidth]{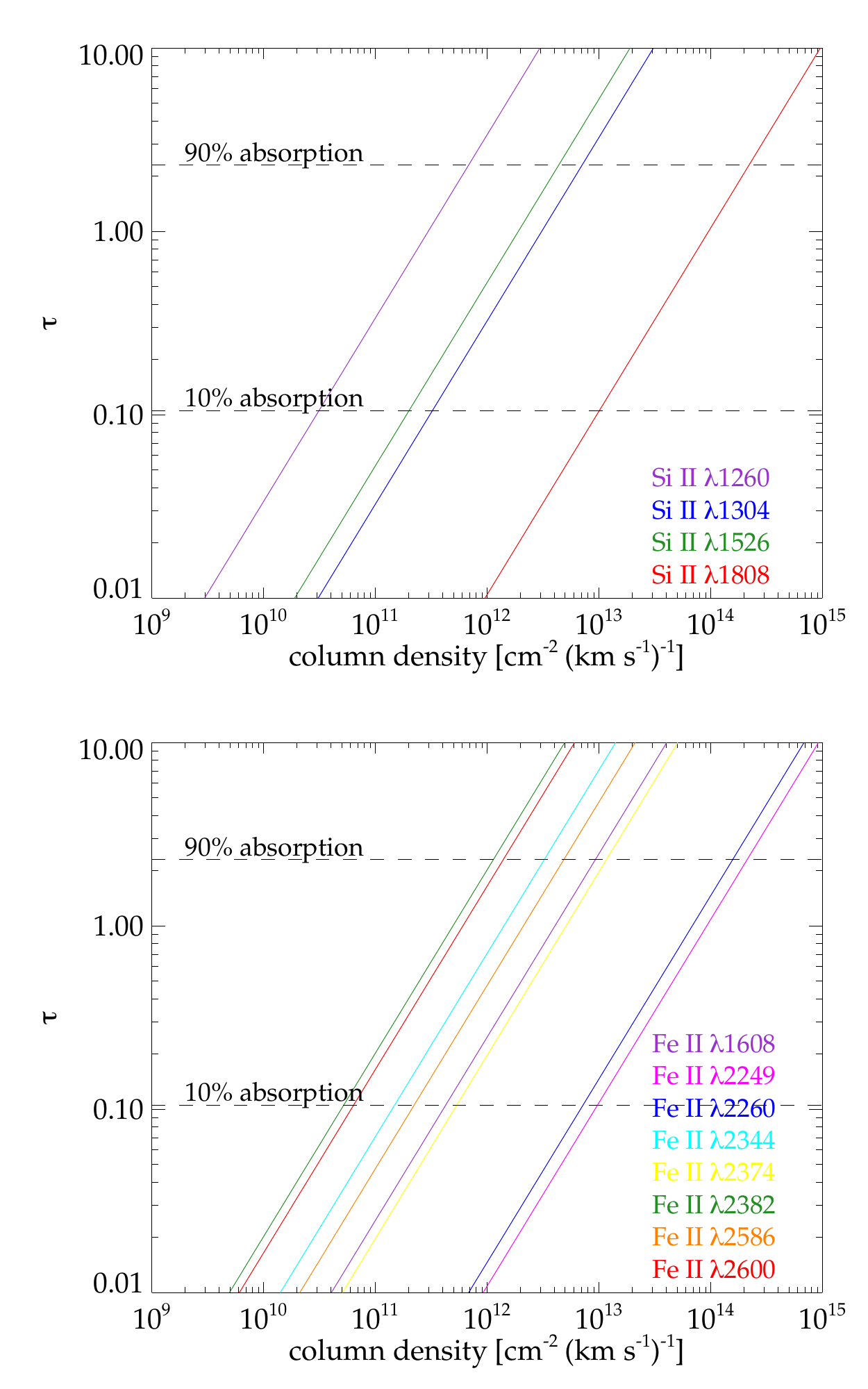}
\caption{
\label{fig:tau}
Optical depth $\tau$ as a function of column density for \Siiia\ and \Feiia\ transitions observed in the ESI spectra. The range of $f\lambda$ values enables robust measurements of column density over several orders of magnitude. For the ESI spectra, column density measurements generally require at least one optically thin line for which the absorption depth is $\lesssim$50\% of the continuum ($\tau\lesssim0.7$), combined with an optically thick transition of $\tau\gtrsim3$ to determine the covering fraction. 
}
\end{figure}

\begin{figure}
\includegraphics[width=\columnwidth]{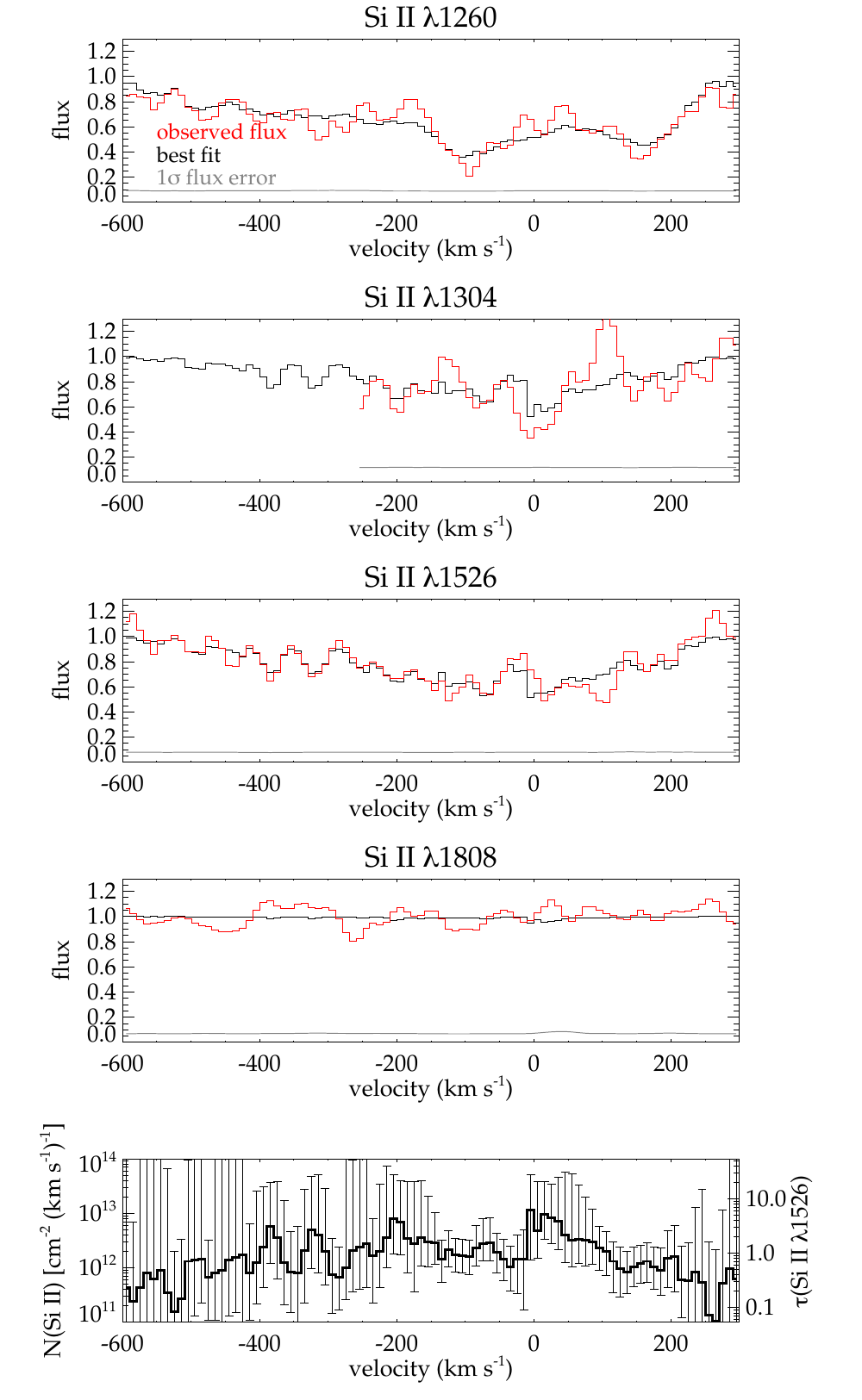}
\caption{
\label{fig:c164_si2}
Column density of \Siiia\ constrained from multiple transitions in the spectrum of CSWA 164. Four transitions of \Siiia\ with different oscillator strengths are available in the spectrum. Profiles of each transition are shown in red (excluding the region where \Siiia$\lambda$1304 is blended with \Oia$\lambda$1302). \Siiia\ column density is derived at each velocity bin from a joint fit to all transitions, with the covering fraction profile fixed to an average of saturated transitions (shown in Figure~\ref{fig:fcov_low}). The best-fit column density profile and its uncertainty are shown in the bottom panel. Line profiles corresponding to the best-fit column density and assumed covering fraction are shown as black lines in the upper panels. In this case the $\lambda$1260 transition is optically thick and therefore traces the covering fraction of low ionization gas. The data are inconsistent with a uniform covering fraction.
}
\end{figure}

In cases where the data include two or more transitions of the same ion, from the same ground state, with different values of $f\lambda$, it is possible to solve Equations~\ref{eq:fcov} and \ref{eq:N} for $N$ and $f_c$. Most spectra discussed here cover multiple transitions of \Siiia, \Feiia, \Niiia, \Aliiia, \Siiva, and \Civa. Of these, \Siiia\ and \Feiia\ typically provide the most robust results as there are several transitions covering a wide range of $f\lambda$. Figure~\ref{fig:tau} illustrates how various transitions of \Siiia\ and \Feiia\ provide excellent constraints over nearly four orders of magnitude in column density from $N \approx 5\times10^{10}$ to $2\times10^{14}$ \Nunit. An example of multiple \Siiia\ transitions in one spectrum is shown in Figure~\ref{fig:c164_si2}.

\subsection{Low ionization gas}\label{sec:low_ions}

\begin{figure}
\includegraphics[width=\columnwidth]{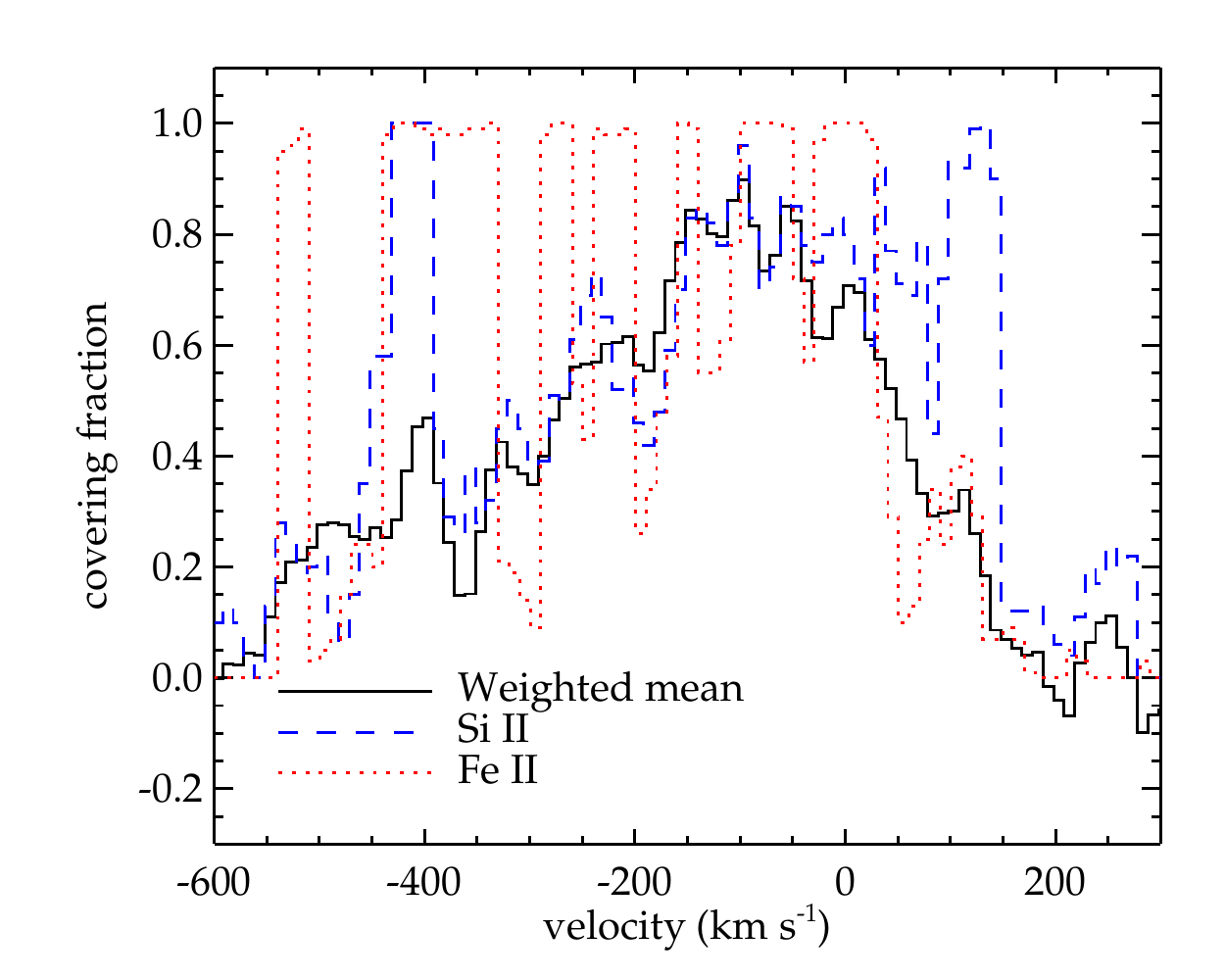}
\caption{
\label{fig:fcov_c128}
Low ion covering fraction of CSWA 128 from three different methods. Covering fractions of \Siiia\ and \Feiia\ are derived as best-fit solutions to Equations~\ref{eq:fcov} and \ref{eq:N}. The solid black line is a weighted mean of various saturated transitions as described in the text. All three methods are in good agreement within the uncertainties, which are smallest for the weighted mean method.
}
\end{figure}

This paper focuses largely on low ionization species (low ions), defined as having ionization potentials straddling 1 Rydberg and which are typically the dominant ionization state in \Hi\ gas. These include \Ciia, \Oia, \Aliia, \Siiia, \Feiia, \Niiia, and \Zniia\ relevant for this work. Low ions from different elements are assumed to be co-spatial and in many cases we can verify that \Siiia\ and \Feiia\ give consistent results for $f_c$ (e.g., Figure~\ref{fig:fcov_c128}). We note that \Siiia, \Feiia, \Niiia, and \Zniia\ in particular have nearly identical ionization potentials (within 1--2 eV) and therefore should be strongly associated. 
Wherever possible we determine $f_c$ for low ions using three different estimates: $f_c$(\Siiia), $f_c$(\Feiia), and a weighted mean of saturated absorption lines. Our methods are essentially identical to those used in \cite{jones2013} and \cite{leethochawalit2016} which we briefly summarize.
For \Siiia\ and \Feiia, we find the values of $N$ and $f_c$ that minimize the least-square residual
\begin{equation}\label{eq:chi2}
\chi^2 = \sum(I_{obs} - I_{N,f_c})^2 / \sigma_{obs}^2
\end{equation}
in each velocity bin, summing over all available transitions (e.g. Figure~\ref{fig:c164_si2}). For the weighted mean approach we approximate Equation~\ref{eq:fcov} as
\begin{equation}\label{eq:sat}
f_c = 1 - I/I_0
\end{equation}
and average the subset of \Siiia$\lambda$1260,$\lambda$1304,$\lambda$1526, \Oia$\lambda$1302, \Ciia$\lambda$1334, \Aliia$\lambda$1670, \Feiia$\lambda$2382, and \Mgii$\lambda\lambda$2796,2803 profiles which appear saturated based on visual inspection. Blended transitions are excluded (e.g., blended regions of \Oia$\lambda$1302 and \Siiia$\lambda$1304). Transitions are considered saturated if their optical depths $\tau\gtrsim3$ such that Equation~\ref{eq:sat} is a good approximation. For example, cases where \Siiia$\lambda$1260,$\lambda$1304,$\lambda$1526 profiles are identical to within $\lesssim$5\% indicate that all three are saturated (see Figure~\ref{fig:tau}). In the example shown in Figure~\ref{fig:c164_si2}, \Siiia$\lambda$1260 is saturated and used to determine $f_c$, whereas the $\lambda$1304 and $\lambda$1526 transitions have $\tau\sim1$ and are not used.

In general the three methods of deriving low ion covering fractions are in good agreement. Figure~\ref{fig:fcov_c128} shows the results for CSWA 128 as an example. The weighted mean of saturated transitions consistently has the lowest uncertainty and so we adopt this as the best measurement of $f_c$ \citep[as done in][]{jones2013,leethochawalit2016}. Another reason for using saturated transitions is that Equations~\ref{eq:fcov} and \ref{eq:N} cannot easily distinguish between cases of low $N$ and high $f_c$, or high $N$ with low $f_c$, when absorption lines are weak. This is especially evident in Figure~\ref{fig:fcov_c128}, where $f_c$(\Feiia) shows increased uncertainty (evidenced by large scatter) but with no strong systematic difference. Likewise $f_c$(\Siiia) shows large scatter where absorption is weak, and excellent agreement with saturated transitions at velocities where $f_c$ is unambiguously high ($\gtrsim0.5$). The weighted mean approach breaks the $N$--$f_c$ degeneracy by assuming low $f_c$, which is thought to be the solution in such cases \citep{pettini2002,quider2010,jones2013}.

\begin{figure*}
\includegraphics[width=\textwidth]{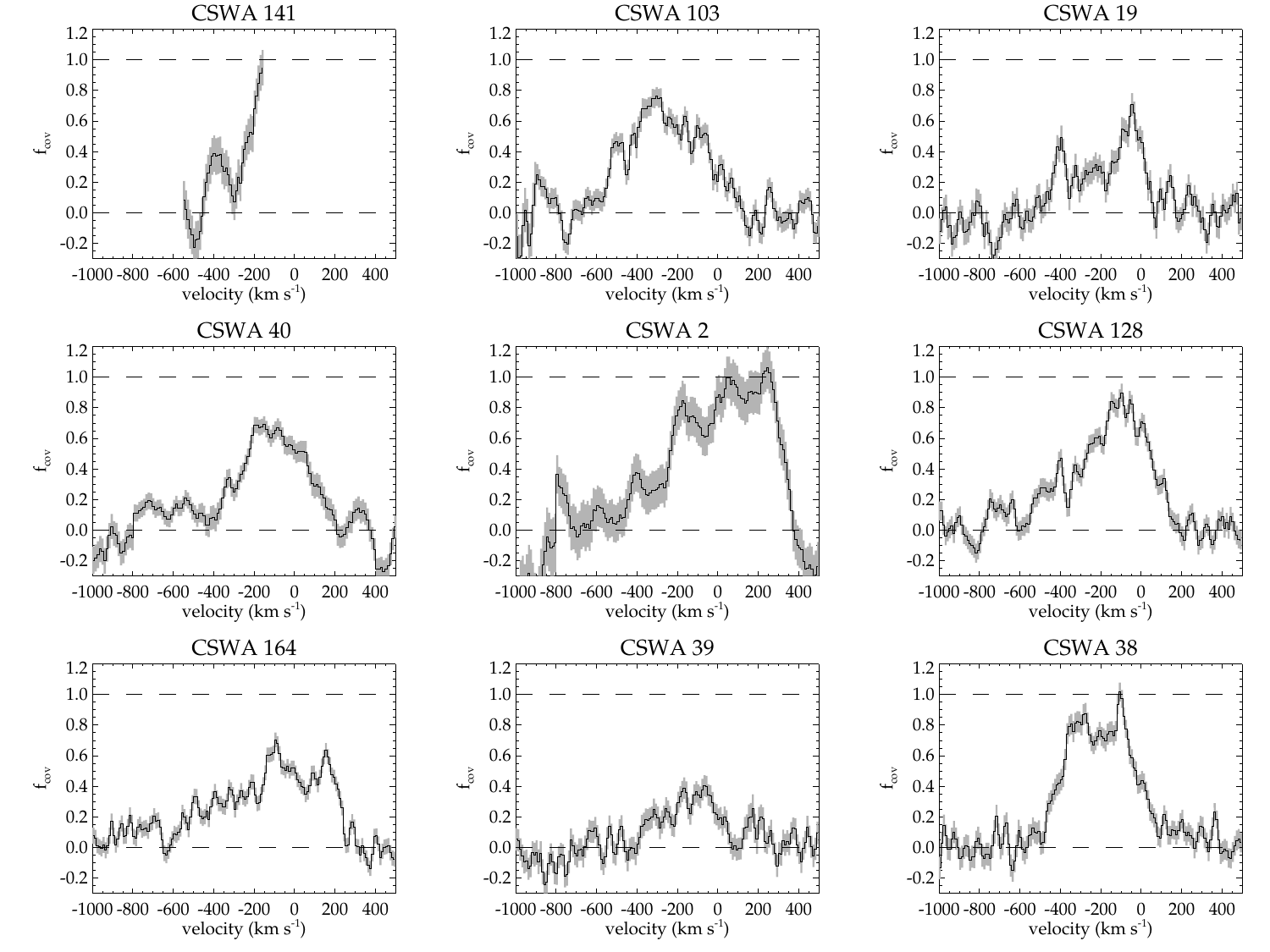}
\caption{
\label{fig:fcov_low}
Low ion covering fractions derived from the weighted mean of saturated transitions. Grey shading shows the formal 1$\sigma$ uncertainty range. The covering fraction of CSWA 141 is derived from \Mgii\ and is valid only over the region shown; other velocities are contaminated by strong emission line filling.
}
\end{figure*}

We are now in a position to summarize the covering fraction and column density measurements of low ion species that we will use throughout this work. Figure~\ref{fig:fcov_low} shows $f_c$ as a function of velocity derived from the weighted mean method. These covering fraction profiles are used to calculate low ion column densities from unsaturated absorption lines via Equations~\ref{eq:fcov} and \ref{eq:N}. Physically, the resulting $N(v)$ corresponds to a column density which covers an areal fraction $f_c(v)$ of the target galaxy, while in our formalism the remaining $1-f_c$ has zero column density. 
For this work we are interested in the aggregate galaxy-integrated column density, defined by integrating the product of $N$ and $f_c$ over velocity:
\begin{equation}\label{eq:Ntot}
N_{tot} = \int N f_c \, dv.
\end{equation}
Effectively $N_{tot}$ corresponds to the {\em average} column density that would be measured from an ensemble of point sources in the galaxy. This quantity has an additional advantage that in the limiting case of optically thin transitions $\tau \ll 1$, Equation~\ref{eq:fcov} simplifies to
\begin{equation}\label{eq:N_tau_fcov}
N \propto \tau \propto f_c^{-1}.
\end{equation}
The product $N f_c \propto N_{tot}$ is thus approximately independent of $f_c$ for weak lines such as \Siiia$\lambda$1808, \Feiia$\lambda$2249, and various \Niiia\ transitions. 
We report integrated column density measurements $N_{tot}$ and corresponding velocity ranges (i.e., bounds of the integral: $\Delta$v) in Table~\ref{tab:transitions}, for transitions most relevant for this work\footnotemark.
\footnotetext{
In some cases we are interested in strong transitions with $\tau \gg 1$ whose column densities cannot be reliably determined due to saturation. We instead report conservative lower limits corresponding to $\tau\geq2$. 
}
Velocity ranges are set to encompass the regions of high $f_c$ where absorption is well-detected. 
We do not integrate Equation~\ref{eq:Ntot} over regions of weaker absorption (i.e. $f_c \approx 0$) as this would essentially add noise, and could introduce systematic error because of the $f_c$-$N$ degeneracy. The adopted velocity ranges encompass a median 73\% of the velocity-integrated covering fraction, varying from 63--85\% for individual sources (except CSWA 141 where the fraction is likely $<$50\%). $N_{tot}$ values in Table~\ref{tab:transitions} are therefore expected to be within $\sim$0.1 dex of the total (integrated across all velocities) and we emphasize that {\it abundance ratios} are robust. 
These low ion covering fractions and abundance ratios derived via $N_{tot}$ constitute the basis of most results discussed in this work.

\subsubsection{Joint fits to multiple transitions}\label{sec:multiple_transitions}

In cases where multiple transitions of a low ion are observed in the spectra, we can jointly analyze all transitions to improve the precision of column density measurements. At each velocity we determine the best-fit $N$ via Equation~\ref{eq:chi2}, with $f_c$ fixed to the values from saturated transitions (Figure~\ref{fig:fcov_low}). Resulting $N_{tot}$ values for \Siiia, \Feiia, and \Niiia\ are given in Table~\ref{tab:transitions_mult} along with the transitions used. Here $N_{tot}$ is integrated over the same velocity range as for individual transitions, to facilitate a direct comparison. $N_{tot}$ values from different transitions of the same ion are in reasonable agreement (within 2$\sigma$ in nearly all cases) given the measurement uncertainties. In contrast, adopting $f_c=1$ would give discrepant results, confirming the non-uniform covering fractions.

Figure~\ref{fig:c164_si2} shows an example of this method applied to \Siiia\ in CSWA 164. Various \Siiia\ transitions probe a range of optical depths $\tau \ll 1$ ($\lambda$1808), $\tau \sim 1$ ($\lambda$1304, $\lambda$1526), and $\tau \gg 1$ ($\lambda$1260). The joint analysis of all transitions enables good measurements of $N(v)$. Relative to the best fit value, lower bounds are tightly constrained by strong lines with $\tau \gtrsim 1$ while upper bounds are subject to larger random uncertainty from the signal-to-noise of optically thin lines. This asymmetry is apparent in many of the uncertainties for \Siiia\ and \Feiia\ in Table~\ref{tab:transitions_mult}. The effect is less pronounced for \Niiia\ where typically all transitions have $\tau<1$.

{
Good agreement between multiple transitions of the same ion additionally confirms that the analysis does not suffer from large systematic errors. Optically thin lines such as the \Niiia\ transitions are highly sensitive to the continuum level, and their mutual consistency verifies that the continuum estimates are reliable.
}

\subsubsection{Neutral hydrogen}

\begin{figure}
\includegraphics[width=0.25\textwidth]{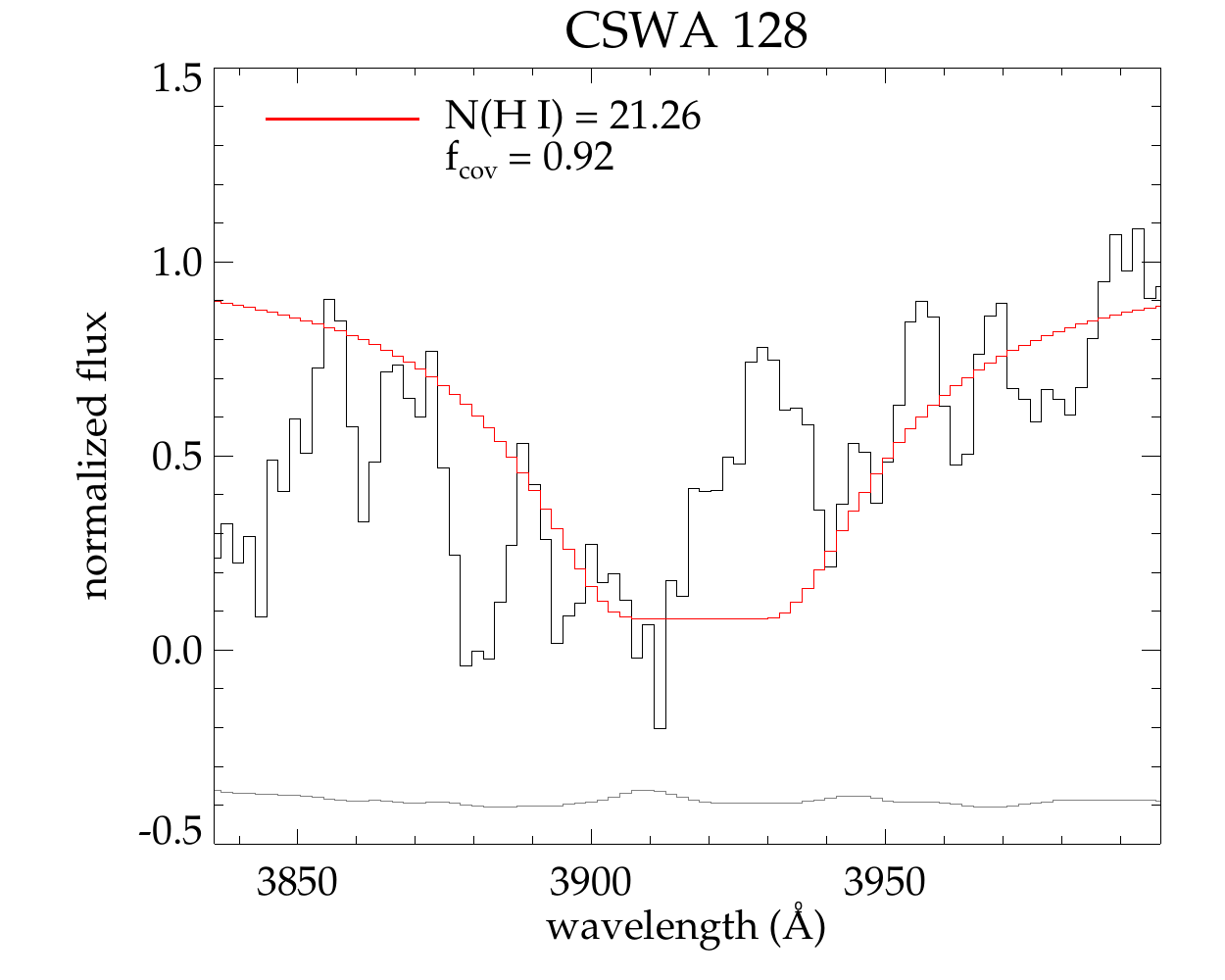}
\hspace{-0.03\textwidth}
\includegraphics[width=0.25\textwidth]{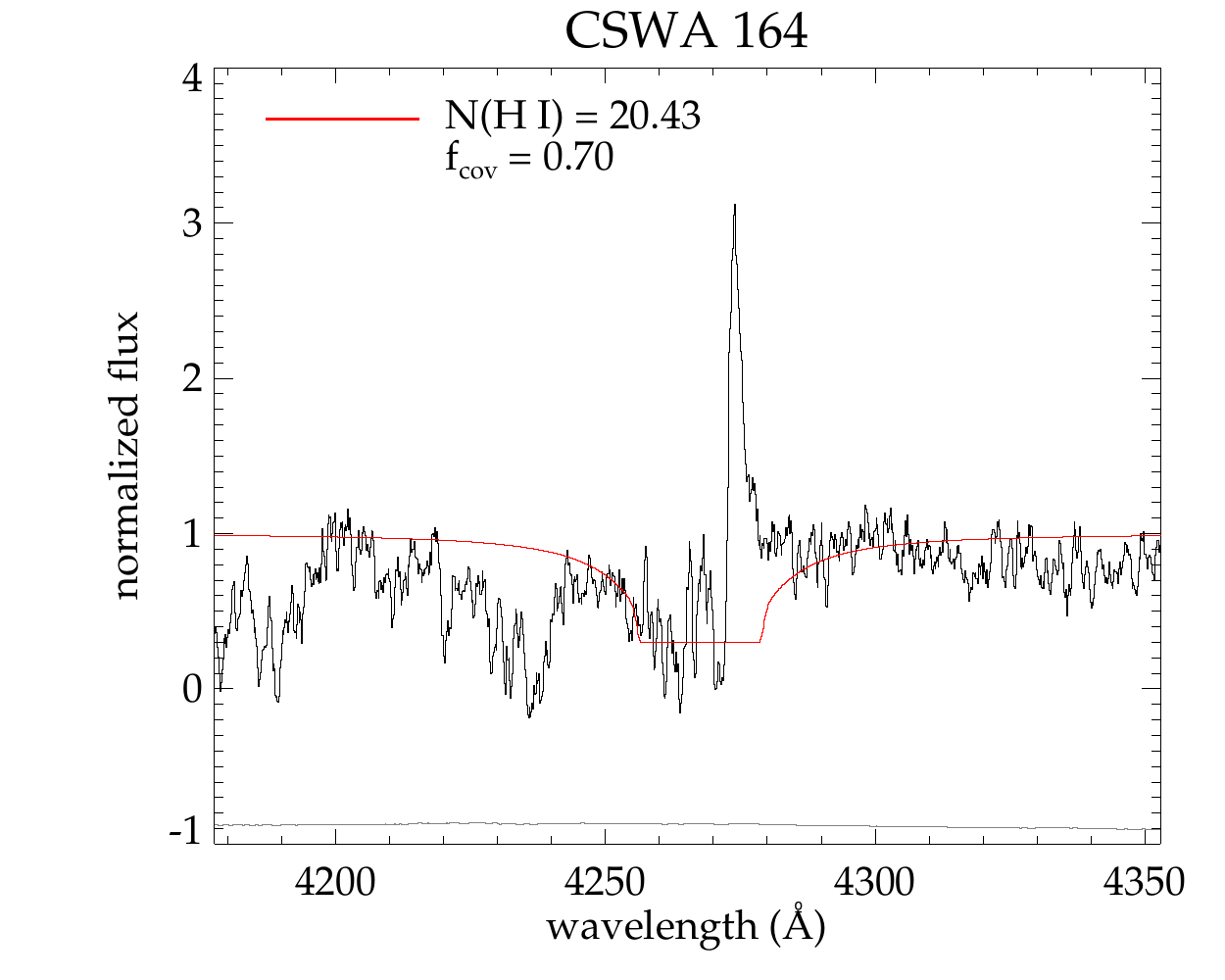}
\hspace{-0.03\textwidth}
\includegraphics[width=0.25\textwidth]{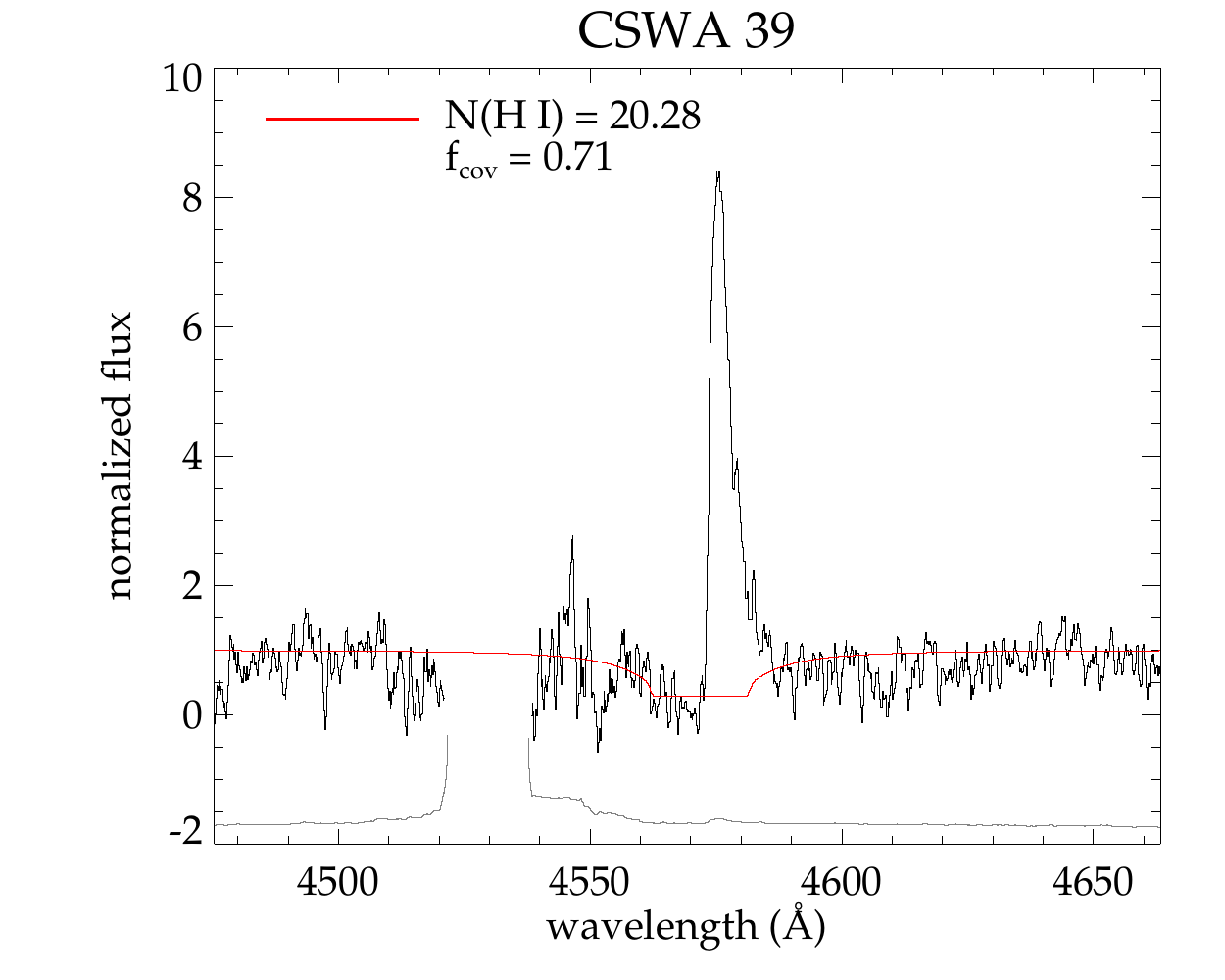}
\hspace{-0.03\textwidth}
\includegraphics[width=0.25\textwidth]{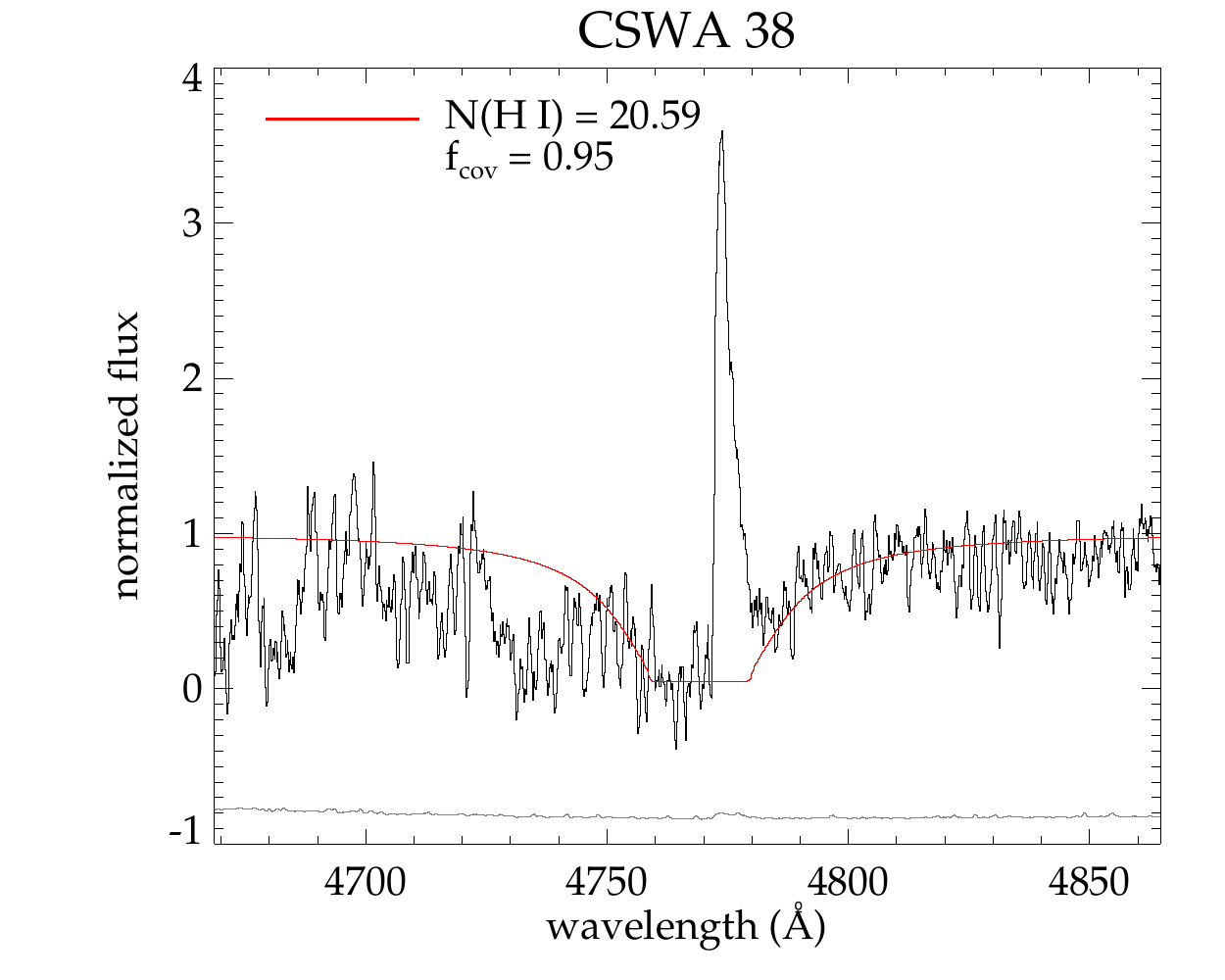}
\caption{
\label{fig:lya_fit}
Voigt profile fits to the region around \Lya\ for galaxies in our sample with suitable spectral coverage. Best fit parameters corresponding to the red lines are indicated in each panel. 1-$\sigma$ error spectra are shown in grey, offset to the bottom of each plot for clarity. 
The spectrum of CSWA 128 is from MMT/BCS \citep{stark2013} while the rest are from Keck/ESI data presented in this paper.
}
\end{figure}

\Hi\ is the dominant low ion and represents the essential reference scale for absolute abundance measurements. Three galaxies in our sample are at sufficiently high redshift ($z>2.5$) to determine \Hi\ column densities from the \Lya\ transition in Keck/ESI spectra. Additionally CSWA 128 has a high quality MMT spectrum reaching the necessary blue wavelengths \citep[described by][]{stark2013}. In these cases we fit the \Lya\ absorption with a Voigt profile. We fix the centroid and Doppler parameter to the centroid and standard deviation of the low-ionization covering fraction profiles shown in Figure~\ref{fig:fcov_low}, allowing \Hi\ column density and covering fraction to vary. Wavelengths affected by \Lya\ emission or by absorption features (e.g. \Siiiia$\lambda$1206 or the \Lya\ forest) are excluded from the fits. Resulting best-fit profiles are shown in Figure~\ref{fig:lya_fit}. As with the metal ion column densities, we define an average line-of-sight $$N_{tot} = N f_c$$ for \Hi\ and list the values in Table~\ref{tab:transitions}. We note that $N_{tot}$ has lower uncertainty than the best-fit $N(\mathrm{H~\textsc{i}})$ due to covariance between $N$ and $f_c$.

The column density $N_{tot}(\mathrm{H~\textsc{i}})$ is derived using a different method than for metal ions and may be subject to different systematic uncertainties. We have assumed that \Hi\ is coincident with low-ionization metals in fixing the kinematics of \Lya\ Voigt profile fits. Best-fit covering fractions provide a sanity check: in all cases the \Hi\ covering fractions are in agreement with $f_c$ derived from low-ionization metals, consistent with a co-spatial distribution of \Hi\ and low-ionization metal absorption. For CSWA 39 the uncertainty is large ($f_c = 0.71^{+0.29}_{-0.32}$ for \Hi); adopting $f_c\simeq0.4$ as for the low ions would increase $N_{tot}(\mathrm{H~\textsc{i}})$ by 0.35 dex, within the uncertainty quoted in Table~\ref{tab:transitions}.

\subsection{Neutral metal transitions}

Most elements that we consider in this work have first ionization potentials which are below 1 Rydberg, such that both the neutral and first ions may be present in \Hi\ gas. For several elements we have access to transitions from both the neutral and singly ionized states (notably Si, Fe, and in the case of CSWA 141 also Mg, with first ionization potentials $\sim$8 eV or 0.6 Rydberg). In general these neutral ions are not detected and their covering fractions are not independently measured. Nonetheless it is informative to constrain their column densities relative to the first ions. We therefore calculate column densities for several neutral ion transitions assuming that they are co-spatial with the low ions (i.e., assuming $f_c$ is identical to that derived in Section~\ref{sec:low_ions} and shown in Figure~\ref{fig:fcov_low}). The results are given in Table~\ref{tab:transitions}. In all cases we find that neutral Mg, Si, and Fe are insignificant compared to their singly ionized abundances. The sample median is $N(\mathrm{Si~\textsc{i}})/N(\mathrm{Si~\textsc{ii}}) = 0.003 \pm 0.003$ with $N(\mathrm{Si~\textsc{i}})/N(\mathrm{Si~\textsc{ii}}) < 0.03$ in all individual cases. We conclude that these neutral species are negligible.

\subsection{Higher ionization states}\label{sec:high_ion}

We now turn briefly to the \Hii\ gas traced by second and higher metal ions. A key question for this work is whether a substantial fraction of the low ion column density arises in an ionized (\Hii) gas phase, in which case relative abundances of low ion species may require ionization corrections. The most useful probes of higher ionization states in the ESI spectra are \Aliiia$\lambda\lambda$1854,1862 and \Siiva$\lambda\lambda$1393,1402 which can be compared with the low ions \Aliia\ and \Siiia. Both \Aliiia\ and \Siiva\ doublets have a factor of $\simeq$2 difference in oscillator strength between the two transitions, such that we can jointly fit the column density and covering fraction using the formalism of Equation~\ref{eq:chi2}. Covering fractions are valuable to test whether ionized gas is physically associated with the low ions (i.e. having similar $f_c$(v)), while column densities inform whether the ionized gas may contribute significantly to the low ions as well as the total metal content.

\begin{figure}
\includegraphics[width=\columnwidth]{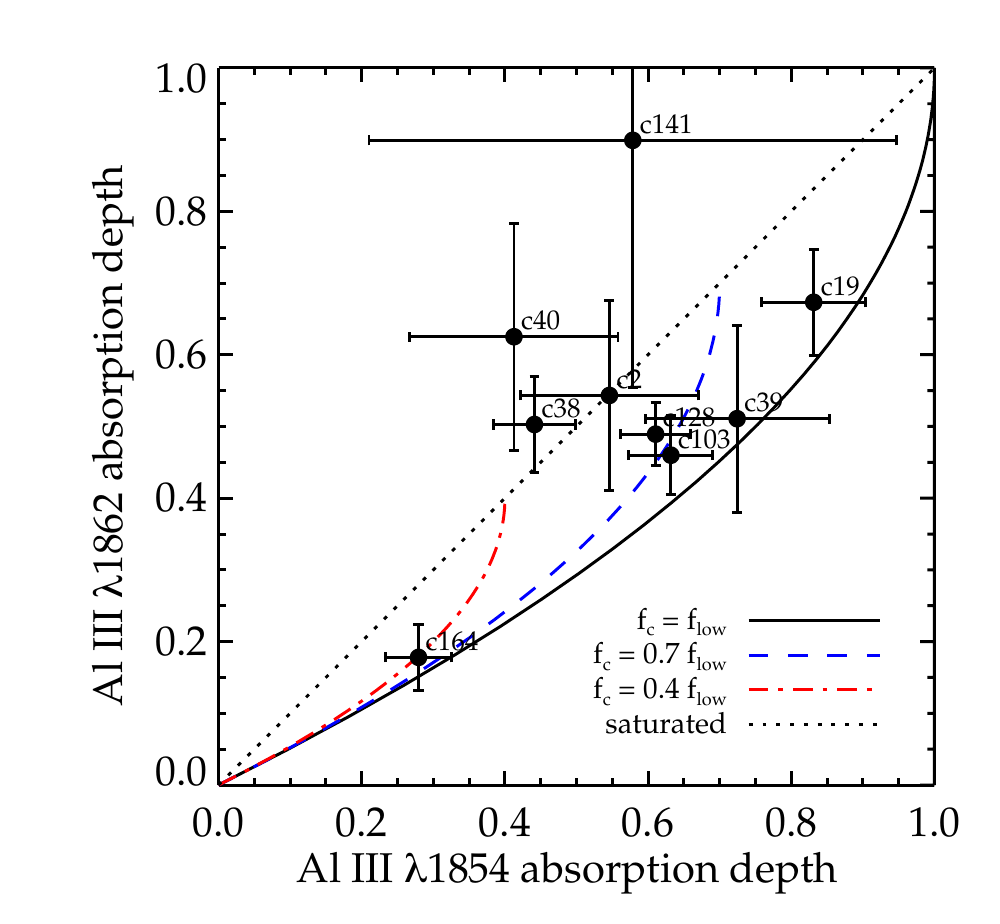}
\caption{
\label{fig:arcs_al3}
Mean depth of absorption in \Aliiia$\lambda\lambda$1854,1862 relative to the low ion covering fraction profile ($f_{low}$), calculated over the velocity ranges given in Table~\ref{tab:transitions} and labeled by CSWA catalog number. The solid black line is the locus expected for the case of identical \Aliiia\ and low ion covering fractions. All data points lie above this line, indicating that $f_c$(\Aliiia) is smaller than $f_{low}$. We note that in CSWA 38, \Aliiia$\lambda$1854 is affected by a sky line residual and we consider this measurement to be unreliable.
Dashed lines show example loci for different covering fractions and for the case of saturated \Aliiia. The data suggest a range $f_c$(\Aliiia)~$\simeq 0.5-1$ relative to $f_{low}$.
}
\end{figure}

Covering fraction profiles of \Aliiia\ and \Siiva\ are generally similar to those of the low ions. However we detect systematic differences in $f_c$ within the aggregate sample. \Siiva\ covering fractions are on average higher than for the low ions, and this difference is statistically significant within individual velocity channels in some cases (notably for CSWA 19 and CSWA 39). 
In contrast, \Aliiia\ covering fractions are systematically {\em smaller} than for the low ions. To quantify this difference, we measure column densities $N_{tot}$(\Aliiia) assuming the low ion $f_c$ and list the results in Table~\ref{tab:transitions}. $N_{tot}$ is higher for the weaker transition $\lambda$1862 in all six cases with reliable measurements (though often at $\lesssim 1\sigma$ significance). This difference indicates a smaller $f_c$(\Aliiia) than for the low ions. As a further test, we measure the average ratio of absorption in \Aliiia\ to the saturated low ion profiles (i.e., averaged over velocity). This provides a good consistency check with better precision than $N_{tot}$, and we recover the same values of $N_{tot}(\lambda1854) / N_{tot}(\lambda1862)$ to better than 0.1 dex. Figure~\ref{fig:arcs_al3} demonstrates that all nine galaxies in the sample have mean ratios indicating that $f_c$(\Aliiia) is comparable or smaller than for the low ions by a factor of $\sim1-2\times$. As a result, values of $N_{tot}$ derived with the low ion covering fraction are systematically underestimated. The differences may also arise from unresolved saturated components, in which case $N_{tot}$ is likewise underestimated. We apply an empirical correction by scaling $f_c$(low ion) by a multiplicative factor, such that both \Aliiia\ transitions give the same column density. For example we find $f_c$(\Aliiia)~$\simeq 0.65$~$f_c$(low ion) for the case of CSWA 128, as can be seen in Figure~\ref{fig:arcs_al3}. Resulting values of $N_{tot}$ are given in Table~\ref{tab:transitions_mult} for all galaxies with reliable constraints.

These differences in covering fraction are explained by separate geometric distributions for the low, intermediate (e.g. \Aliiia), and high (e.g. \Siiva) ions. Composite galaxy spectra show that low ions are highly correlated with \Hi\ \Lya, and essentially uncorrelated with the high ions \Siiva\ and \Civa\ \citep{shapley2003,jones2012,du2018}. 
Our covering fraction analysis indicates that the lack of correlation is due to different ionization stages occupying distinct spatial regions, as opposed to variation in the relative column density of low and high ions. 
This conclusion is further reinforced by high resolution spectra of quasar absorption systems, in which high ion absorption components typically have negligible low ion content \citep[e.g.,][]{wolfe2000}. On the other hand \Aliiia\ absorption is often associated with low ions. 
We infer that up to $\sim$60\% (i.e., the typical ratio of \Aliiia\ to low ion $f_c$) of the low ion absorption in our galaxy sample may be associated with moderately ionized gas traced by \Aliiia. The higher covering fraction of \Siiva\ implies a larger volume filling factor for highly ionized gas clouds compared to \Hi.

\section{Gas kinematics}
\label{sec:kinematics}

The kinematics of ultraviolet line profiles in high redshift galaxies are well known to reflect widespread outflows in addition to a truly interstellar systemic component \citep[e.g.,][]{pettini2000, shapley2003, steidel2010, jones2012, law2012}. The vast majority of these studies rely predominantly on the strongest interstellar features which are typically saturated, and therefore trace the covering fraction (i.e., geometry) of the gas rather than the mass. Here we use optically thin lines to quantify kinematic properties of our sample with respect to the {\em mass distribution} traced by column density profiles.

\subsection{The majority of detected gas is outflowing}

For the purposes of this section, we examine column density profiles derived from individual transitions which are unsaturated, well-detected, and free of contamination. Two to three of the best such transitions in each spectrum are averaged to construct a mean low ion column density profile. Table~\ref{tab:kinematics} summarizes the mean velocity of the column density profile $\bar{v}$, the dispersion $\sigma_v$ (corrected for spectral resolution), and the transitions used. CSWA 141 is excluded from this analysis due to the limited velocity range of our measurements.

The negative mean velocities in Table~\ref{tab:kinematics} indicate a net outflow of the gas. Here we attempt to separate the absorption into outflowing and systemic (at rest with respect to the stars) components in order to determine their relative dominance. 
We define the net outflowing column density as
\begin{equation}\label{eq:N_out}
N_{out} = \int_{-\infty}^0 N(v) \, dv - \int_0^{\infty} N(v) \, dv,
\end{equation}
i.e., the difference between the integrated column density below and above the systemic velocity. In practice we integrate over the velocity range shown in Figure~\ref{fig:fcov_low}, again excluding CSWA 141 due to the limited velocity coverage. $N_{out}$ comprises a total fraction $f_{out}$ of the total column density. We find a sample mean $f_{out} = 0.81 \pm 0.19$ for the low ion column density, indicating that the majority of the detected gas is outflowing. This number strictly applies only to the heavy elements contained in low ionization gas, and not to the molecular nor very hot ($T \gtrsim 10^6$ K) phases nor unenriched gas which are not probed by these data. Nonetheless it is clear that outflows dominate the low ionization gas phase traced by near-UV absorption. 
Assuming that interstellar gas comprises the remaining $\sim$20\%, this implies that the outflowing mass in low ionization metals is larger by a factor 
\begin{equation}\label{eq:M_out}
\frac{M_{out}}{M_{ISM}} \approx 4 \left( \frac{R_{out}}{R_{ISM}} \right)^2,
\end{equation}
where $R_{out}$ and $R_{ISM}$ are the mean galactocentric distances of the outflowing and interstellar components. We note that this factor could be very large if the outflowing gas is located far from its origin galaxy.

Finally, we compare results from the column density profiles with an equivalent analysis of the covering fraction from saturated low ion absorption lines. The covering fraction profiles give a sample mean $f_{out} = 0.53 \pm 0.10$, smaller than but consistent with the column density results. We reiterate that column density is a better tracer of the gas {\em mass} than the profiles of saturated transitions.

\subsection{Saturated line profiles overestimate the bulk velocity dispersion}

\begin{figure}
\includegraphics[width=0.52\columnwidth]{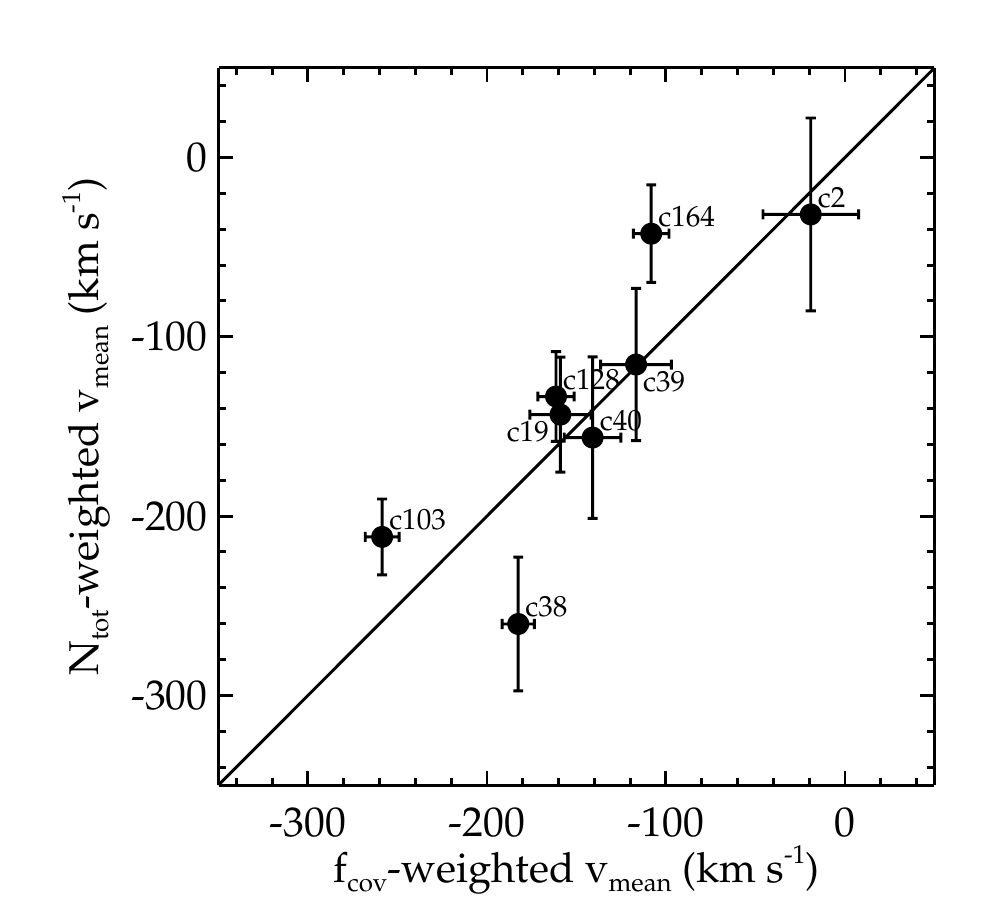}
\hspace{-0.06\columnwidth}
\includegraphics[width=0.52\columnwidth]{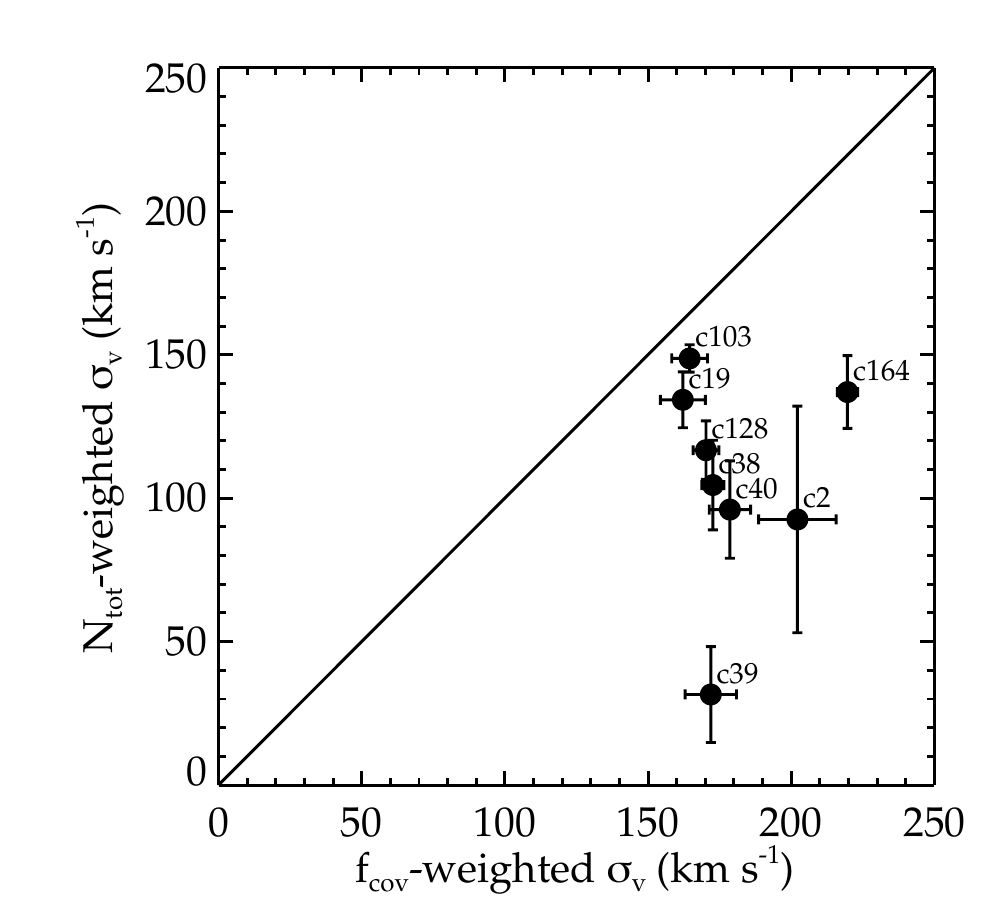}
\caption{
\label{fig:kinematics_lowion}
Comparison of low ion gas kinematics weighted by the $f_c$ and $N_{tot}$ profiles: mean velocity ({\it left}) and velocity dispersion ({\it right}). $f_c$ weighting reflects the geometric distribution of gas traced by the strongest interstellar features (typically saturated), while $N_{tot}$ weighting reflects the column density and hence mass distribution determined from weaker optically thin transitions. Mean bulk velocities are consistent for both the $f_c$ and $N_{tot}$ weighting. Velocity dispersions are systematically $\sim$75 \kms\ smaller for the $N_{tot}$ weighting (median), indicating that the mass is concentrated over a smaller range of velocity than indicated by saturated strong absorption lines.
}
\end{figure}

We now examine whether strong absorption line profiles are representative of the underlying column density distribution. For a direct comparison, we calculate $\bar{v}$ and $\sigma_v$ from the low ion covering fraction profiles shown in Figure~\ref{fig:fcov_low} (i.e., saturated line profiles) using identical methods as for the column density. Results from both distributions are given in Table~\ref{tab:kinematics}, and are compared in Figure~\ref{fig:kinematics_lowion}. 
The mean velocities from both methods are in good agreement demonstrating that strong ISM lines accurately trace the average bulk velocity of the gas. However, velocity dispersions are systematically lower for the column density profiles compared to the geometric covering fraction. This is true for every individual galaxy in the sample, with statistical significance ranging from 2--7$\sigma$ per galaxy. The median velocity dispersion of covering fraction profiles is higher by 63\% or 75 \kms\ than for the column density profiles.

The difference in velocity dispersion between the two profiles arises naturally, since saturated lines trace the full gas velocity range rather than the narrower peaks of column density (or equivalently, optical depth). Mean velocities may also be different and we expect some scatter between $\bar{v}$ derived from the two profiles, although we find good agreement in general. The uncertainties in Table~\ref{tab:kinematics} and Figure~\ref{fig:kinematics_lowion} indicate a low intrinsic scatter of $\sim10$ \kms. We conclude that kinematic studies based on saturated strong ISM lines accurately recover the mean bulk velocity of the gas as traced by its mass distribution, but significantly overestimate the velocity spread by $\sim$60\% for this sample. This leads directly to our finding above that saturated line profiles underestimate the fraction of absorbing material associated with net outflow.

\subsection{{\rm \Aliiia} kinematics are comparable to the low ions}
\label{sec:kinematics_al3}

It is useful to compare the kinematics of low ions with \Aliiia, to further check whether a substantial fraction of the low ion column density may be associated with a moderately ionized \Hii\ phase traced by \Aliiia. 
We determine the mean velocity $\bar{v}$ and dispersion $\sigma_v$ of the \Aliiia\ column density profile for each arc using the same method as for the low ions. Since \Aliiia\ covering fractions are poorly constrained, we calculate $N_{tot}$(\Aliiia) using the low ion covering fraction profiles with the caveat that this systematically underestimates $N_{tot}$ (as discussed in Section~\ref{sec:high_ion}). While this affects the numerical values, any qualitative differences between \Aliiia\ and the low ions should be preserved.

\begin{figure}
\includegraphics[width=0.52\columnwidth]{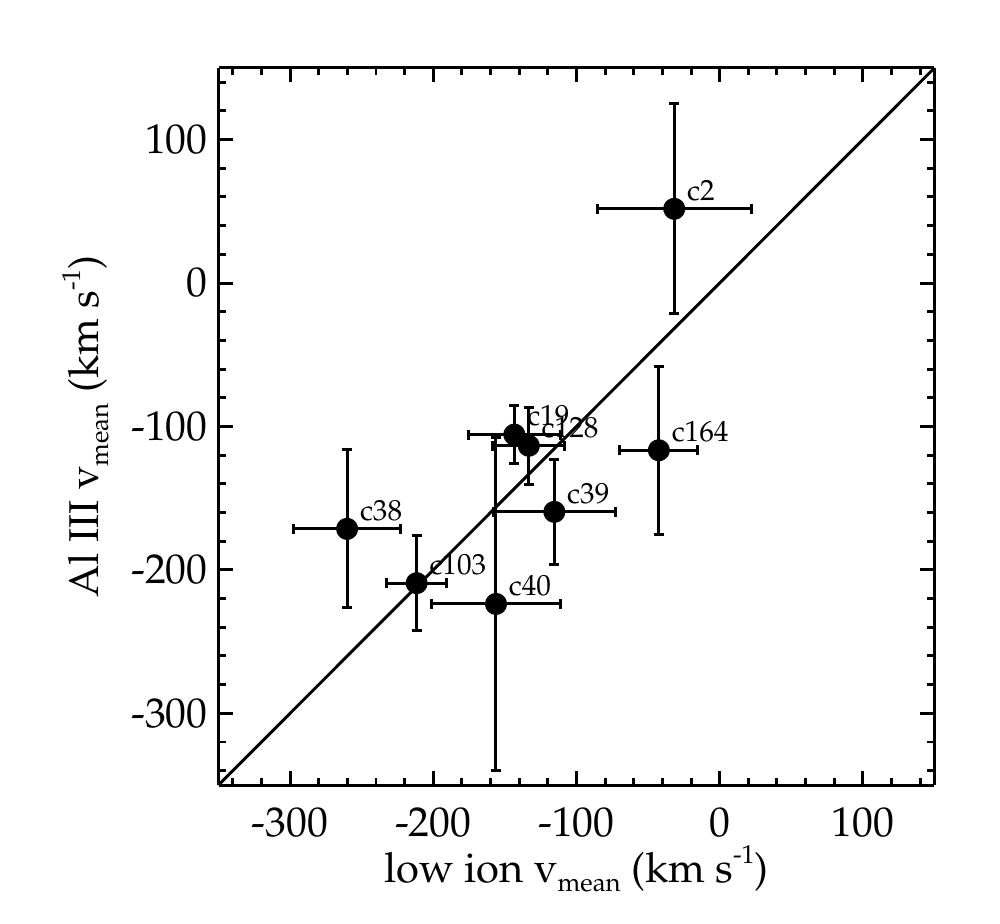}
\hspace{-0.06\columnwidth}
\includegraphics[width=0.52\columnwidth]{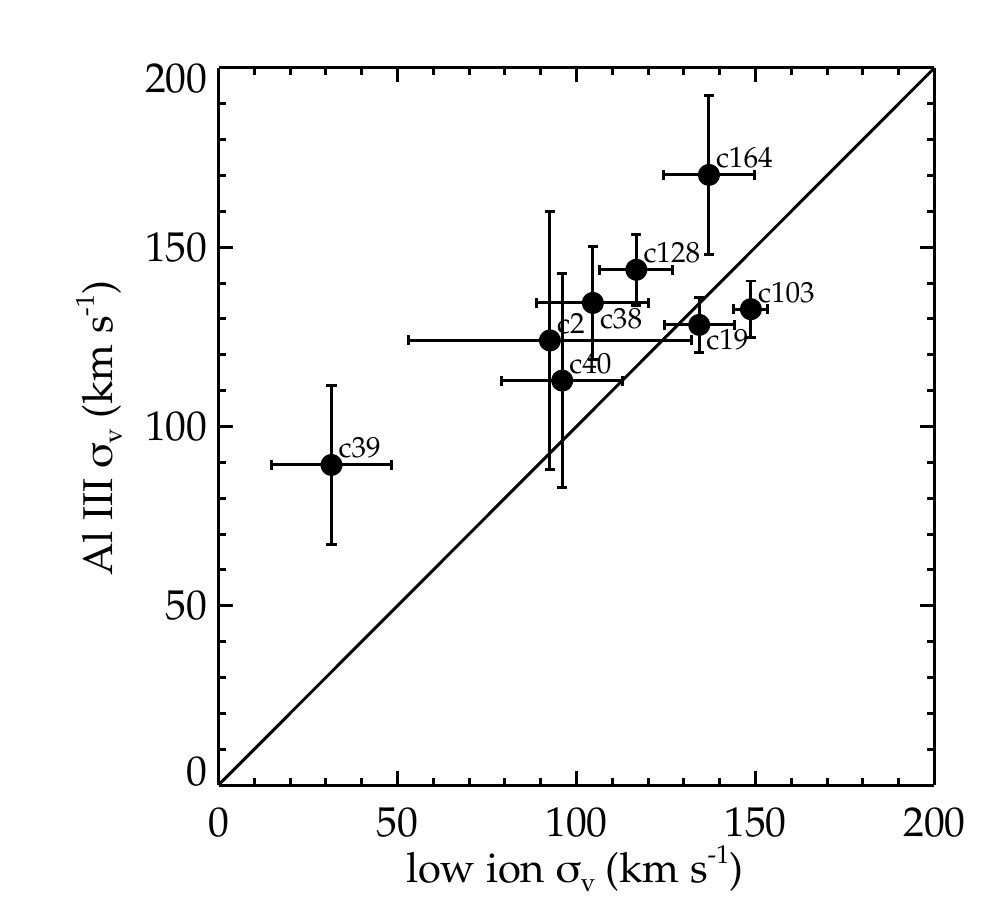}
\caption{
\label{fig:kinematics_al3}
Comparison of \Aliiia\ and low ion gas kinematics: $N_{tot}$-weighted mean velocity ({\it left}) and velocity dispersion ({\it right}). Mean bulk velocities are consistent for both the low ions and \Aliiia. Velocity dispersions for \Aliiia\ are on average higher than the low ions, but this difference is not highly significant.
}
\end{figure}

The results are summarized in Table~\ref{tab:kinematics} and compared with low ion kinematics in Figure~\ref{fig:kinematics_al3}. Mean velocities are in good agreement and their uncertainties are consistent with zero intrinsic scatter between $\bar{v}$ for the low ions and \Aliiia. Velocity dispersions are correlated and higher on average for \Aliiia, with a sample mean difference of $22\pm8$ \kms\ compared to the low ions. This is in qualitative agreement with a study of DLA kinematics by \cite{wolfe2000}, who found a significant correlation between velocity widths $\Delta v_{\textrm{\Aliiia}}$ and $\Delta v_{\textrm{low ion}}$ with the majority of systems having larger $\Delta v_{\textrm{\Aliiia}}$. Given the marginal $\sim 3\sigma$ significance of the difference in our data, we conclude that the kinematics of \Aliiia\ and low ions are generally similar with tentative evidence of a broader velocity range for \Aliiia.

\subsection{Velocity widths correspond to extreme DLAs}
\label{sec:kinematics_dlas}

\begin{figure}
\includegraphics[width=\columnwidth]{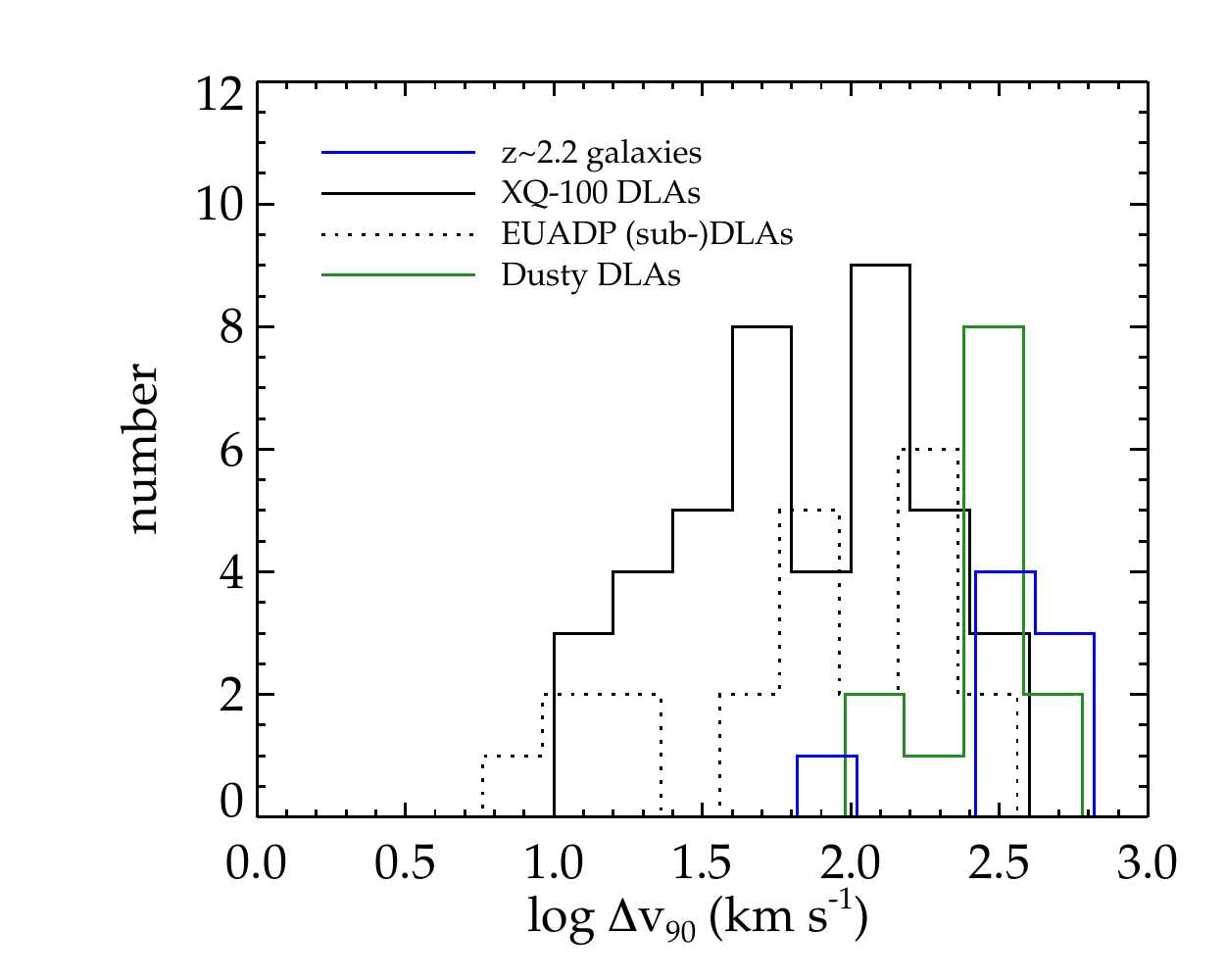}
\caption{
\label{fig:kinematics_dla}
Velocity width distributions of the ESI galaxy sample compared to DLA absorption systems in quasar spectra. We show histograms of low ion $\Delta V_{90}$ from the XQ-100 survey \citep{berg2016}, EUADP sample \citep{quiret2016}, and dusty DLAs selected on the basis of 2175 \AA\ dust attenuation features \citep{ma2017}. Galaxies in our sample have broader velocity distributions than typical DLAs, lying at or beyond the extreme tail of the XQ-100 and EUADP samples. The dusty DLAs have comparable velocity widths. 
We estimate $\Delta V_{90} = 3.29\sigma_v$ for the galaxy sample, noting that redshifted outflowing gas on the far side of the galaxies would further increase $\Delta V_{90}$ by up to 0.3 dex, exacerbating the discrepancy with typical DLAs.
}
\end{figure}

The velocity width of low ion absorption can be directly compared with { (sub-)}DLA systems at similar redshift. Since \Hi\ column densities of our sample are characteristic of DLAs, we expect similar systems tracing galaxy outflows to be present in quasar absorber samples (though they may be rare). 
Previous studies have quantified the velocity width using the $\Delta V_{90}$ statistic, defined as the velocity range encompassing the 5--95 percentile of total optical depth (i.e., column density; \citealt{prochaska1997}). Although $\Delta V_{90}$ is not well determined for the galaxies studied here, we can estimate $\Delta V_{90} \approx 3.29 \sigma_v$ as appropriate for a Gaussian column density profile. This yields a mean and median $\log{\Delta V_{90}} \,\textrm{(\kms)} = 2.55$ for the $N_{tot}$-weighted $\sigma_v$. 
Figure~\ref{fig:kinematics_dla} demonstrates that these widths are at the extreme upper end of { (sub-)}DLA values, well above typical DLAs and comparable to the most metal-rich and dust-depleted systems \citep[with metallicities $\gtrsim 0.1$ solar;][]{ledoux2006,quiret2016,ma2017}. Conservatively adopting $\Delta V_{90} > 2 \sigma_v$ implies typical $\log{\Delta V_{90}} > 2.3$, still at the extreme end of typical DLA values ($\log{\Delta V_{90}} \simeq 1.0$--2.5) and coincident with the most metal-rich systems.

Large velocity widths in the galaxy sample reflect a combination of interstellar gas and kinematically dominant outflows, whereas typical DLAs apparently do not span such a broad range of motion. Furthermore the galaxy spectra sample only one side of the outflow (with negative apparent velocities) and are missing the positive velocity component, hence we expect that the total gas velocity widths are {\em underestimated} for the galaxies. The { (sub-)}DLA population with similarly broad kinematics may likewise be tracing powerful galactic outflows. If so, the high level of enrichment and depletion seen in these extreme systems likely reflects entrained host galaxy ISM.

\section{Chemical abundances}
\label{sec:abundances}

Our column density measurements for multiple elements provide information on chemical abundance patterns, significantly increasing the number of such measurements available for galaxies at $z\gtrsim2$. Ultimately we seek to characterize the abundance patterns, chemical enrichment histories, and the likely descendant stellar populations at $z\simeq0$ based on chemical abundance tagging. 

This section focuses on the low ions which provide relative abundances of multiple elements arising from the same (co-spatial) gas. The following analysis is based largely on \Siiia, \Feiia, and \Niiia\ column densities which are reliably measured for most individual objects. Additional ions in Table~\ref{tab:transitions} provide further constraints, notably \Zniia\ and \Aliia. 
Our approach is as follows. First we discuss the low ion abundance patterns and compare with other classes of objects studied in the literature. Subsequently we assess possible systematic differences between low ion and total abundance ratios due to ionization, finite resolution, and dust depletion. We conclude with inferences on the gas phase abundances, as well as total abundances accounting for depletion.

\subsection{Low ion ratios: a distinct abundance pattern}
\label{sec:abundances_low}

\begin{figure*}
\includegraphics[width=0.33\textwidth]{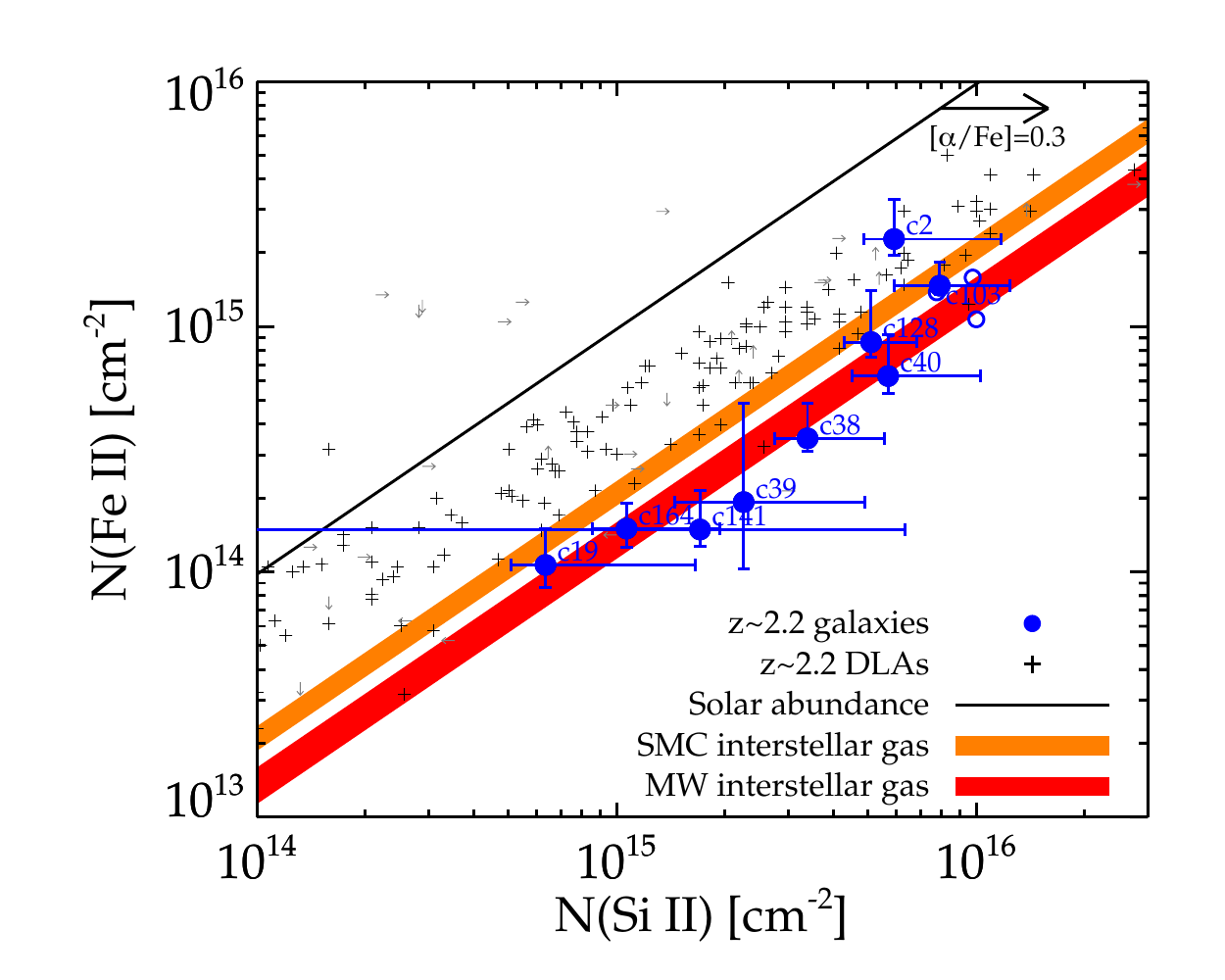}
\includegraphics[width=0.33\textwidth]{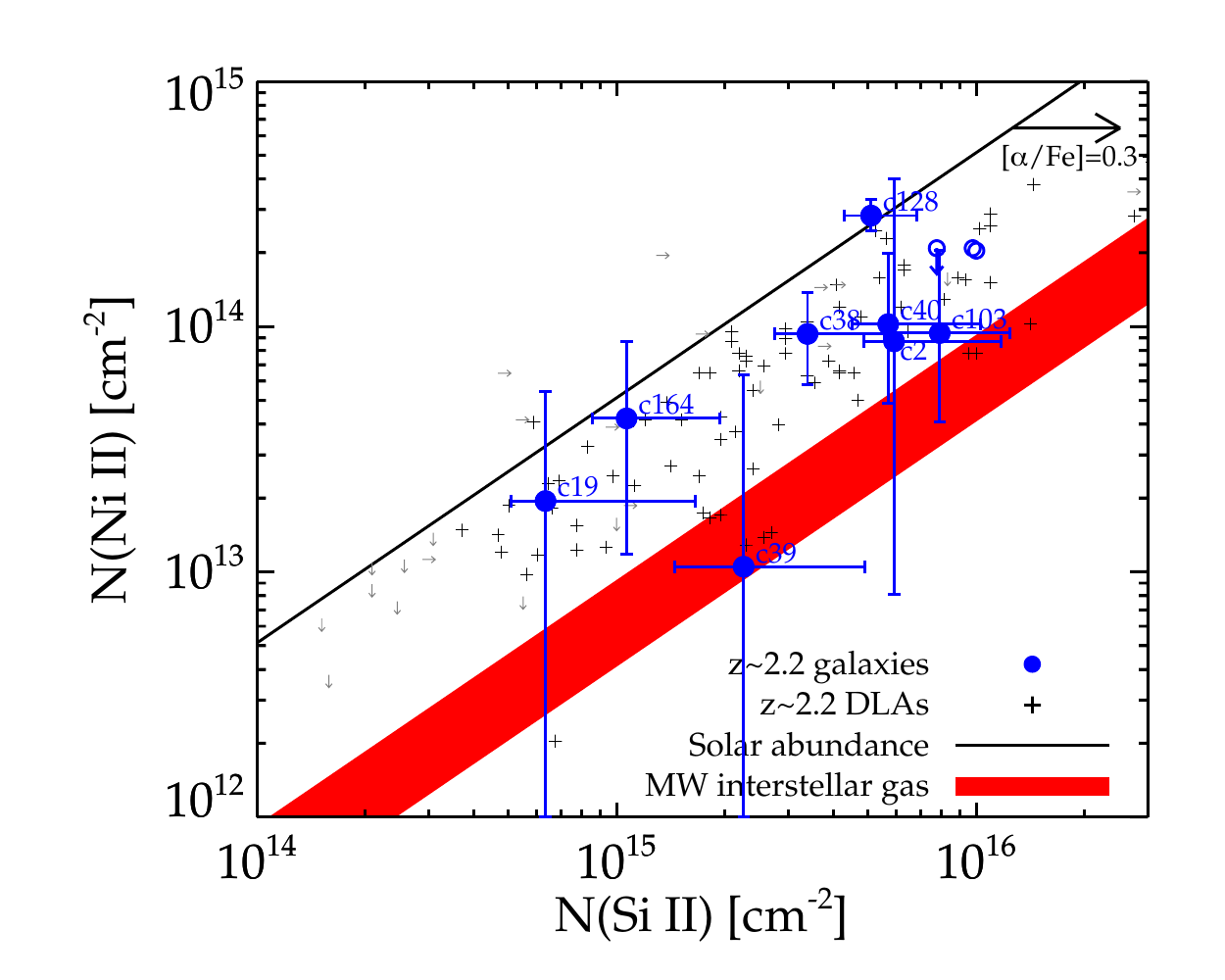}
\includegraphics[width=0.33\textwidth]{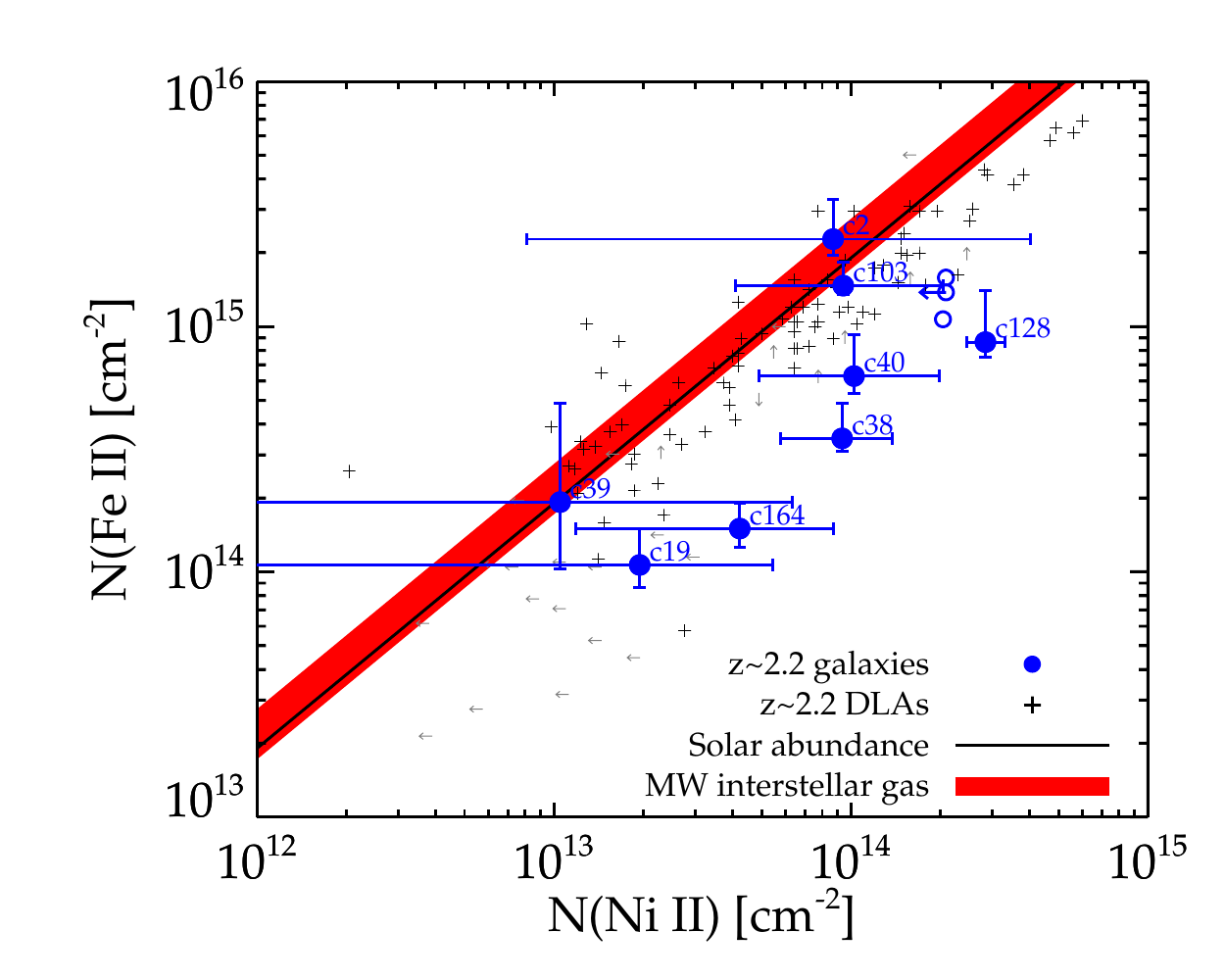}
\caption{
\label{fig:abundances_low}
Column densities $N_{tot}$ of the low ions \Siiia, \Feiia, and \Niiia\ for the ESI lensed galaxy sample, { (sub-)}DLAs drawn from the compilation of \citet[][with gray arrows showing cases with upper or lower limits]{quiret2016}, ISM sightlines in the Milky Way and Small Magellanic Cloud \citep{jenkins2009,tchernyshyov2015}, and solar abundance ratios. Literature comparison samples are discussed further in the text. The galaxy data do not match any of the comparison samples: Ni is overabundant compared to ISM sightlines, while Fe is underabundant compared to DLAs and $\alpha$-enhanced stars (denoted by arrows showing the effect of +0.3 dex enhancement in the $\alpha$-element Si). However, the galaxy data are in good agreement with previous measurements of low ion column densities in lensed star forming galaxies at comparable redshifts \citep[see text;][]{pettini2002,quider2009,dessauges-zavadsky2010}, shown as open blue circles. 
None of the low ion ratios are consistent with solar abundance patterns.
}
\end{figure*}

\begin{figure*}
\centerline{
\includegraphics[width=0.45\textwidth]{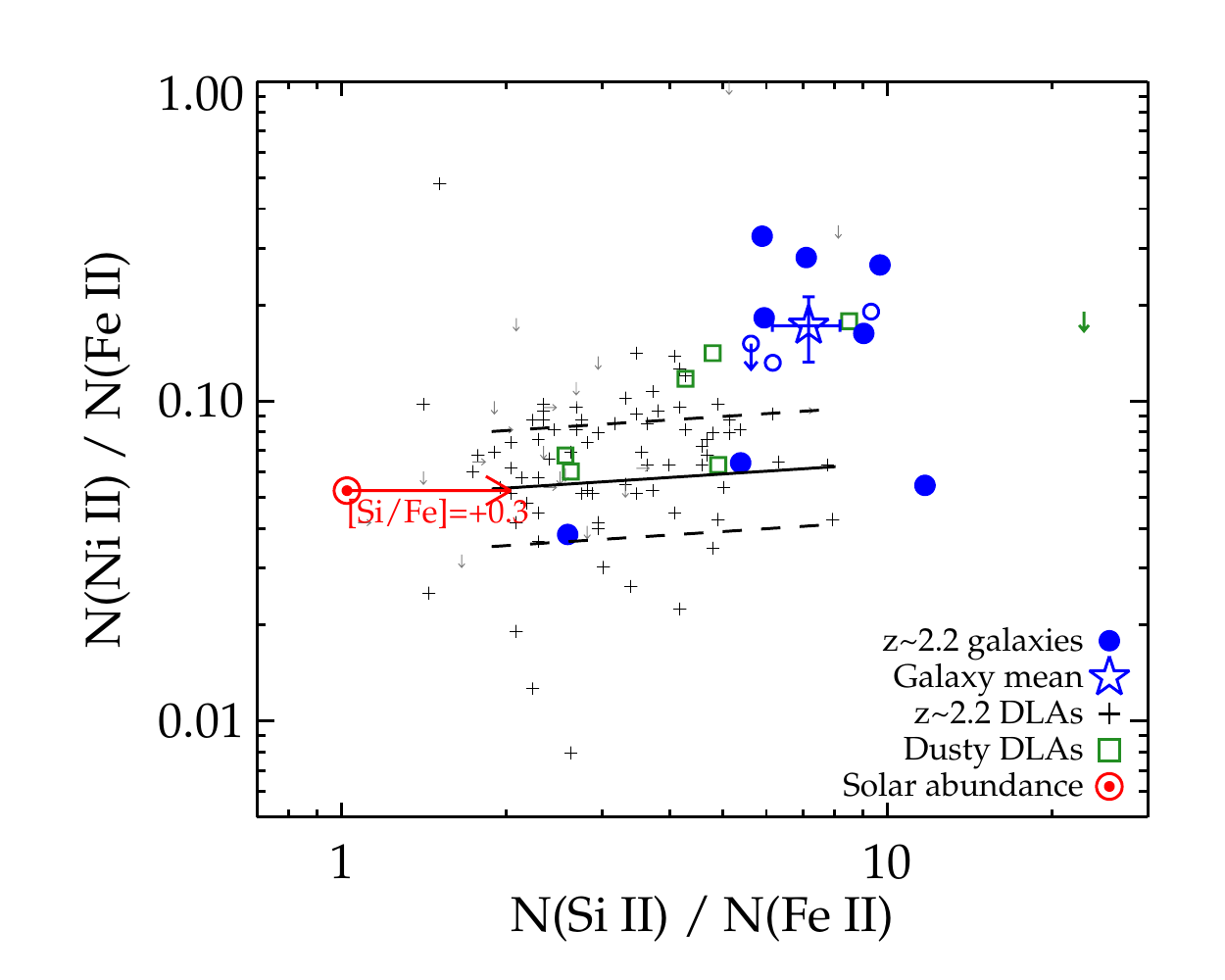}
\includegraphics[width=0.45\textwidth]{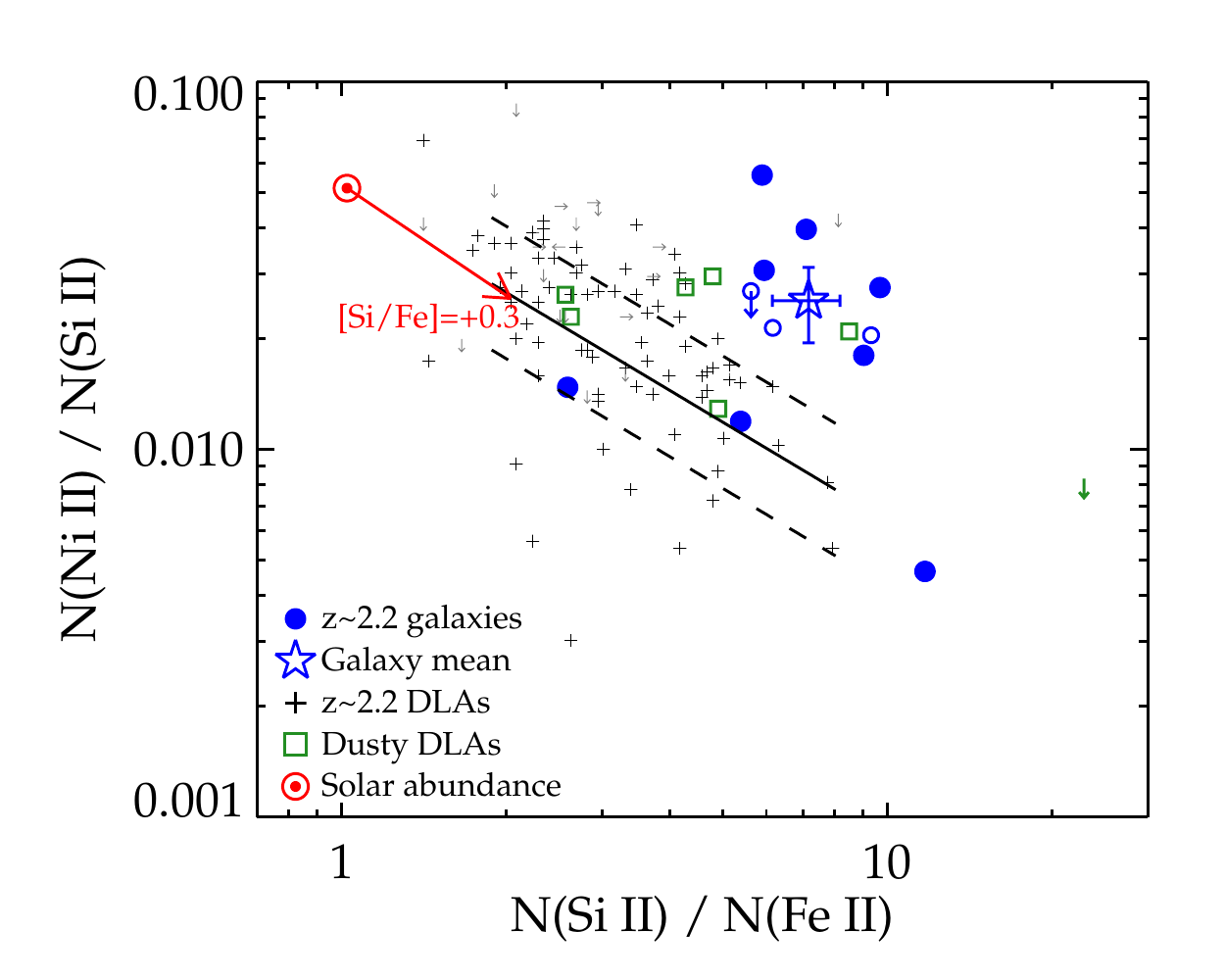}
}
\caption{
\label{fig:abundance_ratios}
Column density ratios of the low ions \Siiia, \Feiia, and \Niiia\ for the ESI lensed galaxy sample, as well as the sample median and uncertainty (star symbols). Typical uncertainties are $\sim$0.2--0.3 dex for individual galaxies. 
In addition to the compilation of { (sub-)}DLAs from \cite{quiret2016} and other lensed star forming galaxies (open blue circles as in Figure~\ref{fig:abundances_low}), we include a sample of ``dusty DLAs'' associated with 2175~\AA\ attenuation features \citep{ma2017}. Solar photosphere and $\alpha$-enhanced abundance patterns are shown for comparison (i.e., 0.3 dex enhancement in [Si/Fe] shown by red arrows). Solid lines show the dust depletion sequence from \cite{decia2016} for [Zn/Fe]~$=0$ to 1 and [Si/Fe]~$=0.26$, Ni depletions as described in the text, and 0.18 dex scatter in Ni shown with dashed lines.
}
\end{figure*}

One of the most striking results of this study is that the low ion column densities reported in Tables~\ref{tab:transitions} and \ref{tab:transitions_mult} follow a different abundance pattern compared to those seen in other well-characterized classes of absorption line systems. Figure~\ref{fig:abundances_low} shows the column densities of low ions \Siiia, \Feiia, and \Niiia\ along with relevant data from the literature. Low ion abundance ratios are shown in Figure~\ref{fig:abundance_ratios}. 
To proceed with comparisons, we quantify typical column density ratios according to the sample mean\footnotemark:
\footnotetext{
CSWA 141 has no measurement of \Niiia\ and is not included in the latter two ratios. Excluding CSWA 141 altogether would result in a negligible change to 
$\log \textrm{N(\Siiia)/N(\Feiia)} = 0.82\pm0.07$.
}
\begin{itemize}
\item log N(\Siiia)/N(\Feiia) $= 0.85 \pm 0.07$ \\
(equivalent to [\Siiia/\Feiia] = 0.84)
\item log N(\Siiia)/N(\Niiia) $= 1.70 \pm 0.12$ \\
([\Siiia/\Niiia] = 0.41)
\item log N(\Niiia)/N(\Feiia) $= -0.88 \pm 0.13$ \\
([\Niiia/\Feiia] = 0.40)
\end{itemize}
where numbers in parentheses are relative to the adopted solar scale. 
Sample median values are within 1$\sigma$ in all cases. Uncertainties in the sample mean reflect the scatter in values for different objects and are compatible with the statistical measurement uncertainty, but do not include possible systematic errors (e.g., from atomic data or blended absorption features).

Among other ions probed by the spectra, \Zniia\ provides useful constraints despite its low detection significance. The sample mean ratios of Zn with other elements are as follows:
\begin{itemize}
\item N(\Zniia)/N(\Siiia) $= 0.0055 \pm 0.0026$ \\
([\Zniia/\Siiia] = 0.69)
\item N(\Zniia)/N(\Feiia) $= 0.030 \pm 0.015$ \\
([\Zniia/\Feiia] = 1.42)
\item N(\Zniia)/N(\Niiia) $= 0.29 \pm 0.15$ \\
([\Zniia/\Niiia] = 1.12)
\end{itemize}
CSWA 141 and CSWA 39 are excluded from the Zn analysis due to low precision (e.g. N(\Zniia)/N(\Feiia)~$=0.05\pm0.06$ for CSWA 141, consistent with the sample mean but with large uncertainty). 
These values are given in linear as opposed to logarithmic units in order to account for negative formal column densities. We caution that \Zniia\ column densities are derived from the $\lambda$2026 transition and may be overestimated due to blending with \Mgia. Based on \Siia\ constraints, \Mgia\ contamination could reduce the derived value of N(\Zniia)/N(\Siiia) by $(7\pm9) \times 10^{-4}$ assuming solar ratios of \Mgia/\Siia. This $\sim$15\% effect is negligible given the uncertainties. \Zniia$\lambda2062$ is not used due to the likelihood of significant blending with \Criia.

The availability of three low ions -- \Siiia, \Feiia, and \Niiia\ -- already provides considerable information about the nature of the absorbing gas for individual galaxies. We compare the ionic abundances with several other classes of objects in the remainder of this section: stars in the solar neighborhood, interstellar gas in the Milky Way and Magellanic Clouds, DLA and sub-DLA absorption systems at similar redshift, and previous studies of star forming galaxies at $z\gtrsim2$. It is apparent in Figures~\ref{fig:abundances_low} and \ref{fig:abundance_ratios} that low ion abundance patterns in $z\gtrsim2$ galaxies follow different distributions than these other types of objects. We expand on this conclusion below with a brief discussion of each comparison sample.

{\em Stellar abundances:} Figures~\ref{fig:abundances_low} and \ref{fig:abundance_ratios} show solar abundance ratios as well as the effect of $\alpha$-element enhancement characteristic of rapidly formed stellar populations \citep[e.g.,][]{nomoto2013}. Our low ion measurements do not follow these stellar abundance patterns. This is perhaps most notable for Ni and Fe where [\Niiia/\Feiia]~$=+0.4$, in contrast with expected stellar [Ni/Fe]~$\approx0$. 
We conclude that the low ions differ from stellar abundance patterns unless the \Feiia\ column densities are systematically underestimated by $\sim$0.4 dex, which would bring agreement with highly $\alpha$-enhanced stellar populations reflecting core-collapse supernovae yields \citep[e.g.,][]{bensby2014}. 

{\em Interstellar abundances:} 
We can make a direct comparison with the ISM of nearby galaxies, where abundances are derived with the same UV absorption line analysis. 
The range of column density ratios found in the Milky Way disk ISM is shown with red shading in Figure~\ref{fig:abundances_low}, corresponding to dust depletion factors $F_* = 0$ to 1 as defined by \cite{jenkins2009}. Typical ratios of [\Siiia/\Feiia] in the (less depleted) Small Magellanic Cloud are also shown for comparison, corresponding to $F_* = 0-1$ in \cite{tchernyshyov2015} or equivalently $F_*$(SMC)~$\simeq-0.3$ to 0.6 in \cite{jenkins2017}. Super-solar [\Siiia/\Feiia] gas abundances of our sample are in excellent agreement with the ISM of the Milky Way and Magellanic Clouds, where a large fraction of Fe is found in solid dust grains. However, Ni is overabundant by $\sim$0.4 dex relative to ISM patterns. The ratio of Ni to Fe is most constraining, since both elements deplete at similar rates resulting in near-solar abundances of [\Niiia/\Feiia]~$\simeq0$ in the ISM. We conclude that the low ions are different from interstellar abundance patterns unless the \Niiia\ column densities are systematically overestimated.

{\em DLA and sub-DLA abundances:} 
As with interstellar abundances, absorption systems observed along quasar sightlines provide a direct comparison of individual ions. We compare with the compilation of DLA and sub-DLA absorption systems reported in Table~B1 of \cite{quiret2016}, which have redshifts and column densities similar to the galaxies presented here. The \cite{quiret2016} sample has mean and standard deviation $z = 2.3\pm0.4$ and $\log{N_{\textrm{\Hi}}} = 20.6 \pm 0.5$ (spanning a total range $\log{N_{\textrm{\Hi}}} = 19-22$). Figures~\ref{fig:abundances_low} and \ref{fig:abundance_ratios} reveal that \Siiia/\Niiia\ ratios are typical of (sub-)DLAs. \Niiia/\Feiia\ and \Siiia/\Feiia\ are typically higher for the galaxies, though consistent with the fringes of the (sub-)DLA distribution. 
As with stars, we conclude that the low ion abundances are different from typical quasar absorption systems unless \Feiia\ column densities are systematically underestimated by $\sim$0.4 dex. This reflects the fact that DLA abundance patterns appear similar to those of old stars in the Milky Way.

{\em Abundances of other galaxies at $z\simeq2-3$:} 
This study represents a four-fold increase in the sample of interstellar absorption line abundances available for high redshift galaxies. Comparable measurements have previously been obtained for the gravitationally lensed galaxies cB58 at $z=2.73$ \citep{pettini2002}, CSWA 1 at $z=2.38$ \citep[a.k.a. the ``Cosmic Horseshoe'';][]{quider2009}, and CSWA 21 at $z=2.74$ \citep[a.k.a. the ``8 o'clock arc'';][]{dessauges-zavadsky2010}. These results are shown as open circles in Figures~\ref{fig:abundances_low} and \ref{fig:abundance_ratios}. Uncertainties are $\sim$0.1 dex, except for \Niiia\ in CSWA 21 which is an upper limit (denoted by blue arrows). We plot the published values for each source; column densities for CSWA 1 should in principle be multiplied by 0.6$\times$ to account for 60\% covering fraction in our formalism. These previous results lie on the upper end of the column density distribution found in this work. Remarkably, low ion abundance ratios from previous work agree with our sample mean values to {\em within $\leq$0.15 dex in all cases}. We conclude that the low ion column densities are in excellent agreement with other extant measurements of star forming galaxies at similar redshift.

\subsubsection{Summary of abundance comparisons}

The low ion column density ratios are inconsistent with abundances of stars, local galaxy ISM, and typical quasar absorption systems. Nonetheless the low ion ratios are in agreement with previous findings for star forming galaxies at similar redshifts. Scatter within the galaxy sample is remarkably low: 0.14 dex RMS in \Siiia/\Feiia, and presumably even lower considering the measurement uncertainties. Low ion abundance patterns measured in our sample therefore represent a distinct, and unique, chemical fingerprint of outflowing gas in $z\simeq2$--3 galaxies.

The non-stellar patterns reinforce conclusions of earlier work that low ions are not representative of the total abundances, due to dust depletion and other possible effects \citep{pettini2002,dessauges-zavadsky2010}. In the following sections we assess the total intrinsic (as opposed to ionic) abundance patterns.

\subsection{Corrections to low ion abundance ratios}

Total abundance patterns of the interstellar medium should correspond to those of newborn stars. The low ion abundance patterns in our sample do not match any common stellar population, suggesting that low ions are not representative of the total abundances. In this section we examine possible causes of this discrepancy. These include: ionization corrections, unresolved saturated absorption components, and depletion onto solid dust grains.

\subsubsection{Ionization corrections}
\label{sec:ionization_corrections}

Ionization corrections are potentially important if a significant fraction of the total low ion column density is associated with ionized (\Hii) gas. 
\Oia\ is a valuable ionization diagnostic as it is strictly coupled to \Hi\ by charge exchange reactions. Saturated \Oia$\lambda$1302 absorption profiles are in good agreement with other low ions in our sample, suggesting a predominantly neutral \Hi\ origin for the low ions. 
On the other hand, a substantial fraction of the low ions may be associated with moderately ionized gas traced by \Aliiia\ in our sample (Section~\ref{sec:kinematics_al3}), in which case ionization corrections could be significant. Several groups have used \Aliiia\ as a diagnostic of ionization corrections in DLA and sub-DLA systems \citep[e.g.,][]{howk1999,dessauges-zavadsky2003, lehner2008,milutinovic2010}, in conjunction with {\sc cloudy} photoionization modeling \citep{ferland2017}. We note that the gas probed in galaxy absorption spectra is unlikely to reflect any idealized scenario and we therefore do not attempt detailed photoionization modeling. Instead we empirically assess the likely ionization corrections based on the relative behavior of different ions. 

\begin{figure}
\includegraphics[width=\columnwidth]{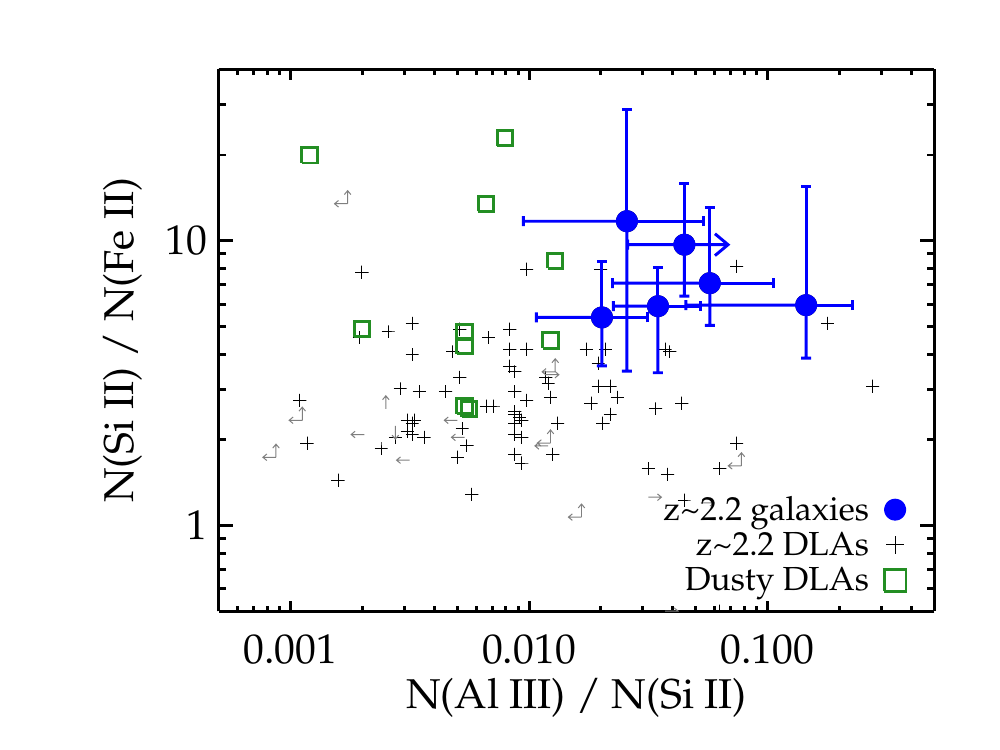}
\caption{
\label{fig:ionization}
Low ion column density ratios as a function of ionization state. Symbols are identical to Figure~\ref{fig:abundance_ratios}. For CSWA 38, N(\Aliiia) is based only on a single transition and we treat this as a lower limit, appropriate if $f_c$ is smaller than for the low ions. 
The low ion ratio \Siiia/\Feiia\ in these samples shows little or no dependence on the amount of ionized gas traced by \Aliiia. In particular, the high \Siiia/\Feiia\ seen in galaxy spectra are not observed in DLAs even when comparable amounts of ionized gas are present, indicating that ionization corrections cannot reconcile abundances in the galaxy spectra with the typical DLA population.
}
\end{figure}

Figure~\ref{fig:ionization} shows the abundance of \Aliiia\ relative to low ions. Ideally we would use \Aliiia/\Aliia\ to assess ionization corrections, but this ratio is available only in a small number of sources ($<$10\% of DLAs in the comparison sample). We therefore use \Aliiia/\Siiia\ as a proxy, noting that this ratio shows relatively little scatter with \Siiia/\Aliia~$\simeq27$ in DLAs and sub-DLAs \citep{vladilo2001,dessauges-zavadsky2003}. This ratio implies $\tau \simeq 5$--10 in the $\lambda$1670 transition for cases which we report as saturated in Table~\ref{tab:transitions}, consistent with our limits and suggesting \Aliiia/\Aliia\ ratios of order unity. Corresponding ionization corrections from \Aliia\ to total Al are $\gtrsim$0.2 dex. While these represent upper limits to the correction required for the low ionization phase, it is clear that significant amounts of moderately ionized gas are present.

If ionization corrections were responsible for the high \Siiia/\Feiia\ and other low ion ratios observed in the galaxy sample, then we would expect similar values in (sub-)DLAs with large fractions of ionized gas (i.e., high \Aliiia/\Aliia\ and \Aliiia/\Siiia). Figure~\ref{fig:ionization} shows that this is not the case: even DLAs with similar \Aliiia\ to low ion ratios do not match our sample. We find similar null results for both \Niiia/\Feiia\ and \Siiia/\Feiia, despite their high sensitivity to ionization \citep[with up to $\sim$0.5 dex corrections possible; e.g.,][]{vladilo2001}. Therefore while significant amounts of moderately ionized gas are present, ionization corrections do not reconcile the galaxy sample with DLA abundance patterns.

\begin{figure}
\includegraphics[width=\columnwidth]{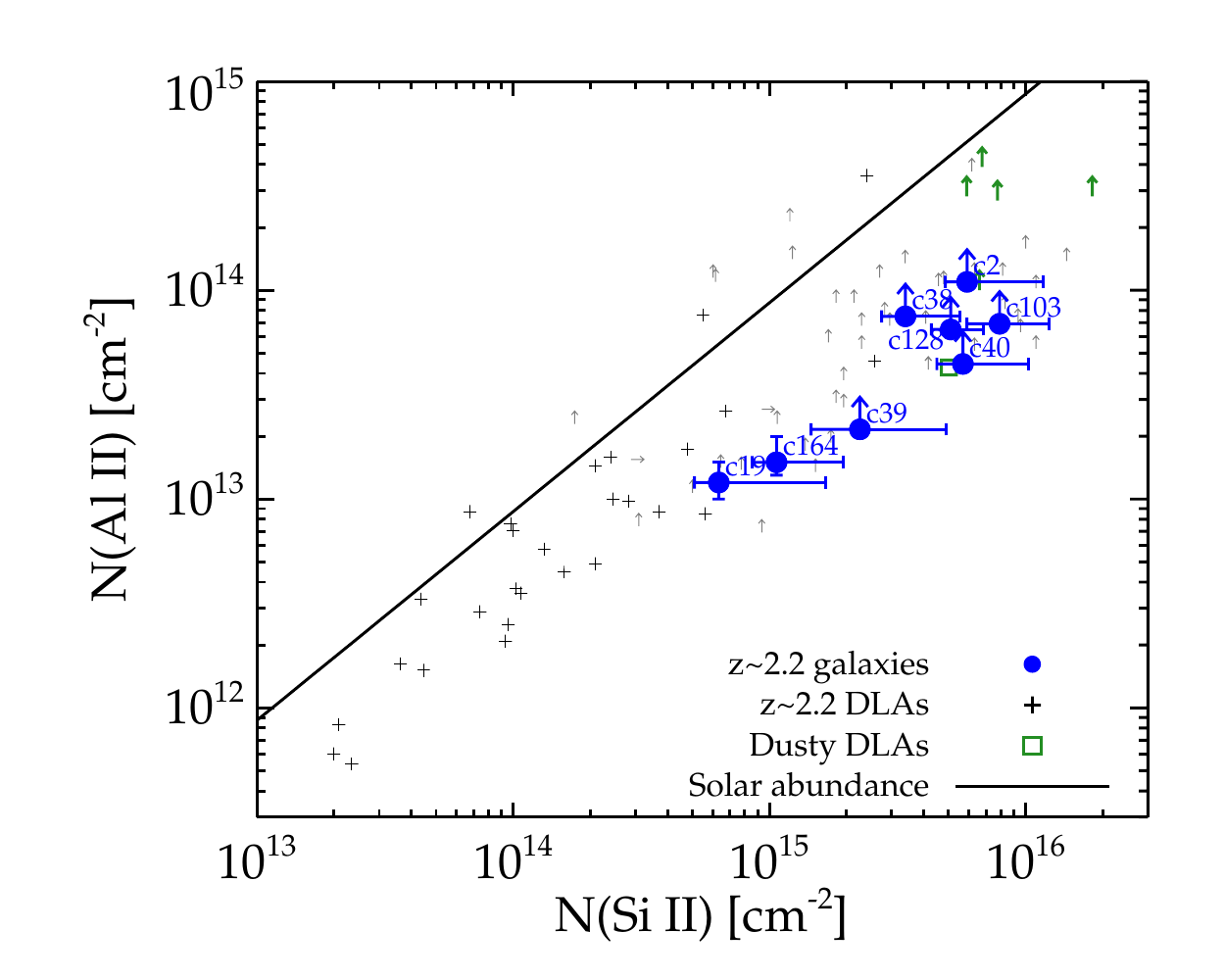}
\caption{
\label{fig:al2_si2}
Column densities of \Siiia\ and \Aliia\ compared with literature samples, similar to Figure~\ref{fig:abundances_low}. In most cases we derive only lower limits for \Aliia\ from the saturated $\lambda$1670 transition (denoted by arrows); this is also true for the literature data at large column densities. In both galaxies where N(\Aliia) is determined directly, we find sub-solar [\Aliia/\Siiia] comparable to the lowest values seen in DLA absorption systems. In all cases our lower limits are consistent with sub-solar [\Aliia/\Siiia]. These results are incompatible with significant stellar ionization.
}
\end{figure}

While ionization alone does not explain the difference in abundance patterns between the galaxy sample and DLAs, we nonetheless consider it likely that ionization corrections are important based on the column densities of \Aliiia. We now assess the possible magnitude of such ionization corrections, which depends on the nature of the ionizing sources. 
The ratio $N$(\Aliia)/$N$(\Siiia) shown in Figure~\ref{fig:al2_si2} is a useful diagnostic: ionization by a stellar radiation source should result in super-solar [\Aliia/\Siiia]. As a fiducial example we consider the two-region model of \cite{vladilo2001} with $\log{N_{\textrm{\Hi}}} = 20.2$, on the low end of our sample. This case results in an increase by $+0.53$ dex to the observed [\Aliia/\Siiia], a considerably larger change than for Si/Fe/Ni low ion ratios (although possibly overestimated as discussed by \citealt{vladilo2001}; \citealt{dessauges-zavadsky2003}). In contrast the two galaxies with measurements of N(\Aliia) show {\em sub-solar} [\Aliia/\Siiia]~$= -0.66$ and $-0.79$, highly inconsistent with significant stellar ionization. We suggest that sub-solar [\Aliia/\Siiia] ratios are instead due to a combination of depletion and even-odd nucleosynthetic effects. 
Furthermore if the sources with lower limits in N(\Aliia) have super-solar ratios, then ionization corrections should be {\em smaller} than our estimates above (that is, the true \Aliiia/\Aliia\ would be smaller than estimated from \Aliiia/\Siiia). 
Stellar ionization therefore does not explain the low ion abundance patterns in our sample, indicating that the absorbing gas is sufficiently dense and/or distant from the sites of active star formation. We consider this fiducial case to be a strongly conservative upper limit on the {\em stellar} ionization corrections (e.g., $<0.08$ dex correction to [\Niiia/\Feiia]).

While stellar ionization appears to be insignificant, a hard incident UV spectrum may explain some of the distinct ionic abundance patterns in our sample. Several groups have modeled the effects of ionization by the $z\simeq2$ extragalactic UV background described by \citet[][and subsequent updates]{haardt1996}. Corrections derived by \cite{howk1999} for ionization parameters $\log{\Gamma} \lesssim -3$ are compatible with our data \citep[and see also, e.g., ][]{dessauges-zavadsky2003,milutinovic2010}. 
Such ionization leads to an increase of up to $\lesssim$0.55 dex to observed [\Niiia/\Feiia]; $\simeq$0 for [\Niiia/\Siiia]; $\lesssim$0.7 for [\Feiia/\Zniia]; and $\simeq$-0.1 for [\Aliia/\Siiia], relative to intrinsic elemental abundances. Lower $\Gamma$ values result in smaller corrections. Higher $\Gamma$ would overproduce \Siiva\ in the galaxies where \Siiva\ is observed, while having little effect on the low ion ratios considered here.

In summary, relatively high \Aliiia\ column densities suggest that ionization corrections from a hard UV spectrum are likely to be important. However, the lack of correlation between low ion ratios and \Aliiia\ suggests that ionization corrections are comparable throughout the sample and approximately independent of the total ionization fraction (i.e., \Aliiia/\Aliia). We note that the DLA and sub-DLA sample compiled by \cite{quiret2016} similarly shows little correlation of \Siiia/\Feiia\ with Al ionization, and systematically lower values of \Siiia/\Feiia\ and \Niiia/\Feiia\ compared to the galaxies at fixed \Aliiia/\Siiia\ (or \Aliiia/\Aliia). Therefore effects other than ionization must contribute to the distinct low ion ratios observed in the galaxy sample.

\subsubsection{Unresolved saturated components}

Due to moderate spectral resolution and finite size of the target galaxies, the data may be affected by saturated components which are spectrally or/and spatially unresolved. This would cause a relative increase in the apparent column density of weaker transitions. Based on the optical depths of transitions used to constrain the column densities, we expect that saturated components would lead to overestimates of [\Niiia/\Feiia], [\Niiia/\Siiia], and [\Siiia/\Feiia]. We can estimate an upper limit on the magnitude of this effect by assuming a solar intrinsic gas-phase [Ni/Fe]~$=0$, noting that the true value may be super-solar if DLA abundance patterns are representative. This implies a correction of at most $\lesssim0.4$ dex to [\Niiia/\Feiia]. By construction any necessary correction to [\Niiia/\Feiia] must equal the sum of corrections to [\Niiia/\Siiia] and [\Siiia/\Feiia]. Therefore we expect $\lesssim0.4$ dex corrections to each of these ratios due to unresolved saturation. This corresponds to intrinsic gas-phase ion abundances [\Siiia/\Feiia]~$\simeq0.4$--0.8, [\Siiia/\Niiia]~$\simeq0.4$--0.8, and [\Niiia/\Feiia]~$\simeq0$--0.4. 
Based on studies comparing high- and low-resolution spectra, we expect that corrections should be smaller than this limit, as the absorption spans a broad velocity range and we have been careful to rely on optically thin lines which should not be highly biased \citep[e.g. $\tau\lesssim0.7$;][]{jorgenson2013,cucchiara2015}. In any case these analyses show that correction for individual transitions should be no more than 0.5 dex, with smaller corrections for ion abundance ratios. 
In summary, unresolved saturation could account for non-solar [Ni/Fe] low ion abundances only in the extreme case, while Si remains significantly super-solar relative to the Fe-peak elements in the gas phase.

\subsubsection{Dust depletion}
\label{sec:dust}

Solid dust grains are an important constituent of the ISM metal budget \citep[and CGM;][]{menard2010}. Within the Milky Way disk, typically $\gtrsim$90\% of Fe and Ni are depleted into the solid state along with the majority of Si \citep[e.g.,][]{savage1996,jenkins1986}, resulting in the non-solar interstellar {\em gas} abundances shown in Figure~\ref{fig:abundances_low}. Smaller depletions are inferred for (sub-)DLA systems ($\simeq0$--90\% of Fe; e.g., \citealt{decia2016}), in some cases comparable to the halo of the Milky Way \citep{savage1996}. Low ion column densities reported here enable a similar assessment of dust grain composition in high redshift galaxies and their associated outflows. 

The ratio of [\Siiia/\Niiia] is especially constraining for depletion, since we infer a super-solar gas phase [Si/Ni]~$\simeq0.4$--0.8 after accounting for the combined effects of ionization ($\lesssim$0.1 dex) and saturation ($\lesssim$0.4 dex). Such large values can plausibly be explained by moderate depletion (by $\simeq0.8$--1.6 dex in Fe and Ni), extreme $\alpha$-enhancement with Type II supernovae dominating the enrichment, or a combination of these effects. 
The standard diagnostic [\Zniia/\Feiia] likewise suggests moderate depletion, although this is weak evidence given the $\sim2\sigma$ significance of \Zniia\ absorption. Strong depletion with [Zn/Fe]~$\gtrsim2$ is ruled out as inconsistent with the weakness of \Zniia\ lines; this corresponds to heavily depleted Milky Way sightlines. 
Sub-solar [\Aliia/\Siiia] in CSWA 19 and CSWA 164 likewise suggests moderate depletion \citep[e.g.,][]{howk1999a}, although it could also reflect an odd-even nucleosynthetic effect. 
Finally, we note that the galaxy low ion abundance patterns in Figures~\ref{fig:abundance_ratios} and \ref{fig:ionization} most closely resemble the ``2DA'' quasar absorbers presented by \cite{ma2017} which are associated with 2175 \AA\ attenuation features, clearly indicating significant dust in these systems. We consider this indirect but compelling evidence that low ion abundances reflect dust depletion in the galaxy outflows.

A key factor in our interpretation is the dust composition in high redshift systems. \cite{decia2016} have proposed that dust depletion patterns in high-redshift DLAs and local galaxy ISM form a continuous sequence. The DLA-like absorption systems studied here should naturally lie on the same sequence. Since \cite{decia2016} do not report results for Ni, we follow their methodology to estimate the Ni depletion sequence using the subset of systems with \Niiia, \Feiia, \Criia, and \Zniia\ measurements tabulated by \cite{quiret2016}. \Criia\ is used as a consistency test. We successfully reproduce the depletion pattern 
\begin{equation}
\delta_{X} = A_X + B_X \times \textrm{[Zn/Fe]}
\end{equation}
for Cr with $A_{Cr} = 0.15\pm0.04$, $B_{Cr} = -1.30\pm0.07$, and scatter $\sigma_{Cr} = 0.10$ dex (in excellent agreement with $A_{Cr} = 0.15$, $B_{Cr} = -1.32$, and $\sigma_{Cr} = 0.10$ given by \citealt{decia2016}). The resulting Ni depletion sequence is characterized by $A_{Ni} = -0.03\pm0.06$, slope $B_{Ni} = -1.19\pm0.11$, and relatively large scatter $\sigma_{Ni}=0.18$. The resulting depletion sequences are shown in Figure~\ref{fig:abundance_ratios} for the range of (depleted) gas-phase [Zn/Fe]~$=0$ to 1, with [Si/Fe]~$=0.26$ as assumed for low-metallicity DLAs by \cite{decia2016}. The resulting depletion sequence is unable to explain low ion column densities in the galaxy sample unless \Niiia\ has been overestimated by $\sim$0.5 dex, echoing the conclusions from Section~\ref{sec:abundances_low} regarding Milky Way ISM. Notably, Figure~\ref{fig:abundance_ratios} shows that this depletion sequence is unable to explain several of the DLAs from \cite{ma2017} which are known to be associated with large dust column densities. As with the galaxies, these dusty DLAs are on average overabundant in \Niiia\ relative to the depletion pattern. This can be explained by additional ionization or saturation effects, combined with depletion of $\sim$0.6 dex in Fe and Ni. Alternatively the dust composition in these systems may be Ni-poor compared to the Milky Way ISM.

In summary, the low ion column densities show evidence of dust depletion, but depletion patterns seen in the Milky Way and inferred for typical DLAs at $z\simeq2$ cannot fully reproduce the observations. DLA-like or Milky Way halo-like dust with Fe depletions of $\sim$0.6 dex can provide a consistent picture in combination with ionization corrections. Another possibility is that the dust composition is fundamentally different from these systems, leading to enhanced gas-phase Ni abundances similar to those reported for dusty DLAs \citep{ma2017}. This would indicate a different origin or processing for the dust in high redshift galaxy outflows compared to normal ISM grain growth processes. This would imply a greater depletion.

\subsection{Intrinsic abundance ratios}
\label{sec:abundances_intrinsic}

In this section we assess the intrinsic heavy element abundance ratios, and evidence for departures from the solar pattern. In particular the $\alpha$ element Si may be enhanced or decreased relative to Fe and Ni, depending on the history of star formation and metal mixing. The target galaxies have high specific star formation rates such that core-collapse supernova are expected to dominate the metal production \citep[Mainali et al. {\em in prep};][]{jones2013,leethochawalit2016}, resulting in super-solar \AFe. This is reflected in abundance patterns of metal-poor DLAs and Milky Way stars which reach [Si/Fe]~$=0.3$--0.4 \citep[e.g.,][]{becker2012,bensby2014}. The observational effect of such an $\alpha$-enhancement is shown in Figures~\ref{fig:abundances_low} and \ref{fig:abundance_ratios}. Zn and Ni may also be moderately enhanced by up to $\sim$0.1 dex, but this is a small effect given the observational uncertainties.

It is clear from Figure~\ref{fig:abundance_ratios} that [Si/Fe] enhancement is highly degenerate with depletion. This arises from the similar depletion of Ni and Fe in addition to their common nucleosynthetic pathway, whereas Si is less depleted. The intrinsic gas phase abundances of [Si/Fe] and [Si/Ni], which are constrained to be $\simeq0.4$--0.8 considering ionization and saturation effects, can readily be explained by a combination of depletion and $\alpha$ enhancement. However additional elements are needed to distinguish these effects. In this regard [Al/Si] is a useful diagnostic of chemical maturity, as the odd-even nucleosynthetic signature is correlated with $\alpha$ enhancement. We infer a galaxy sample median [\Aliia/\Siiia]~$\geq -0.6$ based on \Aliia\ column density measurements and lower limits. We expect total [Al/Si] to be similar to or higher than [\Aliia/\Siiia], accounting for ionization and depletion. 
For comparison, we consider Milky Way stellar abundance trends from SDSS APOGEE data \citep{blanton2017}. Based on a running median of abundance ratios as a function of metallicity, we find that [Al/Si]~$\geq -0.6$ implies maximum $\alpha$ enhancement [Si/Fe]~$\leq0.3$, and [Ni/Fe]~$\leq0.1$. The corresponding metallicity limit is [Fe/H]~$\geq-1.7$, although we may expect a higher metallicity scale if the descendants of our galaxy sample are more massive than the Milky Way.

We consider the above result [Si/Fe]~$\leq0.3$ to be a stringent limit, provided that metal-poor Milky Way stars are representative of early nucleosynthesis in our galaxy sample. The limit [Fe/H]~$\geq-1.7$ is compatible with our available measurements of [\Feiia/\Hi]~$\simeq-1.4$, which are likely underestimated due to depletion. Nebular abundance measurements suggest [O/H]~$\geq-0.7$ (with CSWA 141 having [O/H]~$\simeq-0.7$ based on the direct T$_e$ method, and higher nebular [O/H] for other sources: \citealt{stark2013,jones2013,leethochawalit2016}). In general, adopting higher metallicities as a constraint gives abundance ratio limits closer to the solar value (e.g. [Si/Fe]~$\leq0.2$, [Ni/Fe]~$\leq0.06$, and [Al/Si]$\geq-0.2$ for Milky Way stars with [O/H]~$\geq-0.7$).

\subsubsection{Joint constraints on depletion and intrinsic abundances}
\label{sec:model}

\begin{figure*}
\centerline{
\includegraphics[width=0.45\textwidth]{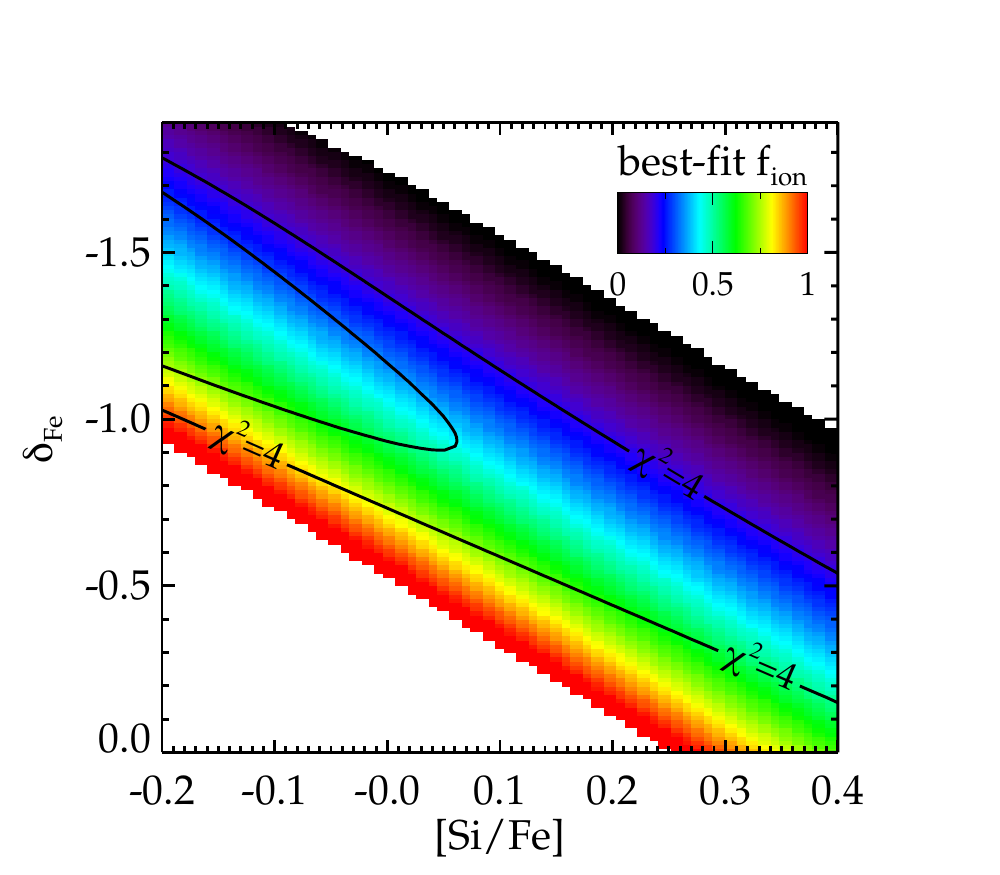}
\includegraphics[width=0.45\textwidth]{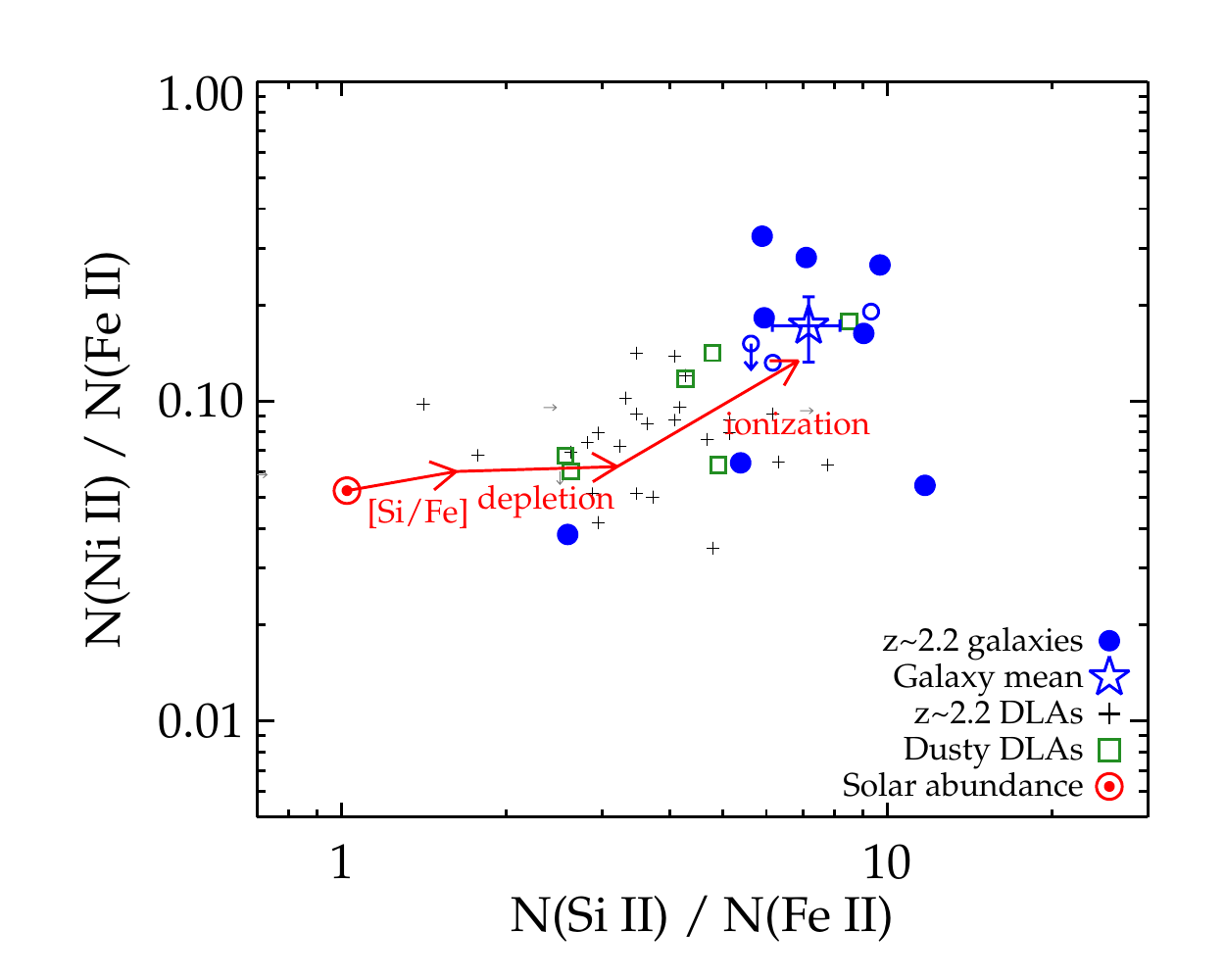}
}
\caption{
\label{fig:model}
Constraints on a model of ionization correction, dust depletion, and intrinsic $\alpha$-enhancement. 
{\it Left:} Model parameter constraints in the plane of $\delta_{Fe}$ (dex of Fe depletion) and [Si/Fe] ($\alpha$-enhancement). The best fitting ionized fraction $f_{ion}$ is shown in color scale for all regions with a solution having $\chi^2 \leq 9$. Contour lines enclose the regions with solutions having $\chi^2 \leq 3$ and $\leq4$ (as labeled). This plot illustrates the strong degeneracy between depletion and $\alpha$-enhancement from the available data, with good solutions existing over a wide range. The data strongly favor a non-zero ionization correction. Non-zero depletion is also favored for realistic values of [Si/Fe] ($\lesssim 0.35$). 
{\it Right:} Arrows show an example model solution with [Si/Fe]~$=0.2$, $\delta_{Fe}=-0.65$ dex of Fe depletion (corresponding to $-0.35$ dex depletion of Si, and gas-phase [Zn/Fe]~$=0.5$), and $f_{ion}=0.45$, resulting in good agreement with the low ion abundances. As with all plausibly good solutions, all three effects are significant in this example. 
Plot symbols are identical to Figure~\ref{fig:abundance_ratios} but with the \cite{quiret2016} DLA comparison sample restricted to the metal-rich DLAs presented by \cite{berg2015}. A substantial fraction of these are proximate DLAs with evidence for significant ionization by a hard quasar spectrum, further supporting the importance of ionization corrections.
}
\end{figure*}

Given the degeneracy between dust depletion and intrinsic abundance ratios (in particular \AFe\ and the even-odd effect traced by [Al/Si]), we now quantify their joint constraints. Our approach is to determine which combinations of depletion, abundance patterns, and ionization corrections can match the observed low ion column densities. We do not include possible effects of saturation, noting that this would decrease the required ionization corrections with little effect on abundances or depletion.

We construct a model as follows. The depletion sequence is taken from \cite{decia2016} and combined with our estimate of Nickel depletion described in Section~\ref{sec:dust}. 
We adopt fiducial ionization corrections corresponding to gas which is (almost) fully ionized by a $z=2$ extragalactic UV background with ionization parameter $\log{\Gamma}=-3$ \citep[as described by][]{howk1999}. The corrections are c([\Siiia/\Feiia])~$=0.55$, c([\Niiia/\Feiia])~$=0.55$, and c([\Zniia/\Feiia])~=$-0.7$. The ionized gas is assumed to constitute a fraction $f_{ion}$ of the total \Feiia\ column density, with the rest coming from a neutral region where the low ions are proportional to their total gas-phase abundances (i.e. with no ionization correction). This two-phase treatment is motivated by the different covering fractions of ionized gas (i.e. \Aliiia) and low ions. The ionized phase is taken to be an extreme case, such that a mix of ionized and neutral phases can reproduce the full plausible range of ionization corrections. 
For abundance patterns, we treat intrinsic [Si/Fe] as a free parameter and set [Ni/Fe]~$=0.3\times$[Si/Fe] to approximate of the mean trend for Milky Way stars. [Zn/Fe] is fixed at the solar value; varying this ratio does not affect the results due to the large uncertainty in \Zniia\ measurements. 

The model has three variables: depletion (parameterized by $\delta$(Fe), the difference between total and gas-phase Fe abundance in dex), ionized fraction $f_{ion}$, and the value of the intrinsic [Si/Fe] abundance. The resulting low ion ratios are compared with sample mean values to construct a goodness-of-fit statistic: 
\begin{equation}\label{eq:model}
\chi^2 = \sum_{\mathrm{X}} 
\frac{(\mathrm{[X/Fe~\textsc{ii}]_{model} - [X/Fe~\textsc{ii}]_{observed}})^2}{\sigma^2(\mathrm{[X/Fe~\textsc{ii}]})}
\end{equation}
where X = \{\Siiia, \Niiia, \Zniia\} and $\sigma$ is the observational uncertainty. \Feiia\ is used as the reference normalization scale. 
Figure~\ref{fig:model} shows results of comparing the model to observed low ion column densities, with several general features evident. It is clear that this simple model can satisfactorily explain the data with a range of solutions having $\chi^2<4$ (i.e. better than $2\sigma$ agreement with all measurements).

Our aim with this model is to quantify the range of physical conditions which can plausibly explain the low ion abundances. Figure~\ref{fig:model} clearly illustrates the degeneracy between $\alpha$ enhancement, depletion, and ionization corrections. The locus of best-fit solutions can be well described by a linear relation, 
\begin{equation}\label{eq:alpha_fe}
\mathrm{\delta_{Fe} = -0.5 + 1.8([Si/Fe] - 0.3)}
\end{equation}
with a range $\pm0.3$ dex in $\delta_{Fe}$ corresponding to $\Delta \chi^2 = 1$. The best-fit ionization corrections are described by $f_{ion} \simeq 0.5$ in the model formalism. 

The prescriptions used in our model are not unique, and we have verified that other plausible assumptions can also reproduce the sample mean properties. For example, SMC or Milky Way depletion patterns \citep{jenkins2009,jenkins2017} can be extrapolated to match the data. Different sets of ionization corrections can match the data, subject to additional constraints discussed in Section~\ref{sec:ionization_corrections}. The ability of various models to explain the data underscores the unknown nature of the ionizing spectrum and dust grain compositions. Nonetheless we can draw some robust general conclusions based on straightforward physical arguments, which are quantitatively supported by the model. 
In particular: 
\begin{itemize}
\item A combination of dust depletion and non-solar \AFe\ are required to explain the data, with larger depletion implying smaller \AFe. This result is driven by the highly super-solar [\Siiia/\Niiia] which is not explained by ionization or saturation (and may even be underestimated if saturation is important). The entire plausible range of [Si/Fe]~$\simeq -0.1$ to 0.3 can be well-fit by the model, in all cases requiring depletion of at least $\sim0.5\pm0.3$ dex in Fe.
\item Non-negligible ionization corrections are required to explain the data. The model is degenerate in the sense that larger ionization corrections imply smaller \AFe\ and/or depletion. This result is driven by super-solar [\Niiia/\Feiia] which is inconsistent with known stellar abundances and depletion patterns. If this ratio is affected by unresolved saturated components, we would infer smaller ionization corrections. Independent of these arguments, \Aliiia\ column densities suggest that ionization corrections are likely to be important.
\end{itemize} 
We have explained these conclusions without reference to \Zniia, appealing instead to those ions with the most precise column densities. Excluding the \Zniia\ constraint reduces the overall $\chi^2$ but otherwise has little effect on Figure~\ref{fig:model}. The points above are therefore robust to our inclusion of \Zniia, which serves as a good consistency check.

\section{Discussion}\label{sec:discussion}

\subsection{Metallicity and $\alpha$/Fe}\label{sec:alpha_abundance}

One of the goals of this work is to estimate the ISM enrichment of various elements, particularly of the $\alpha$-capture and Fe-peak groups. 
Total metallicity constraints are available for the four sources with measurements of \Hi\ column density from \Lya. These sources are representative of the sample in terms of average low ion abundance ratios (i.e. consistent with the mean values in Section~\ref{sec:abundances_low}). Their mean gas-phase ion abundance is log N(\Feiia)/N(\Hi) $= -6.06\pm0.10$, or 
\begin{equation}
\mathrm{[Fe~\textsc{ii}/H~\textsc{i}]} = -1.56\pm0.10
\end{equation}
in solar units. We can express the intrinsic metallicity as 
\begin{equation}\label{eq:FeH}
\begin{split}
\mathrm{
[Fe/H]} \, & \mathrm{ = [Fe~\textsc{ii}/H~\textsc{i}] - \delta_{Fe} + I.C.(Fe~\textsc{ii}/H~\textsc{i}) } \\
           & \mathrm{ \approx [Fe~\textsc{ii}/H~\textsc{i}] - \delta_{Fe} }, 
\end{split}
\end{equation}
where [\Feiia/\Hi] is the gas-phase value and [Fe/H] is total abundance. The ionization correction term I.C.(\Feiia/\Hi) is expected to be negligible since [\Feiia/\Hi]~$\simeq$~[Fe/H] even in highly ionized gas \citep[within $\lesssim0.1$ dex; e.g.,][]{dessauges-zavadsky2003}. Likewise the term for I.C.(\Siiia/\Niiia) is expected to be small, and we can write the intrinsic abundance ratio as
\begin{equation}\label{eq:SiNi}
\begin{split}
\mathrm{
[Si/Ni]} \, & \mathrm{ = [Si~\textsc{ii}/Ni~\textsc{ii}] - (\delta_{Si} - \delta_{Ni}) + I.C.(Si~\textsc{ii}/Ni~\textsc{ii}) } \\
            & \mathrm{ \approx [Si~\textsc{ii}/Ni~\textsc{ii}] + 0.44\, \delta_{Fe} }. 
\end{split}
\end{equation}
The relation
\begin{equation}\label{eq:d_SiNi}
\mathrm{
\delta_{Si} - \delta_{Ni} \approx -0.44\, \delta_{Fe}
}
\end{equation}
is appropriate for the adopted DLA depletion sequence of Section~\ref{sec:dust}. 
We note that Equation~\ref{eq:SiNi} describes the same locus as the best-fit model of Equation~\ref{eq:alpha_fe}.

\begin{figure}
\includegraphics[width=\columnwidth]{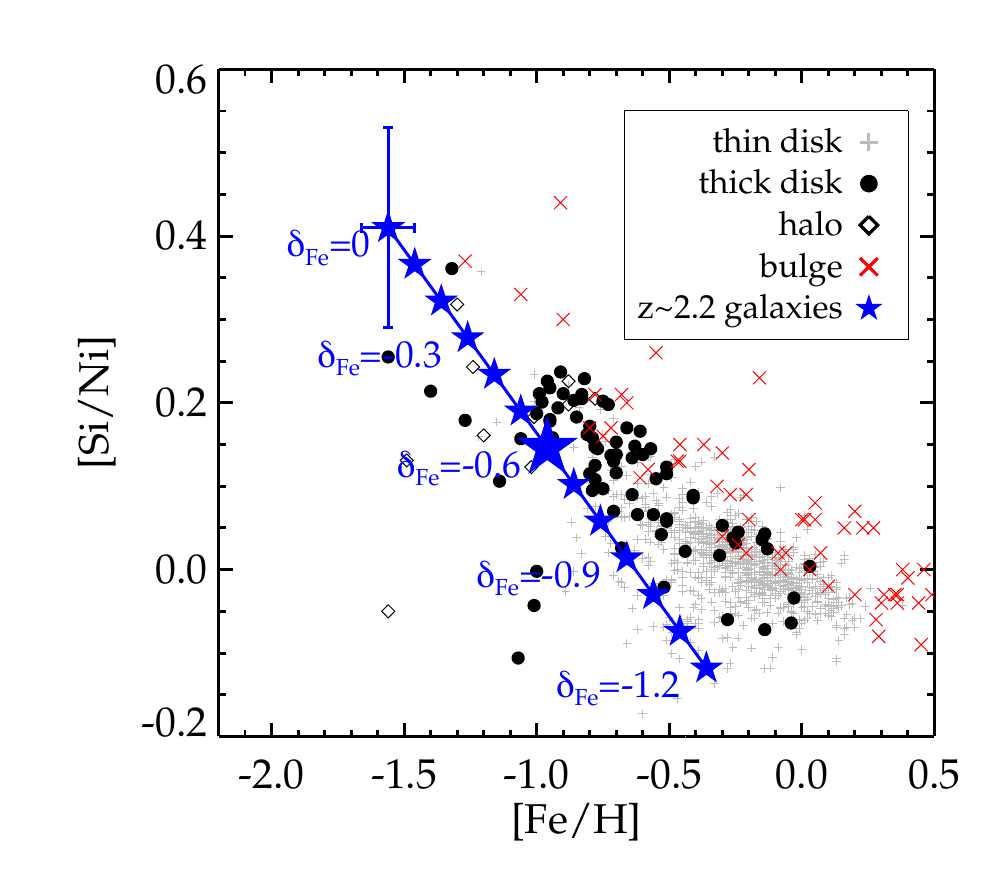}
\caption{
\label{fig:stellar_comparison}
Comparison of Milky Way stellar abundance patterns with interstellar abundances in the $z\simeq2.2$ galaxy sample (shown for a range of Fe depletion $\delta_{\mathrm{Fe}}$, with a larger symbol for the fiducial $\delta_{\mathrm{Fe}}=-0.6$). Galaxy abundances correspond to Equation~\ref{eq:abundances}, with 1$\sigma$ statistical error bars shown on the uppermost point. 
Stellar data are from \cite{adibekyan2012} for the Milky Way thin disk, thick disk, and halo, and from \cite{bensby2013} for the Milky Way bulge.
}
\end{figure}

Equations~\ref{eq:FeH} and \ref{eq:SiNi} together characterize the total abundance ([Fe/H]) and $\alpha$ enhancement ([Si/Ni]) in terms of dust depletion. Substituting our column density measurements into these equations, the typical abundances of the sample are
\begin{equation}\label{eq:abundances}
\begin{split}
& \mathrm{
[Fe/H]  \approx -1.56 - \delta_{Fe}      \pm 0.10, } \\
& \mathrm{
[Si/Ni] \approx 0.41  + 0.44 \delta_{Fe} \pm 0.12  }
\end{split}
\end{equation}
(taking the available [\Feiia/\Hi] measurements as representative; quoted uncertainties are the statistical error in the mean). We have chosen this parameterization to minimize systematic error and covariance: [Fe/H] and [Si/Ni] rely on independent ion ratios which are relatively insensitive to ionization. The expected corrections are $\lesssim0.1$ dex, comparable to measurement uncertainties. 
We illustrate the locus of [Fe/H] and [Si/Ni] as a function of depletion in Figure~\ref{fig:stellar_comparison}.

Independent of further analysis, we can characterize the range of abundance patterns allowed by our measurements. It is clear from the above equations and from Figure~\ref{fig:stellar_comparison} that negligible depletion would imply metallicities of $\lesssim -1$ in both [Fe/H] and [$\alpha$/H]. This is significantly lower than expected from nebular abundances measured for our sample (as well as expected [O/H]~$\simeq-0.4$ based on the mass-metallicity relation at $z=2.2$; e.g., \citealt{erb2006,sanders2015}), unless the outflows are highly diluted by near-pristine gas. 
At the other extreme, large depletion would imply sub-solar \AFe\ ratios. The range $\delta_{\mathrm{Fe}}=-1$ to $-2$ seen in the Milky Way disk corresponds to [Si/Ni]~$=0$ to $-0.5$. Higher depletions are ruled out by gas phase [\Zniia/\Feiia] abundances. Furthermore, basic chemical evolution arguments suggest that intrinsic \AFe\ should be super-solar due to short timescales available for enrichment by type Ia supernovae. Typical star formation timescales sSFR$^{-1}$ and the $\sim$3 Gyr age of the universe both suggest \AFe~$\gtrsim0.1$ \citep[e.g.,][]{thomas2005}, and likely higher based on the sSFR. 
These considerations therefore suggest intermediate depletions with the most plausible range being $\delta_{\mathrm{Fe}} \simeq -0.6\pm0.3$. This gives sample mean abundance patterns ranging from [Fe/H]~$\simeq-1.3$ to $-0.7$ and [Si/Ni]~$\simeq0.3$ to 0, with statistical uncertainties of order 0.1 dex.

\subsection{Comparison to Milky Way stellar abundances}\label{sec:stellar_comparison}

Abundance patterns of the galaxy sample are directly compared with the Milky Way stellar population in Figure~\ref{fig:stellar_comparison}. The track of abundance patterns for different depletions closely follows that of the thick disk. Thick disk stars are offset by $\sim$0.1 dex higher in [Si/Ni] than the galaxy sample mean. Within the uncertainties, mean galaxy abundance patterns may be similar to the thick disk or bulge, or lower by up to $\sim$0.2 dex in [Si/Ni] or equivalently 0.5 dex in [Fe/H].

Integral field spectroscopy of the CSWA sample shows that the star-forming gas is characterized by high local velocity dispersions and low ratios of rotation to velocity dispersion ($V/\sigma$), with $\sim$50\% showing signatures of major merger activity \citep{leethochawalit2016,jones2013}. Kinematic information therefore suggests that ongoing star formation is associated with a thick disk or bulge component. If the enrichment timescale is similar to that of the Milky Way thick disk, then the associated \AFe\ (i.e., [Si/Ni]) suggests that moderate depletions of $\delta_{Fe} \simeq -0.4$ to $-0.9$ are appropriate for the sample. The corresponding gas metallicity is [Fe/H]~$=-1.2$ to $-0.7$.

We do not necessarily expect agreement between the $z\simeq2.2$ galaxy ISM and abundance patterns of Milky Way stars. However, average stellar masses of our sample are similar to those expected for Milky Way progenitors (and abundance matching suggests typical $z=0$ descendant masses $\log \Mstar \simeq 10.8 \, \Msun$ based on \citealt{moster2013}). Given their similar mass scales, the {\em star forming gas} in our sample galaxies should have similar \AFe\ compared to Milky Way stars at fixed [Fe/H] \citep[$\lesssim0.05$ dex; e.g.,][]{choi2014,gallazzi2005}, provided the star formation histories are similar. 
In contrast, outflows may entrain a significant amount of nearly pristine gas resulting in lower metallicity. The $\lesssim0.3$ dex agreement in metallicity between outflow-dominated low ion gas and Milky Way stars, seen in Figure~\ref{fig:stellar_comparison}, suggests a limited and possibly negligible mass of entrained pristine material ($\lesssim$50\% of the total outflow mass).

\subsection{Comparison to nebular oxygen abundance}
\label{sec:nebular_comparison}

Oxygen is the most accessible gas-phase metallicity diagnostic for galaxies at $z=2$--3 owing to its strong optical emission lines. A major concern of emission line studies is the uncertainty in absolute abundance scale \citep[e.g.,][]{kewley2008}, although recent work has made progress with direct measurements and calibrations at high redshifts \citep{jones2015,sanders2016}. Emission lines are additionally sensitive to the ionizing spectrum which depends on stellar [Fe/H] and therefore \AFe, which is likely systematically different at high redshifts than in local star forming galaxies. Independent confirmations of the abundance scale and \AFe\ patterns are clearly needed, and our interstellar absorption data provide such complementary measurements.

In several cases we can make a direct comparison with nebular oxygen abundances in our sample, which suggest typical intrinsic [O/H]~$\simeq-0.4$ in the young stars and ISM \citep[with uncertainty $\sim0.2$ dex in the absolute abundance scale;][]{leethochawalit2016,jones2013}. CSWA 141 is an outlier in terms of having strong high-ionization nebular emission lines and low  metallicity (as well as low stellar mass and high sSFR), with [O/H]~$=-0.7$ measured via the direct method from \Oiii$\lambda$4363 \citep{stark2013}. Overall our sample is in good agreement with the mass-metallicity relation at these redshifts \citep[e.g.,][]{sanders2015}. 

Of the elements probed by our absorption spectra, Si represents the best comparison with nebular [O/H]. We note that non-detections of \Oia$\lambda$1355 absorption in our spectra provide lower limits [\Oia/\Hi]~=~[O/H]~~$\gtrsim-1.8$ for those sources with \Lya-based measurements of \Hi. While not particularly constraining, these limits are robust to ionization corrections and depletion. The same sources have mean [\Siiia/\Hi]~$=-0.7\pm0.2$. This is lower than nebular [O/H]~$\simeq-0.4$ though consistent at the $<$2$\sigma$ level, and indeed previous studies have found reasonable agreement between nebular $\alpha$-element abundances and interstellar [\Siiia/\Hi] \citep{pettini2002,dessauges-zavadsky2010}. However, [\Siiia/\Hi] is susceptible to large changes from both depletion and ionization (and we suggest that anomalous [\Niiia/\Feiia] abundances in previous studies may indicate non-negligible contributions from ionized gas, as in our sample). The effects are opposite in that ionization {\it increases} [\Siiia/\Hi] while depletion {\it decreases} it. Indeed, the example solution shown in Figure~\ref{fig:model} has $-0.35$ dex depletion of Si and 0.28 dex ionization correction to \Siiia, such that the effects approximately cancel in this case. In general the combined systematic effects may be large, however, and we caution that they should be considered carefully. Our best-fit model locus (Equation~\ref{eq:alpha_fe}) corresponds to intrinsic
\begin{equation}\label{eq:SiH}
\mathrm{
[Si/H] = -0.98 - 0.44 \delta_{Fe}.
}
\end{equation}
Comparison with the nebular oxygen abundance then gives
\begin{equation}\label{eq:SiO}
\mathrm{
[Si_{IS}/O_{neb}] \simeq -0.6 - 0.44 \delta_{Fe}, 
}
\end{equation}
explicitly noting that the abundances are from separate interstellar (IS) and nebular (neb) regions. It is clear that Equation~\ref{eq:SiO} is not compatible with solar ratios unless depletion is large or the outflow is diluted with entrained metal-poor gas. 
We discuss the relative abundances of Si and O further in the following section.

\subsection{O/Fe and Si/O}

The degree of $\alpha$ enhancement and [O/Fe] in particular has garnered recent interest for its effect on the ionizing stellar spectra of star forming galaxies at $z\simeq2$--3. For example, \cite{steidel2016} found a large value of [O/Fe]~$\simeq0.6$ from stacked spectra of $z\simeq2.4$ galaxies with properties similar to our sample. 
A direct comparison of [O/Fe] with our findings requires an estimate of intrinsic [Si/O] ratios. Both O and Si are $\alpha$ elements predominantly generated by core-collapse supernovae, and their ratio is often assumed to be near-solar. If we assume solar [Si/O]~$=0$, a value [O/Fe]~=~[Si/Fe]~$\simeq0.6$ is only compatible with our measurements if depletion is negligible. This implies [Fe/H]~$\simeq-1.6$ and [$\alpha$/H]~$\simeq-1$, lower than nebular metallicities of both our sample and that of \cite{steidel2016} by 0.6 dex. If abundances are indeed characterized by solar [Si/O] and high \AFe~$=0.6$, then our measurements imply that the predominantly outflowing ISM seen in absorption must be diluted by a factor of $4\times$ with pristine gas, and must have negligible dust content.

\begin{figure*}
\centerline{
\includegraphics[width=0.8\columnwidth]{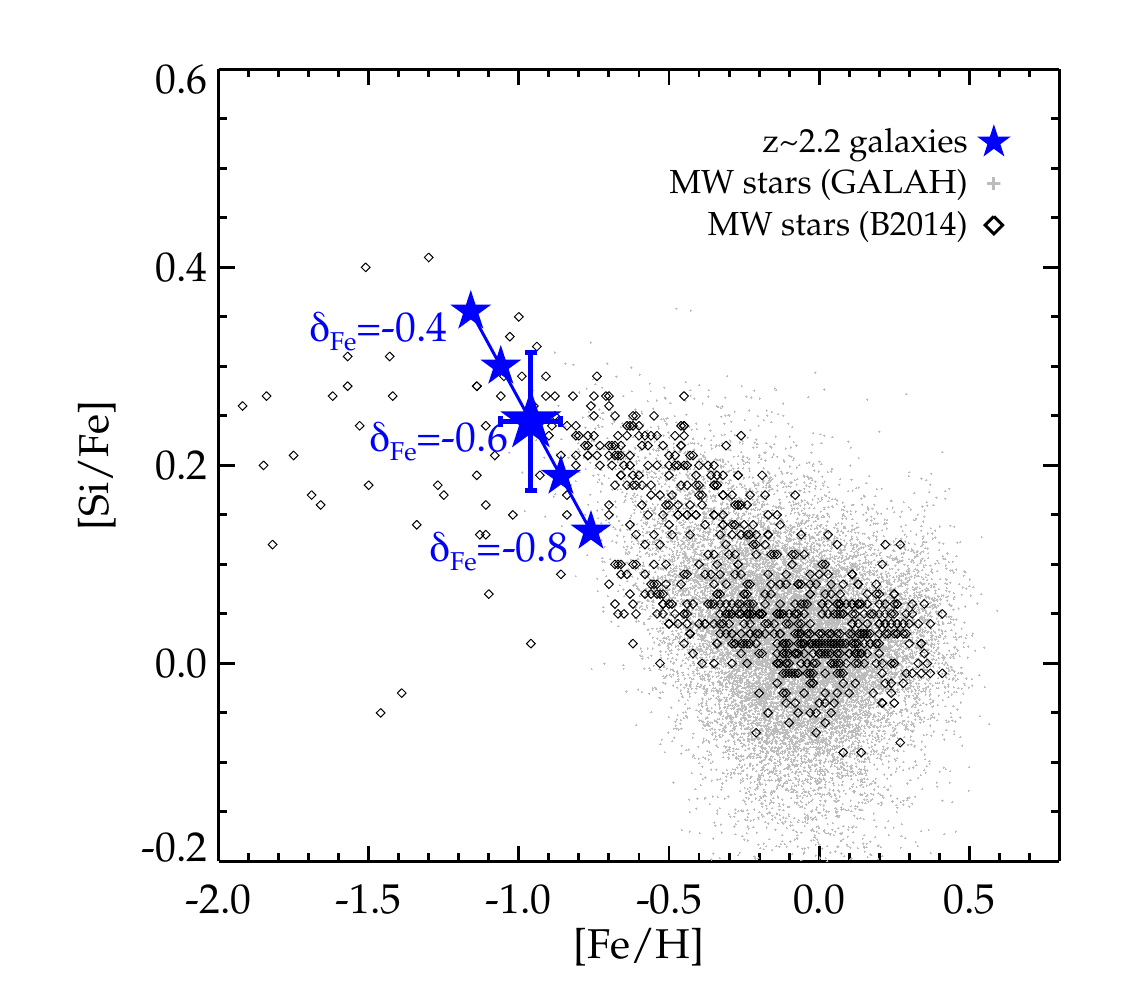}
\includegraphics[width=0.8\columnwidth]{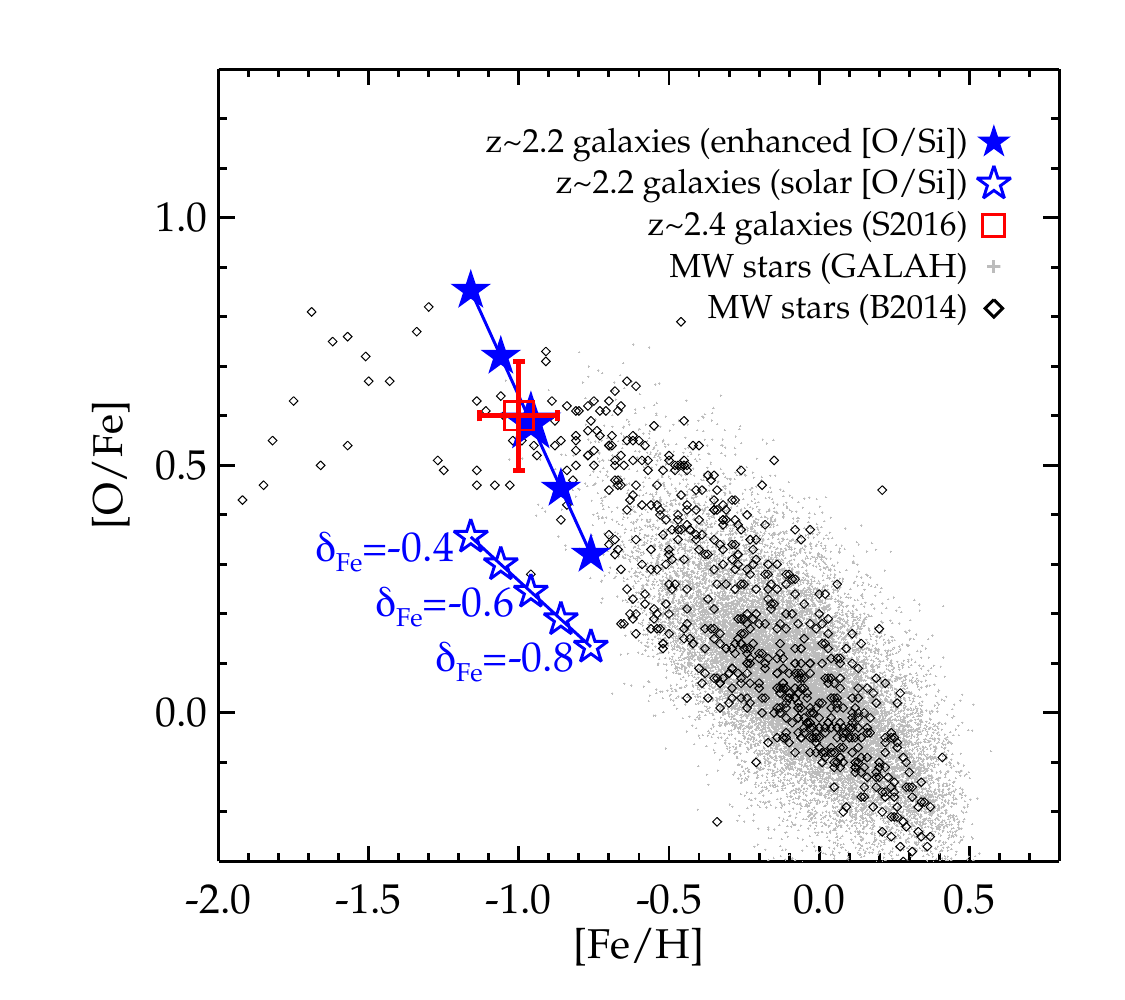}
}
\caption{
\label{fig:alpha_fe}
Comparison of Milky Way stellar \AFe\ abundance patterns with interstellar abundances in the $z\simeq2.2$ galaxy sample (blue stars, shown for a range of Fe depletion $\delta_{\mathrm{Fe}}=-0.4$ to $-0.8$ as labeled). Stellar measurements are from \citet[][B2014]{bensby2014} and the GALAH survey \cite[second data release;][]{buder2018}. 
{\em Left:} [Si/Fe] as a function of [Fe/H]. 
Depletion of $\delta_{\mathrm{Fe}}\simeq-0.6$ gives abundance patterns consistent with the moderately metal-poor thick disk of the Milky Way. Error bars on the fiducial $\delta_{\mathrm{Fe}}=-0.6$ point show the measurement uncertainty. 
{\em Right:} 
[O/Fe] as a function of [Fe/H]. We show separate tracks for the $z\simeq2.2$ sample assuming (1) a solar [O/Si] ratio (open stars, consistent with APOGEE abundance patterns), or (2) super-solar [O/Si] as described in the text (filled stars, consistent with B2014 and GALAH abundance patterns). 
The red square represents nebular [O/H] and stellar [Fe/H] measured by \citet[][S2016]{steidel2016} for stacked spectra of galaxies with properties similar to our sample. We find excellent agreement with S2016 for the case of super-solar [O/Si] and depletions $\delta_{\mathrm{Fe}}\simeq-0.6\pm0.1$. 
}
\end{figure*}

Motivated by recent stellar surveys, we consider non-solar [Si/O] abundance patterns illustrated in Figure~\ref{fig:alpha_fe}. 
Results for the solar ratio with [O/Fe]~=~[Si/Fe] are shown as unfilled symbols in Figure~\ref{fig:alpha_fe}, and are compatible with SDSS APOGEE measurements for $\alpha$-enhanced stars in the Milky Way \citep{hayes2018}. 
In contrast, data from the GALAH survey \citep{martell2017,buder2018} and from \cite{bensby2014} indicate that $\alpha$-enhanced stars in the solar neighborhood have [O/Fe]~$\simeq 2.4\times$[Si/Fe]. 
As with Figure~\ref{fig:stellar_comparison}, our measurements give Si and Fe abundances similar to the Milky Way thick disk if depletions are $\delta_{\mathrm{Fe}} \simeq -0.6$. However, the implied [O/H] differs by 0.3--0.4 dex depending on the assumed [O/Si] pattern. Which, if either, is correct? 
As discussed above, [O/Si]~$\simeq0$ implies negligible depletion and large mass loading by near-pristine gas. Alternatively large depletion factors $\delta_{\mathrm{Fe}}<-1$ dex can bring interstellar and nebular [O/H] into agreement, but imply sub-solar \AFe~$<0$. We note also that if [O/Si] is near solar, then our data imply significantly different abundance patterns than found by \cite{steidel2016}. 
On the other hand, super-solar [O/Si] values give a remarkably consistent picture between interstellar measurements, our estimates of nebular abundance [O/H]~$\simeq-0.4$, and the independent measurements by \cite{steidel2016}, with $\delta_{\mathrm{Fe}}=-0.6$ (Figure~\ref{fig:alpha_fe}). We note that \cite{steidel2016} find nebular [O/Si]~$=0.63$ from their stacked spectra, implying depletions $\delta_{\mathrm{Si}}\simeq-0.3$ dex (and $\delta_{\mathrm{Fe}}\simeq-0.6$) if the intrinsic [O/Si]~$\simeq0.3$. Such good agreement in depletion may indicate survival of dust grains as they are accelerated to the outflow velocities we measure in Section~\ref{sec:kinematics}, or alternatively that dust in outflows and \Hii\ regions undergoes similar processing.

\subsection{Dust to gas ratio}

Abundances and depletion factors can be tested for consistency by considering the mass ratio of dust to gas, or dust to metals. This quantity is also of interest for estimating total gas masses of distant galaxies, given the relative ease of detecting thermal dust continuum compared to atomic or molecular gas emission lines \citep{scoville2014,scoville2016}. 
We can express the dust-to-gas mass ratio (DGR) as a function of Fe depletion via
\begin{equation}\label{eq:DGR}
\mathrm{DGR} = \frac{(1-10^{\delta_{\mathrm{Fe}}})}{f_{\mathrm{Fe}}} \frac{56}{1.4} \mathrm{\frac{N(Fe)}{N(H)}}
\end{equation}
where 56 and 1.4 are the approximate mean ion masses (in atomic mass units) for Fe and for the entire gas phase, respectively. N(Fe) and N(H) refer to the total abundances (i.e. the metallicity). 
The term $1-10^{\delta_{\mathrm{Fe}}}$ is the fraction of Fe found in solid grains, and $f_{\mathrm{Fe}}$ is the Fe mass fraction of solid grains. The value of $f_{\mathrm{Fe}}$ is of order $\sim 0.1$ (e.g., 0.22 for the solar metallicity chondrite composition described by \citealt{draine2007}, and we expect lower $f_{\mathrm{Fe}}$ for environments with super-solar \AFe). 

For a fiducial abundance [Fe/H]~$=-1$ and $\delta_{\mathrm{Fe}}=-0.6$ for our sample (e.g., Figure~\ref{fig:alpha_fe}), Equation~\ref{eq:DGR} gives
$$\mathrm{DGR} = 10^{-3} \frac{0.1}{f_{\mathrm{Fe}}}.$$ 
A change of $-0.1$ dex in $\delta_{\mathrm{Fe}}$ corresponds to $+0.13$ dex in DGR. 
This is lower than DGR~$\simeq10^{-2.2}$ found in the Milky Way or in the SINGS galaxies with reliable SCUBA data discussed by \cite{draine2007}, but within the range found by \cite{engelbracht2008} for galaxies of comparable gas-phase metallicity. DGR exhibits a large scatter at fixed metallicity which may be due in part to mixing of relatively enriched and pristine material. If so, this may provide further joint constraints on the dust content, total metallicity, and entrainment of pristine gas in outflows. \citealt{draine2007} suggest that DGR~$\approx0.01 \frac{Z}{Z_{\odot}}$ for unmixed ISM, compatible with our measurements and thereby indicating limited entrainment of pristine gas.

We draw two notable conclusions from the DGR. First, our measurements are consistent with the range seen in local galaxies of similar metallicity, providing a good consistency check and suggesting that outflow composition is similar to the ISM. Second, we infer low DGR compared to the Milky Way ISM, suggesting caution in estimating gas masses based on dust continuum emission. If our measurements are representative of the total ISM, then we expect a factor $\sim$5 difference in total gas mass per unit dust emissivity relative to the Milky Way. 
Estimates of gas mass using a Milky Way-like calibration factor would be underestimated by the same factor of 5 (and hence, e.g., a true gas fraction $f_{gas}=0.6$ would be misclassified as $f_{gas}=0.23$). Sub-millimeter observations of thermal dust emission from our sample would further test this scenario, in combination with independent dynamical gas mass estimates \citep[e.g.,][]{leethochawalit2016}.

\subsection{Spatial extent of absorbing gas}\label{sec:spatial_extent}

The spatial distribution of outflowing material is a crucial parameter for determining mass loss rates, and whether outflows exceed the local escape velocity. Although we lack direct spatial information for our sample, we can consider several independent constraints. Foremost, bulk velocities $v\simeq-150 \, \kms$ imply galactocentric distances 
$$\mathrm{ d = \frac{v}{-150 \kms} \frac{t}{100\, Myr} \times 15\, kpc } $$
or $\sim$45 kpc if the material travels at constant velocity over the characteristic star formation timescales sSFR$^{-1} \simeq$~300 Myr. We expect the majority of column density to be within this radius. This is supported by measurements along transverse sightlines to $z\simeq2.2$ galaxies, in which the vast majority of \Hi\ and low ion absorption is concentrated within radii $<$30 kpc \citep{steidel2010}. 

We can estimate the spatial extent of outflows in a statistical sense by comparison with quasar absorption systems. 
{ 
We first consider the broad population of systems identified by \Hi\ absorption.
}
\Hi\ column densities in the galaxy spectra correspond to DLA absorbers, which have line density $dN_{DLA}/dz \simeq 0.2$ per unit redshift at $z=2.2$, or $\sim10^{-4}$ per Mpc \citep{sanchez-ramirez2016}. Defining $f_{DLA}$ as the fraction of DLAs arising in galaxy outflows similar to our sample, the cross sectional area of outflowing gas $\sigma_{abs} = \pi R_{abs}^2$ is related to galaxy number density as
\begin{equation}\label{eq:cross_section}
n_{gal} = \sigma_{abs}^{-1} f_{DLA} \frac{dN_{DLA}}{dz}.
\end{equation}
The typical stellar mass of our galaxies suggests comoving volume densities $n_{gal} \simeq 10^{-3}$ Mpc$^{-3}$ \citep{tomczak2014}, hence
\begin{equation}\label{eq:radius}
R_{abs} \simeq 0.2 \sqrt{f_{DLA}} \, \mathrm{Mpc}.
\end{equation}
{
However, we have shown that the kinematics and abundance ratios seen down-the-barrel in galaxy spectra are not typical of the DLA population, suggesting $f_{DLA}$ must be small. 
This is supported by a range of studies indicating that the majority of DLAs
at $z\sim2$ are not associated with massive galaxies \citep[e.g.,][]{fumagalli2010,kulkarni2010,perez-rafols2018}. 
Absorption kinematics are particularly constraining in this regard. All galaxies except one in our sample have characteristic $\Delta v_{90} \gtrsim 250$ \kms. Since we observe only the blueshifted and systemic gas, the value towards a background source is likely doubled: $\Delta v_{90} \gtrsim 500$ \kms. None of the 41 DLAs within the complete XQ-100 sample show such broad velocities, and only a few exhibit the distinct ion abundance patterns of our sample. If we suppose that one of the XQ-100 DLAs is associated with the absorption systems seen in the galaxy spectra, then $f_{DLA} = 1/41$ gives a characteristic radius $R_{gal} = 30$ kpc. Equivalently, if the absorption we observe arises from within $R\leq30$ kpc from the galaxies, then we expect 2.5\% of DLAs to be associated with such outflows. The scarcity of similar systems among DLA populations is therefore explained if the bulk of absorbing material is at impact parameters within a few tens of kpc. 
}

\begin{figure}
\centerline{
\includegraphics[width=\columnwidth]{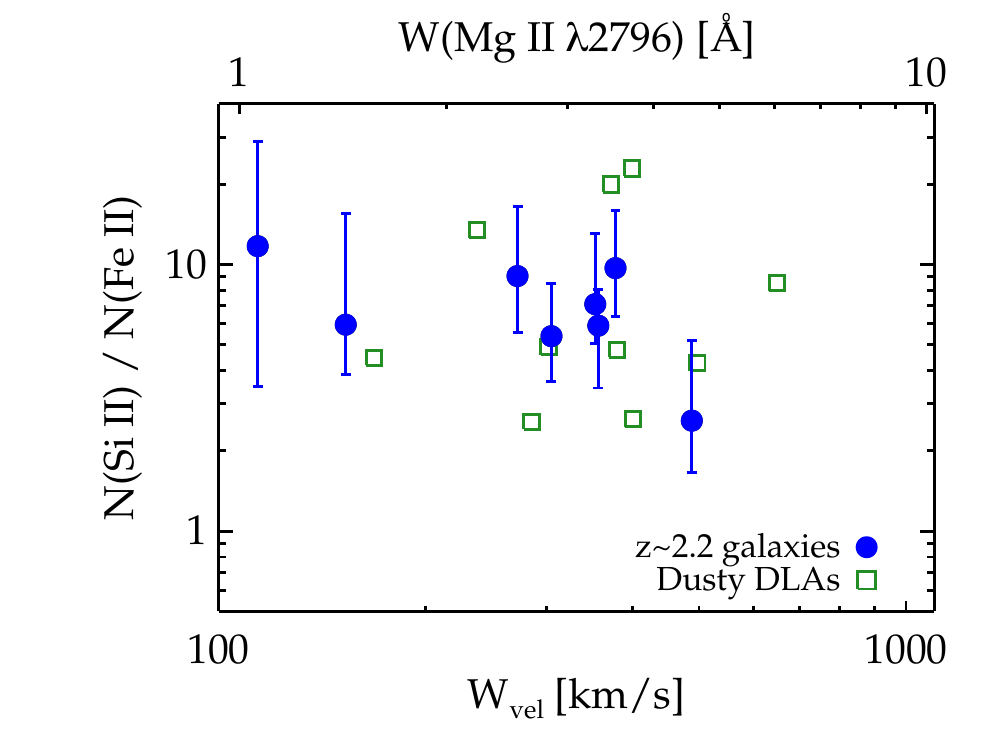}
}
\caption{
\label{fig:W_MgII}
Equivalent width of saturated low ion transitions in our sample. We define $W_{vel}$ from the covering fraction profile which is in turn directly proportional to the equivalent width of saturated transitions (Equations~\ref{eq:W_vel} and \ref{eq:W_lambda}). \Mgii\ rest-frame equivalent widths are given on the upper axis as an example, and the \cite{ma2017} ``dusty DLA'' sample is shown for comparison. Absorption profiles in the galaxy sample are comparable to these dusty DLAs, with typical W(\Mgii$\lambda$2796)~$\simeq3$~\AA. Such strong \Mgii\ absorbers are $\sim$10 times less abundant than DLAs at these redshifts \citep{zhu2013}. Metal ion equivalent widths are therefore useful for identifying the rare subset of quasar absorption systems which trace galactic outflows.
}
\end{figure}

{
We next consider which subset of quasar absorption systems are most likely to be associated with galaxy outflows. Metal ion equivalent widths are a powerful discriminant, as the chemical enrichment and broad velocities of outflows give rise to large equivalent widths compared to typical DLAs. 
Of the various metal transitions, \Mgii$\lambda\lambda$2796,2803 has been extensively studied and provides a good basis for comparison with quasar absorbers. Moreover, a variety of studies have associated strong \Mgii\ absorption systems at $z\sim1$ with galactic outflows \citep[e.g.,][]{zibetti2005,bouche2007,menard2009,bordoloi2011,lundgren2012}. 
We estimate \Mgii\ equivalent widths of our sample from velocity-integrated low ion covering fraction profiles, defining 
\begin{equation}\label{eq:W_vel}
W_{vel} = \int f_{cov} \, dv.
\end{equation}
The values for our sample are shown in Figure~\ref{fig:W_MgII}. $W_{vel}$ is versatile in that a saturated transition at wavelength $\lambda$ has equivalent width given by 
\begin{equation}\label{eq:W_lambda}
W = \frac{W_{vel}}{c} \times \lambda.
\end{equation}
While \Mgii\ is not directly measured in most cases, we expect the strong $\lambda$2796 transition to be optically thick based on other low ion column densities (confirmed in the case of CSWA 141). Therefore we use Equation~\ref{eq:W_lambda} to calculate W(\Mgii$\lambda$2796), with results given on the upper axis in Figure~\ref{fig:W_MgII}. 
This demonstrates that low ion profiles in our sample correspond to rest-frame W(\Mgii $\lambda$2796)~$\simeq 3$~\AA. As we only observe the blueshifted absorption, equivalent widths seen towards a background source would be higher \citep[$\sim$5 \AA; e.g.][]{steidel2010}. 
Analogous to Equation~\ref{eq:radius}, we can now constrain the spatial extent of outflows based on the incidence of strong \Mgii\ absorbers seen in quasar spectra. Adopting a conservative threshold W(\Mgii $\lambda$2796)~$\geq 3.5$~\AA\ (with line density $dN_{MgII}/dz \simeq 0.01$ at $z=2.2$; \citealt{zhu2013}) gives
\begin{equation}\label{eq:radius_MgII}
R_{abs} \simeq 50 \sqrt{f_{MgII}} \, \mathrm{kpc}.
\end{equation}
Here $f_{MgII}$ is the fraction of such \Mgii\ absorbers arising in outflows similar to our sample. 
\cite{bouche2012} have carried out a search for star forming galaxies within $\sim$100 kpc of strong \Mgii\ absorbers at $z\simeq2$, with a success rate of 4/20. The impact parameters 6--26 kpc of associated galaxies further suggest $f_{MgII}\sim0.2$ and $R_{abs} \simeq 25$ kpc. 
Comparison with Equation~\ref{eq:radius} shows clearly that these galactic outflows comprise at most a few percent of the DLA population, considering only the fact of their low ion equivalent widths. Furthermore we reaffirm that the outflows are confined to at most $\sim$50 kpc and quite possibly much smaller radii.
}

{
If the reasoning of this section is correct, we expect that the subset of quasar absorption systems with properties similar to our sample (e.g. in terms of column density, abundance ratios, velocity spread, and equivalent width of low ions) are associated with star forming galaxies at small angular separation. 
In Section~\ref{sec:f_ej} we discuss further evidence for small characteristic impact parameters from an independent chemical yield argument.
}
The closest analogs in Figures~\ref{fig:kinematics_dla}, \ref{fig:abundance_ratios}, and \ref{fig:W_MgII} are the 2175~\AA\ dust absorbers described by \cite{ma2017}. Recently \cite{ma2018} report Hubble Space Telescope followup of one such system which reveals the $z=2.1$ host galaxy at an impact parameter of only 5.5 kpc, supporting this hypothesis. 
We propose that other similar quasar absorption systems are associated with hosts at scales $R\lesssim30$ kpc or equivalently $\lesssim$ 3\farcs5, which can be further tested via targeted searches for host galaxy emission lines using integral field spectrographs \citep[e.g.,][]{bouche2012}.
{
We emphasize that these systems represent only a small percentage of the quasar absorber population, but can be efficiently identified on the basis of metal absorption properties. 
}

\subsection{Mass loss and recycling rates}\label{sec:mass_loss}

Mass loss rates can be estimated from the column density and velocity profiles of outflowing material. The accuracy is limited by uncertainty in the geometry of the outflow, in particular its distribution in galactocentric radius which we discuss above. Here we will adopt a toy model shell geometry characterized by a shell radius $R$, width $\Delta R$, and number density $n = \rho / \mu$. Here $\rho$ is the mass density and $\mu\simeq1.4$ AMU is the mean ion mass assuming the outflow is dominated by atomic or ionized H and He. 
At radius $R$, the flux of mass in a given time interval $dt$ is
\begin{equation}\label{eq:dM_out}
dM = 4 \pi R^2 n \mu v\, dt
\end{equation}
where $v$ is the outflow velocity. The mass loss rate is 
\begin{equation}\label{eq:Mdot}
\dot{M} = 4 \pi R^2 n \mu v = 4 \pi R \frac{R}{\Delta R} N \mu v
\end{equation}
with column density $N = n\times\Delta R$ for the shell geometry. This is equivalent to the formalism of \cite{pettini2000} with $R/\Delta R = 3$. 
{ We note that while a spherically symmetric geometry is appropriate for an ensemble average, individual galaxy outflows are presumably not isotropic. Consequently the outflow major axes for individual galaxies will extend beyond the radial distances in Equations~\ref{eq:dM_out} and \ref{eq:Mdot}, by an amount dependent on the degree of collimation.
}

Column density $N$ in Equation~\ref{eq:Mdot} represents the total of all ionic and molecular species. In practice we can redefine $N$ in terms of a single species and its abundance. Hydrogen would be a natural choice as it dominates the mass. In our case, the velocity and column density profiles are more robustly constrained for metal ions. \Feiia\ is the most uniformly well determined ion and the best available tracer of \Hi. Adopting \Feiia\ as a reference yields
\begin{equation}\label{eq:Mdot_Fe}
\begin{split}
\dot{M} = & - \frac{R}{\mathrm{5\, kpc}} \, \frac{R}{\Delta R} \, \mathrm{\frac{N_{Fe II}}{10^{14.5} \, cm^{-2}}} \, \mathrm{\frac{N_{Fe}}{N_{Fe II}}} \, 10^{\mathrm{-(1+[Fe/H])}} \\
& \frac{v}{-150 \mathrm{\kms}} \times 10\, \mathrm{M_{\odot}\, yr^{-1}}
\end{split}
\end{equation}
where the negative value implies mass loss. 
The main uncertainty is due to galactocentric distance mean ($R$) and range ($\Delta R$). We expect these to be $\lesssim$30 kpc as discussed above, and larger than the characteristic galaxy sizes ($\gtrsim$3 kpc), hence we can obtain results to within an order of magnitude. The product of depletion (via Fe/\Feiia) and metallicity correction terms is actually well defined from our results in Equation~\ref{eq:FeH}, and represents a factor of $\sim4\times$. Estimating the remaining terms to be of order unity, mass loss rates of {\em low ionization gas alone} are $\dot{M} \sim 40$ $\sfrunit$. This is $\sim3\times$ larger than star formation rates of the sample. The implied mass loading factors $\eta=\dot{M}/$SFR are therefore at least of order unity and may be substantially higher when accounting for ionized gas phases, or if the galactocentric distances are $>$5 kpc. 

We now turn briefly to the question of how much enriched material is being recycled back to the galaxy via a ``galactic fountain'' process \citep[e.g.,][]{bregman1980}. It appears that relatively little metal mass is recycling, given the low column densities at positive velocity (inflowing). From the analysis in Section~\ref{sec:kinematics}, the typical mass flux $\dot{M} \propto N v$ associated with inflowing velocities is 0.13 that of the outflowing velocities. A systemic ISM component at $v\simeq0$ would further reduce the ratio of inflow/outflow flux. On the other hand, relative inflow rates could be larger if the inflowing material is predominantly at larger distances. We would then have
\begin{equation}\label{eq:Min}
\frac{\dot{M}_{in}}{\dot{M}_{out}} \leq 0.13 \left( \frac{R_{in}}{R_{out}} \right)^2
\end{equation}
with the upper bound corresponding to zero ISM component. We note that this result is derived from the low ions only. Larger velocity spreads for the higher ions (Figure~\ref{fig:kinematics_al3}) permit higher recycling rates if the inflowing component is more highly ionized. We conclude that the metal recycling rate is not more than $\sim$10\% of the metal outflow rate, unless the inflowing gas is significantly farther from the galaxy or more highly ionized.

\subsection{Fraction of metals ejected}\label{sec:f_ej}

The impact of outflows on galaxy chemical evolution depends on both the mass loss rate and the relative {\em fraction} of metals which are ejected in outflows. The fraction of metals lost is better constrained than total mass loss, as the metal budget can be derived from a galaxy's stellar population. In particular \cite{peeples2014} find that $\sim L^*$ galaxies at $z=0$ -- encompassing the expected descendants of our sample -- retain only $\sim$20-25\% of their total metals while the rest are ejected over the course of their formation histories. Our measurements provide an independent view at the peak epoch of metal production.

We can address the ejected metal fraction most readily with the $\alpha$-element Si. Production of Si is dominated by core collapse supernovae on approximately instantaneous time scales. For comparison purposes we adopt the same Type II supernovae yields as \cite{peeples2014}, such that a star formation rate of 1 $\sfrunit$ corresponds to 0.0014 $\sfrunit$ of Si production. The typical production rate for galaxies in our sample is then $\dot{M}_{Si,SFR} = 0.02\, \sfrunit$ (median). 
Rewriting Equation~\ref{eq:Mdot} in terms of the Si mass loss rate, we have
\begin{equation}\label{eq:Mdot_Si}
\begin{split}
\dot{M}_{Si} & = 4 \pi R \frac{R}{\Delta R} \mathrm{N_{Si}} \, \mathrm{\mu_{Si}} \, v \\
& = - \frac{R}{\mathrm{5\, kpc}} \, \frac{R}{\Delta R} \, \mathrm{\frac{N_{Si II}}{10^{16} \, cm^{-2}}} \, \mathrm{\frac{N_{Si}}{N_{Si II}}} \, \frac{v}{-150 \mathrm{\kms}} \\
& \;\;\;\; \times 0.022\, \sfrunit.
\end{split}
\end{equation}
We measure a median $\mathrm{N_{Si II}}=3.4\times10^{15}$ cm$^{-2}$ and infer Si depletions of $\sim$0.25 dex from the discussion above (i.e., $\mathrm{\frac{N_{Si}}{N_{Si II}}} = 1.8$ accounting only for the low ions and associated dust). 
Adopting these values together with $v = -150$ \kms, we can express the median instantaneous fraction of Si metal loss as the ratio of ejection in outflows to production in supernova:
\begin{equation}\label{eq:f_ej}
f_{ej,Si} = \frac{\dot{M}_{Si,ej}}{\dot{M}_{Si,SFR}} = 0.6 \frac{R}{\mathrm{5\, kpc}} \, \frac{R}{\Delta R}.
\end{equation}
In other words, $\sim$60\% of the $\alpha$-element metal budget is {\it ejected} in the low ion and solid phases alone! While \Siia\ and \Siiva\ are sub-dominant compared to \Siiia, both \Siiiia\ and a more highly ionized outflow phase could further increase $f_{ej,Si}$. Given the limited remaining metal budget, however, Equation~\ref{eq:f_ej} affirms the importance of low ions and dust in tracing a large fraction of the total outflow mass.

Equation~\ref{eq:f_ej} shows clearly that a substantial fraction and perhaps the {\it majority} of all $\alpha$ elements generated by supernovae are being ejected in outflows, with a low ionization phase dominating the column density. This is in good agreement with the FIRE cosmological simulations discussed by \cite{muratov2017} who explicitly quantified the thermal state of heavy elements. At similar stellar masses and $z\simeq2.2$, these authors find that a majority of metals are ejected from galaxies, with $\sim$70\% of CGM metals in a low ionization phase \citep[see in particular Figure~12 of][]{muratov2017}. These simulations have mass loading factors of $\sim$3--5 for stellar masses and redshifts similar to our sample \citep{muratov2015}, similar to our results in Section~\ref{sec:mass_loss}. Notably, the simulations also reproduce observed mass-metallicity relations via this mass and metal loss. 
While the large fraction found in Equation~\ref{eq:f_ej} may seem extreme, it is in fact expected both from simulations and from the $z=0$ metal budget accounting by \cite{peeples2014}.

Radial distribution factors are left as variables in Equation~\ref{eq:f_ej} since they are not directly measured. However, we can derive interesting limits on the spatial extent and other outflow properties from the simple metal budget argument used here. While not strictly necessary at all times, the total metal content of galaxies generally increases over time which requires $f_{ej}<1$. Applying this condition gives a limit to the characteristic radius
\begin{equation}\label{eq:R_max}
{\mathrm R \times \frac{R}{\Delta R}} < 8\, \mathrm{kpc}.
\end{equation}
Such small radial scales support our conclusions from Section~\ref{sec:spatial_extent} and explain why the outflow properties in our sample are so rarely seen in quasar absorber surveys. This result motivates a need for sightlines at small impact parameters ($\lesssim10$ kpc) in order to probe the majority of outflowing mass. 
By the same argument, depletion factors must be $\delta_{Si} \lesssim0.5$ (and corresponding $\delta_{Fe}\lesssim1.1$) to keep $f_{ej}<1$, with the precise constraint limited by the unknown radial extent. The presence of metals in phases other than the low ions and solids is likewise constrained to be small (i.e. column density similar or less than the low ions). These constraints further justify our focus on the low ion phase as likely representative of the total outflow composition.

In summary, accounting of the metal budget -- comparing observed column densities with expected stellar yields -- provides strong constraints on the radial extent, dust depletion, and any unseen ionization states of outflowing material. 
Our measurements indicate a theoretically-supported picture in which the majority of heavy elements are ejected in a predominantly low ionization outflow, which regulates galactic chemical evolution.

\subsection{Diversity within the galaxy sample}

We have focused the discussion largely on mean characteristics of the sample, and we now consider the extent to which individual galaxies deviate from the average in various properties. 
Low ion column density ratios are remarkably consistent within measurement uncertainties across the sample (Figure~\ref{fig:abundances_low}). 
Mean low ion velocities vary but are uniformly within 0 to $-200$ \kms\ relative to systemic, with covering fraction profile widths spanning a narrow range $\sigma_v \simeq 150$ to 200 \kms (Figure~\ref{fig:kinematics_lowion}). 
In contrast to the uniform kinematics and abundance ratios, the total column densities vary by a factor of $\sim$10 within the sample. The covering fraction profiles also vary by a factor of $\sim$2 (Figure~\ref{fig:fcov_low}).

Homogeneous low ion abundance ratios suggest that intrinsic abundance patterns, depletion factors, and ionization factors are similar throughout the sample. This extends to previous studies which we have shown to be remarkably consistent with our sample properties \citep[i.e., the galaxies studied by][]{pettini2002,quider2009,dessauges-zavadsky2010}. 
The most significant variations are in the total inferred interstellar mass and its covering fraction. These are correlated in the sense that galaxies with higher covering fractions have higher column densities. The correlation is partially but not entirely due to our definition of $N_{tot} \propto f_c$. We might expect the low ion column density to depend on ionization state, but this does not appear to be significant: the ionization indicator [\Niiia/\Feiia] shows no strong correlation with total column densities (Figure~\ref{fig:abundances_low}). 

While we find that our sample is relatively homogeneous in many properties of the interstellar and outflowing medium, we can also give insight into the systematic trends of strong absorption properties found in stacked spectra. Perhaps most notable is the variation in equivalent width of saturated low ion transitions, which is most strongly correlated with \Lya\ equivalent width \citep{shapley2003,jones2012,du2018}. This is clearly driven by variations in the gas covering fraction: we see a wide range in low ion covering fraction within the sample (sufficient to explain the variation in stacked spectra; Figure~\ref{fig:fcov_low}) yet very little range in the velocity width of strong low ion transitions (Table~\ref{tab:kinematics}). This supports the conclusions of our earlier work \citep{jones2013,leethochawalit2016} with a larger sample. Variations in covering fraction also contribute partially to the range of column densities $N_{tot}$ as noted above. Additional trends with demographic properties such as galaxy stellar mass, SFR, optical extinction, and surface density are undoubtedly present but our sample is not yet large enough to characterize these dependencies. Nonetheless these data provide clear guidance for the interpretation of statistically significant trends from samples of hundreds to thousands of galaxies at these redshifts.

\section{Summary}\label{sec:summary}

We have examined the kinematics, geometric covering fraction, and chemical composition of the interstellar medium in a sample of 9 galaxies at redshifts $z=1.4$--2.9. Galaxies in our sample are gravitationally lensed but otherwise representative of the star forming population at these redshifts. Our analysis takes advantage of optically thin absorption lines probed by high quality spectra, necessary to study the mass distribution and relative abundances of different elements in both the gaseous and solid ISM phases, as well as to guide interpretation of results from lower-resolution composite or individual galaxy spectra. 
Our main results are as follows.

\begin{itemize}

\item {\it Covering fraction:} 
The low ionization gas exhibits a wide range of geometric covering fractions. $f_c$ varies as a function of velocity, with maximum covering fractions ranging from $f_c \simeq 0.4$ to 1 for individual galaxies in our sample. Most galaxies have non-uniform covering at all velocities ($f_c < 1$). Variations in covering fraction are able to explain demographic trends measured from stacked spectra of galaxies at similar redshifts, such that a lower covering fraction is correlated with higher \Lya\ equivalent width and lower reddening (i.e., dust column density). Correlations with \Lya\ and reddening arise naturally from the presence of \Hi\ and dust, respectively, in the low ionization phase.

\item {\it Kinematics:} 
We use optically thin transitions to measure gas column densities as a function of velocity. In contrast to the range of covering fractions, low ionization gas kinematics show only modest variation within the sample. Approximately 80\% of the low ion phase is associated with a net outflow, with absorption detected at speeds of at least 400 \kms\ in every galaxy. Mean velocities along the line of sight are $\simeq -150$ \kms. We find little or no significant metal recycling ($\lesssim 10$\%, if inflows and outflows are similar in distance and ionization). Strong absorption transitions are systematically $\sim2\times$ broader in velocity width than the column density distribution but have consistent centroids, such that saturated transitions can be used to accurately measure mean bulk velocities.

\item {\it Composition:} 
Low ion column density ratios provide measurements of the metallicity, dust depletion, and nucleosynthetic abundance patterns. Column density ratios in the outflowing gas show remarkably little dispersion within the sample, yet they are distinct from other well-characterized astrophysical sources (e.g. DLAs, nearby galaxies' ISM, and stellar abundances). 
We infer that the outflowing medium is characterized by sub-solar metallicity, $\alpha$ enhancement, and moderate dust depletion ([Fe/H]~$\simeq -0.9$, [Si/Fe]~$\simeq 0.2$, and $\delta_{Fe} \simeq -0.6$ dex). Depletion is highly important in determining abundance patterns from gas-phase transitions: we find that the majority of Fe and Ni atoms are in the solid state, along with approximately half of the Si. 
Dust-to-gas ratios are lower than in the Milky Way ISM, suggesting caution in calibrating thermal dust emission to estimate ISM gas masses. 
While these results have some degeneracy, the abundance patterns are in good agreement with metal-poor and $\alpha$-enhanced stars in the Milky Way thick disk and bulge. We consider this evidence for {\it in situ} thick disk and/or bulge formation at these redshifts (corroborated by kinematics of the star forming gas available for a subset of the CSWA sample). 
Our interstellar data are in good agreement with recent measurements of nebular and stellar abundances from composite spectra by \cite{steidel2016}, provided that [O/Si] is highly super-solar.

\item {\it Mass and metal outflow rates:} 
We determine mass loss rates and mass loading factors to within an order of magnitude accuracy for the low ionization phase, limited by uncertainty in the radial distribution of outflowing gas. Typical mass loss rates are $\sim40\, \sfrunit$ with corresponding mass loading factors $\dot{M}_{out}/\mathrm{SFR} \sim 3$. Total mass loss rates may be somewhat larger when accounting for highly ionized gas. Comparing the instantaneous rates of metal production by Type II supernovae versus metal loss in outflows, we find that of order half of all metals produced are being ejected in the low ionization and solid phases alone, consistent with censuses of the metal budget for descendent $z=0$ galaxies. 
Requiring the metal production rate to be greater than the ejection rate, we find that the characteristic outflow radii must be $\lesssim 10$ kpc. Such small impact parameters are rarely sampled with background sightlines, explaining the scarcity of quasar absorption systems with similar physical properties. 

\end{itemize}

This work represents the first characterization of the ISM and outflow properties based on {\em column densities} measured for a sample of galaxies at $z\simeq2-3$, when star formation and associated feedback are most active. Such information is essential for understanding the role of feedback and large-scale outflows in regulating gas content, star formation, and metallicity of galaxies and their surrounding medium. We envision a number of feasible avenues to further extend and improve upon our results in the near future. 
First, knowledge of the chemical compositions will benefit from column density measurements of additional ions especially at blue wavelengths ($\lambda_{rest} < 1260$ \AA), most notably \Hi\ which is available only for 4 of 9 galaxies in our sample. Metal ions with a range of ionization and depletion factors will help to more accurately correct for these effects, and to better probe abundance patterns of multiple nucleosynthetic groups (e.g. the iron peak, $\alpha$ capture, and secondary elements). \Nia, \Suiia, \Piia, \Criia, \Mniia, and \Zniia\ are all achievable with moderately deep spectroscopy. Volatile elements such as S and Zn are of particular interest (and we find that even low signal-to-noise measurements of \Zniia\ are valuable).

Mass loss rates and metal loss fractions can be improved substantially with better information on the spatial distribution of outflows (i.e., their radial extent). We find that the majority of the column density is likely within galactocentric distances of order 10 kpc. One possible way forward is to specifically target these small impact parameters with background sightlines (e.g., QSOs); such alignments are rare and current surveys typically do not probe this regime. 
In our view a more promising approach is to spatially map fluorescent emission from fine structure transitions of \Siiia* and \Feiia* which originate in the outflowing gas \citep[e.g.,][]{jones2012}. 
Recently the first such measurement at $z>1$ shows a \Feiia* half-light radius of $\simeq$4 kpc \citep[$\sim$0\farcs5; ][]{finley2017}, supporting our conclusions. Areal magnification from gravitational lensing is needed to adequately sample these scales with seeing-limited instruments. Our lensed sample is ideal for mapping fine structure emission with the newly commissioned Keck Cosmic Web Imager (KCWI) which we are now pursuing.

Finally, while this work quadruples the previously available sample, we ultimately seek equivalent data for larger samples of galaxies spanning a range of redshift and physical properties. Enlarged samples are needed to firmly establish demographic trends such as the mass loading factor as a function of stellar mass. Characterizing these scaling relations will allow us to better understand origins of the stellar-to-halo mass relation, mass-metallicity relation, and other fundamental properties of the galaxy population. 
In the short term, gravitational lensing continues to be a productive approach. Bright lensed galaxy samples are growing thanks to both wide area sky surveys \citep[e.g. the Dark Energy Survey and Pan-STARRS;][]{nord2016,chambers2016} and targeted lensing surveys \cite[e.g. eMACS and RELICS;][]{ebeling2013,salmon2017}. 
Non-lensed galaxies may become accessible with $\sim$100 hour integration times using multi-object spectrographs, and 30-meter class optical telescopes will more easily reach the bright field galaxy population. However we note that current ultra-deep surveys such as VANDELS \citep{pentericci2018} and the MUSE UDF \citep{bacon2017} lack the spectral resolution and blue wavelength coverage needed for this work. 
We emphasize that upcoming 30-meter class telescopes will require instrumentation with moderate to high spectral resolving power and good throughput in the near-UV ($R\gtrsim5000$ and $\lambda<4000$ \AA, ideally reaching the atmospheric cutoff) to optimally address chemical abundance patterns at $z\simeq2$.

\begin{acknowledgements}

We thank Ramesh Mainali for providing measurements of the stellar masses; George Becker, Sarah Loebman, and Crystal Martin for several enlightening conversations; and the referee for several suggestions which considerably improved the discussion and clarity of the manuscript.
RSE acknowledges financial support from the European Research Council
from an Advanced Grant FP7/669253. 
DPS acknowledges support from the National Science Foundation  through the grant AST-1410155. 
TJ acknowledges support from NASA through Hubble Fellowship grant HST-HF2-51359.001-A awarded by the Space Telescope Science Institute, which is operated by the Association of Universities for Research in Astronomy, Inc., for NASA, under contract NAS 5-26555. 
This work is based on data obtained at the W. M. Keck Observatory, which is operated as a scientific partnership among the California Institute of Technology, the University of California, and the National Aeronautics and Space Administration. The Observatory was made possible by the generous financial support of the W. M. Keck Foundation. 
We wish to acknowledge the very significant cultural role and reverence that the summit of Maunakea has within the indigenous Hawaiian community.  We are most fortunate to have the opportunity to conduct observations from this sacred mountain, and we respectfully say mahalo. 

\end{acknowledgements}

\begin{appendix}

\section{Systemic redshifts}\label{sec:zsys}

In this section we briefly describe the spectral features used to measure systemic redshifts for individual sources. Redshifts were measured from line centroids determined from Gaussian fits, compared to reference vacuum transition wavelengths from the NIST Atomic Spectra Database\footnote{http://physics.nist.gov/asd} \citep{kramida2016}. In cases of emission line doublets, we simultaneously fit both lines assuming a common redshift and velocity dispersion, to maximize precision. This applies to O {\sc iii}] $\lambda\lambda$1661,6, \Ciii\ $\lambda\lambda$1907,9, and \Oii\ $\lambda\lambda$3727,9. Similarly, in several cases we fit stellar photospheric absorption lines simultaneously with a common redshift and velocity dispersion in order to improve the significance and fidelity of redshift measurements. For this analysis we use the relatively unblended Si {\sc iii} $\lambda$1294, Si {\sc iii} $\lambda$1417, S {\sc v} $\lambda$1501, N {\sc iv}$\lambda$1718, and C {\sc iii} $\lambda$2297 features.

In several cases we also measure redshifts of fine structure emission lines arising from Si {\sc ii}* and Fe {\sc ii}* (where ``*'' denotes fluorescent emission to excited fine structure ground states). These lines are thought to arise from de-excitation into the excited ground state following absorption by interstellar gas \citep[e.g.,][]{jones2012,prochaska2011}. Since resonant absorption transitions of Si {\sc ii} and Fe {\sc ii} typically show a much larger velocity extent than stars and \Hii\ regions, the corresponding fine structure transitions do not necessarily trace the systemic velocity. However their centroids are typically consistent with the systemic redshift \citep[within $\sim 100$ \kms\, e.g.,][]{kornei2013,jones2012,shapley2003} as expected from outflow models \citep{prochaska2011}. In this sample we find fine structure line centroids within 50 \kms\ of the adopted systemic redshift in all cases with good signal-to-noise, with mean and median offsets of $<10$ \kms. These lines therefore provide a consistency check on the accuracy of redshifts derived from noisier features.

\begin{figure}
\centerline{
\includegraphics[width=0.8\columnwidth]{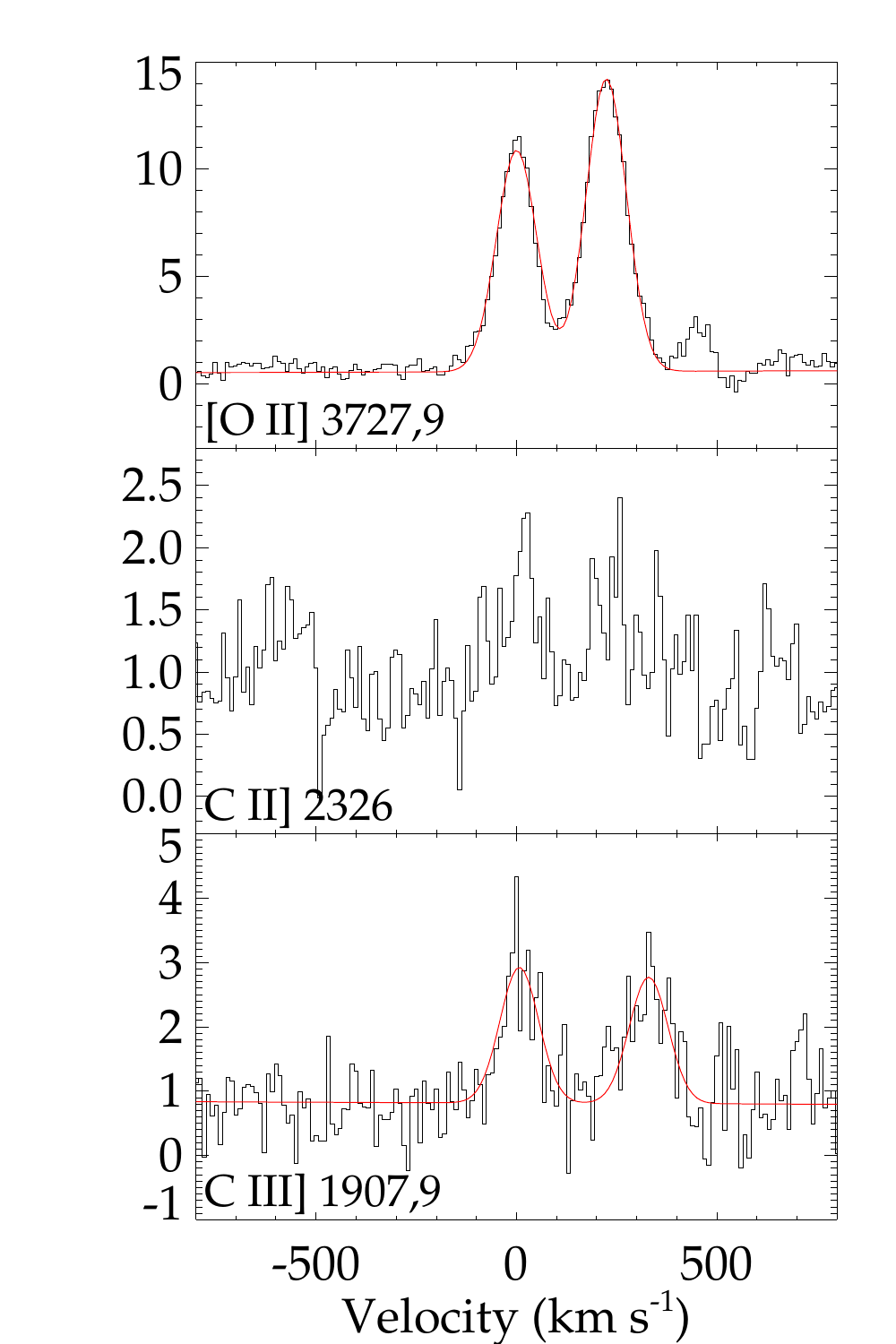}
}
\caption{
\label{fig:c141_zsys}
Features used to derive the systemic redshift of CSWA 141. Red lines show the best-fit double Gaussian functions to \Oii\ and \Ciii.
}
\end{figure}

\begin{figure}
\centerline{
\includegraphics[width=0.8\columnwidth]{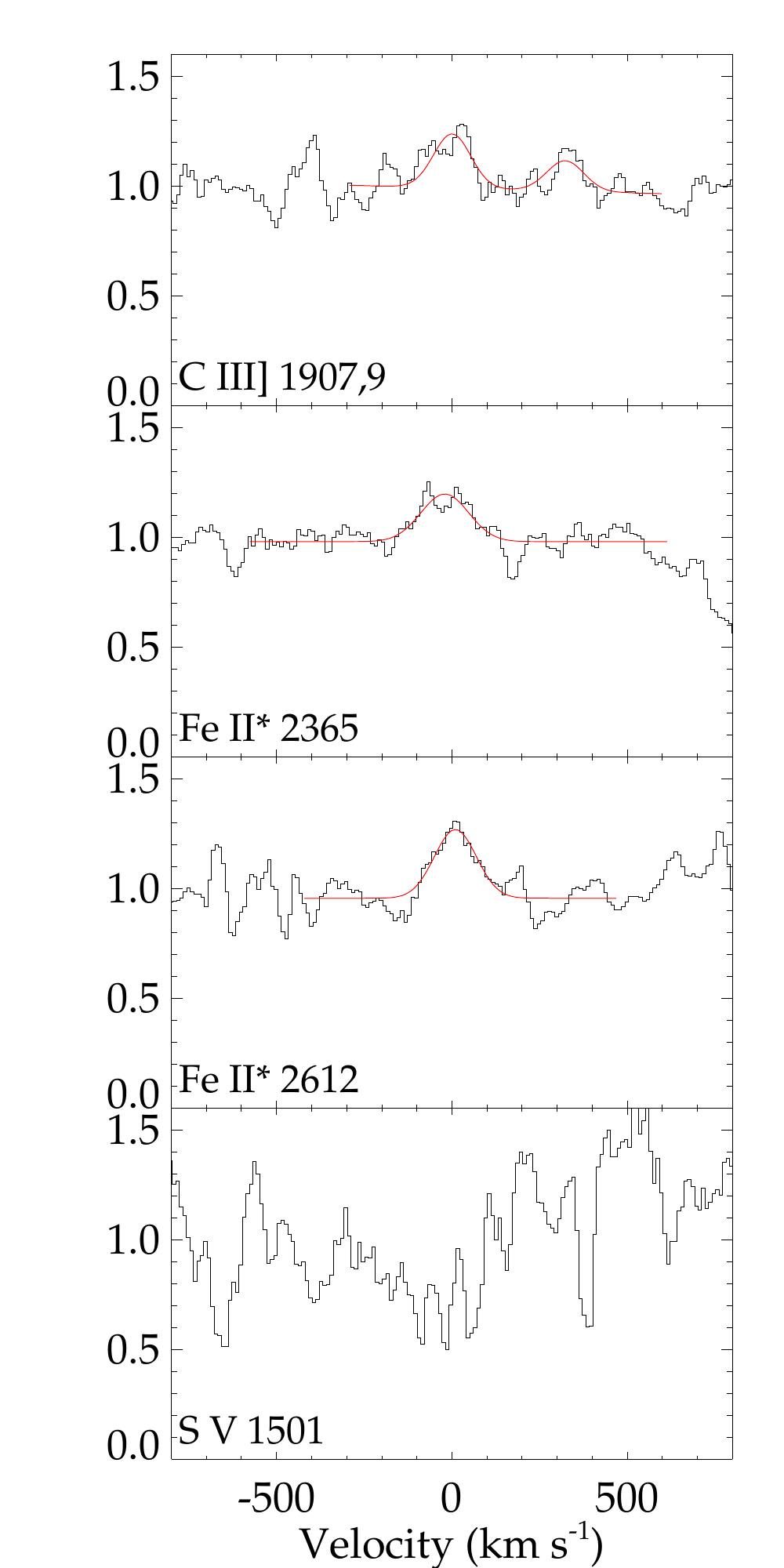}
}
\caption{
\label{fig:c103_zsys}
Features used to determine the systemic redshift of CSWA 103.
}
\end{figure}

\begin{figure}
\centerline{
\includegraphics[width=0.8\columnwidth]{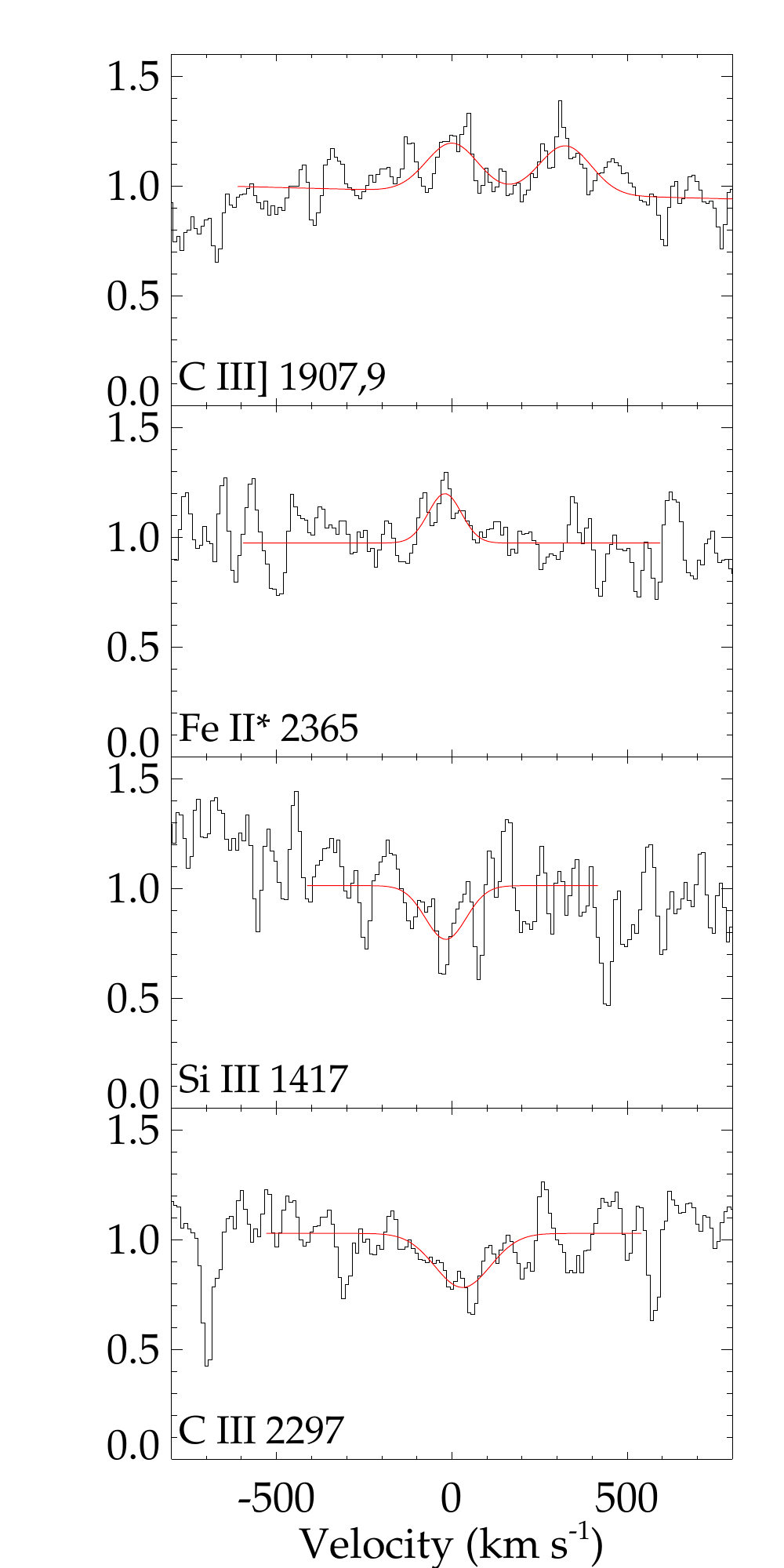}
}
\caption{
\label{fig:c19_zsys}
Features used to determine the systemic redshift of CSWA 19.
}
\end{figure}

\begin{figure}
\centerline{
\includegraphics[width=0.8\columnwidth]{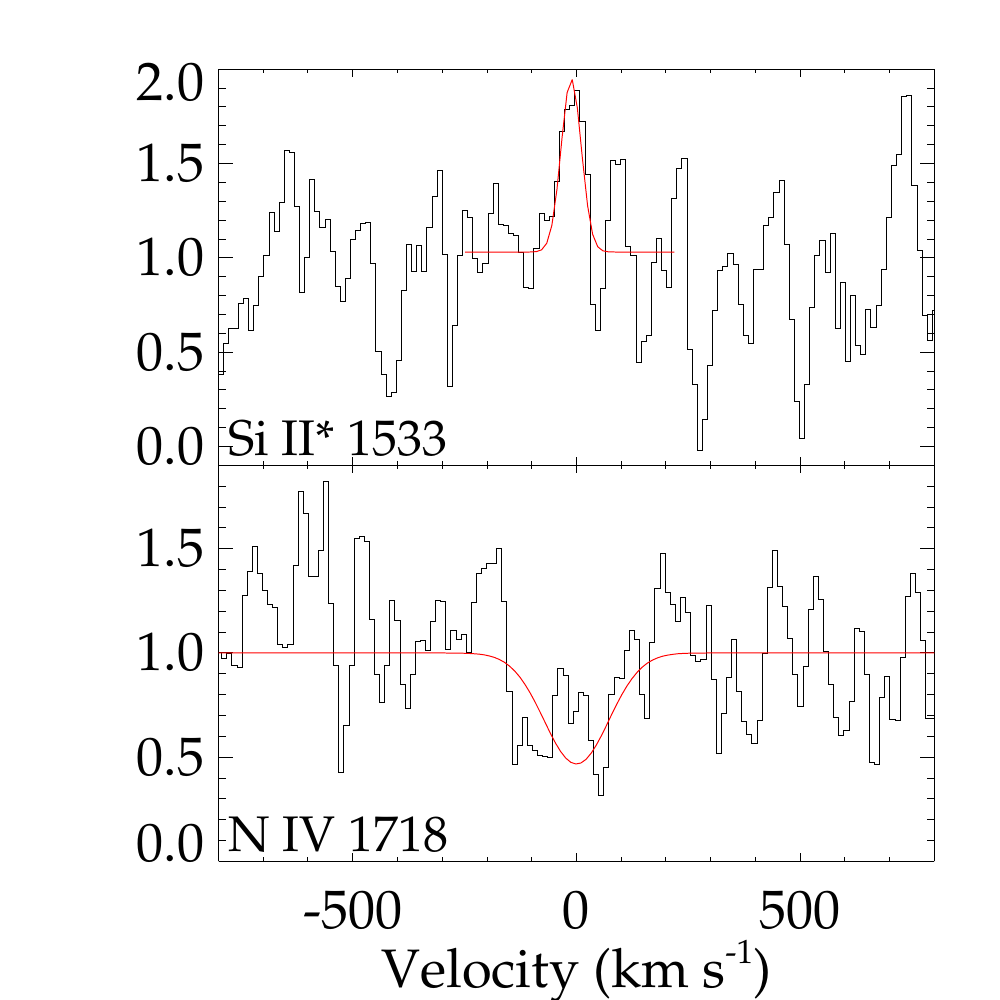}
}
\caption{
\label{fig:c40_zsys}
Features used to determine the systemic redshift of CSWA 40.
}
\end{figure}

\begin{figure}
\centerline{
\includegraphics[width=0.8\columnwidth]{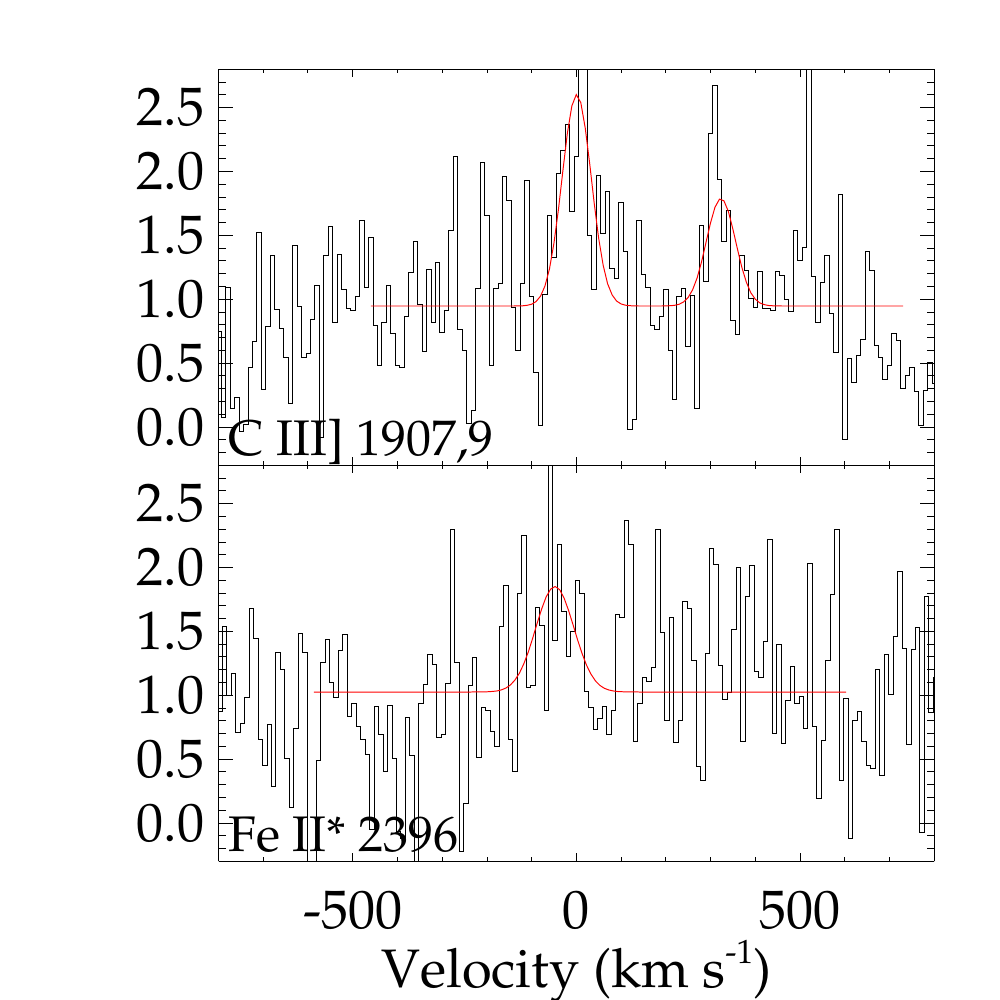}
}
\caption{
\label{fig:c2_zsys}
Features used to determine the systemic redshift of CSWA 2.
}
\end{figure}

\begin{figure}
\centerline{
\includegraphics[width=0.8\columnwidth]{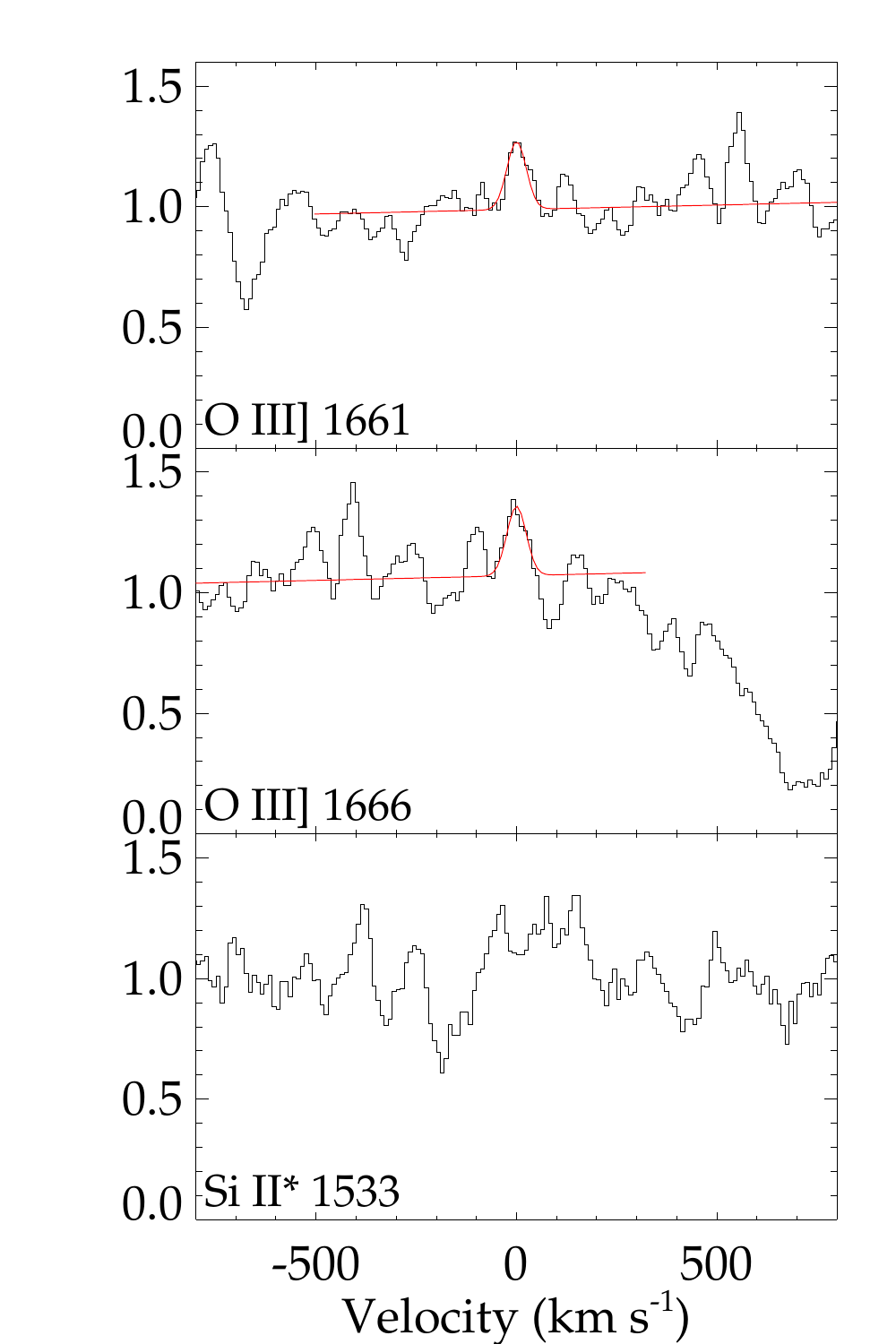}
}
\caption{
\label{fig:c128_zsys}
Features used to determine the systemic redshift of CSWA 128. Absorption at $>500$ \kms\ relative to O {\sc iii}] $\lambda$1666 is due to Al {\sc ii} $\lambda$1670.
}
\end{figure}

\begin{figure}
\centerline{
\includegraphics[width=0.8\columnwidth]{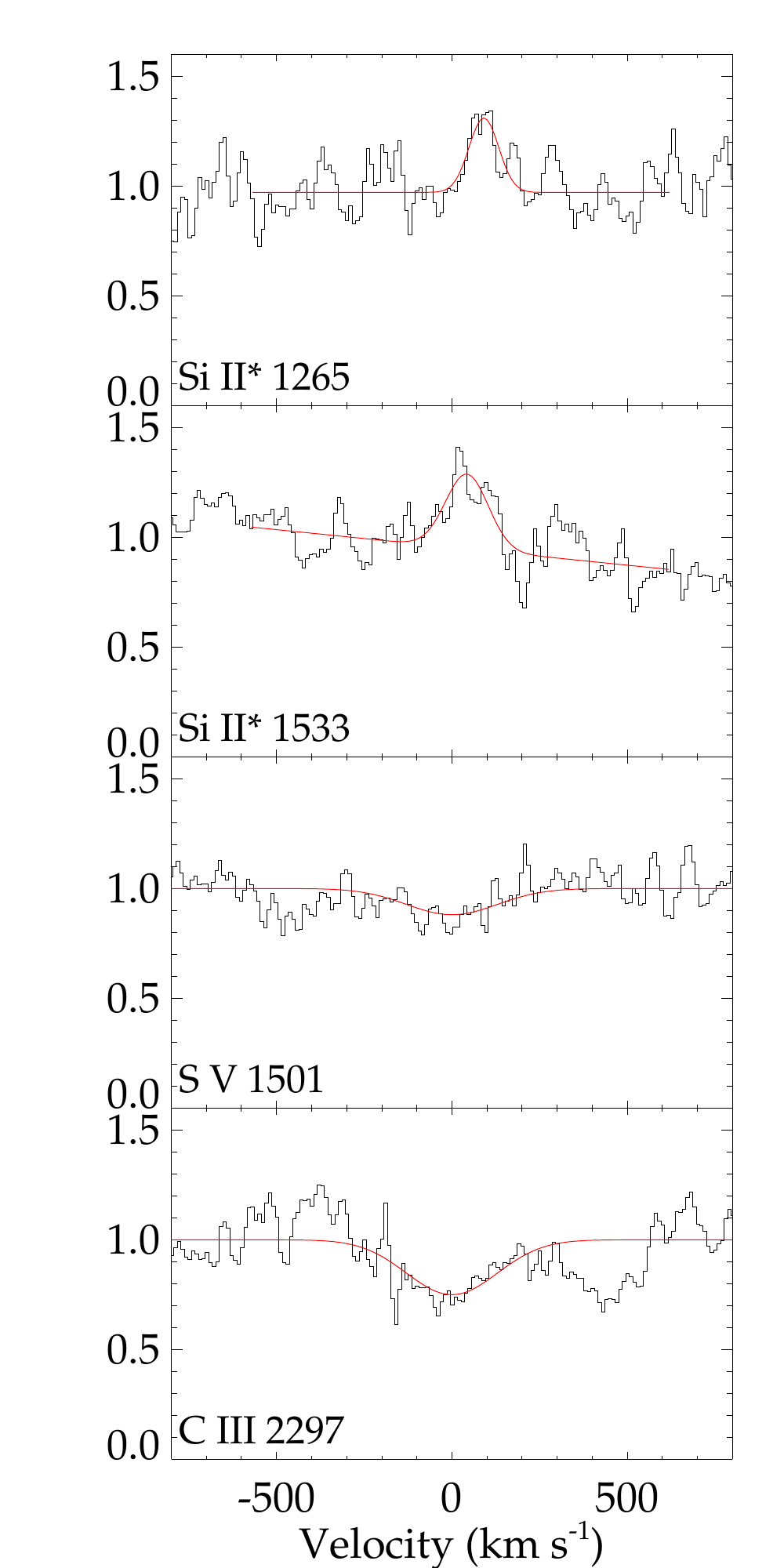}
}
\caption{
\label{fig:c164_zsys}
Features used to determine the systemic redshift of CSWA 164. The continuum slope near Si {\sc ii}* $\lambda$1533 is due to broad stellar P-cygni C {\sc iv} $\lambda\lambda$1548,51 absorption.
}
\end{figure}

\begin{figure}
\centerline{
\includegraphics[width=0.8\columnwidth]{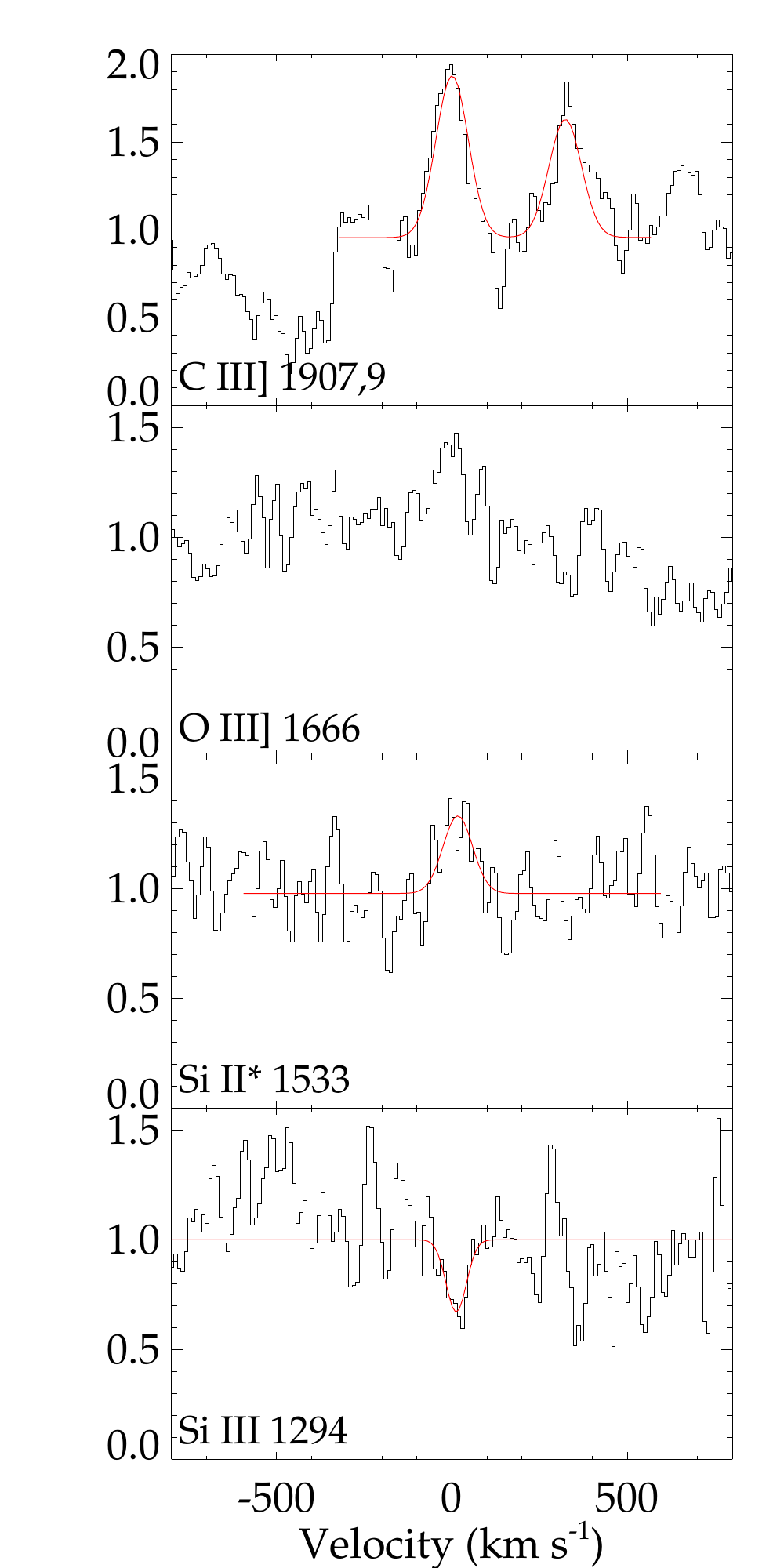}
}
\caption{
\label{fig:c39_zsys}
Features used to determine the systemic redshift of CSWA 39.
}
\end{figure}

\begin{figure}
\centerline{
\includegraphics[width=0.8\columnwidth]{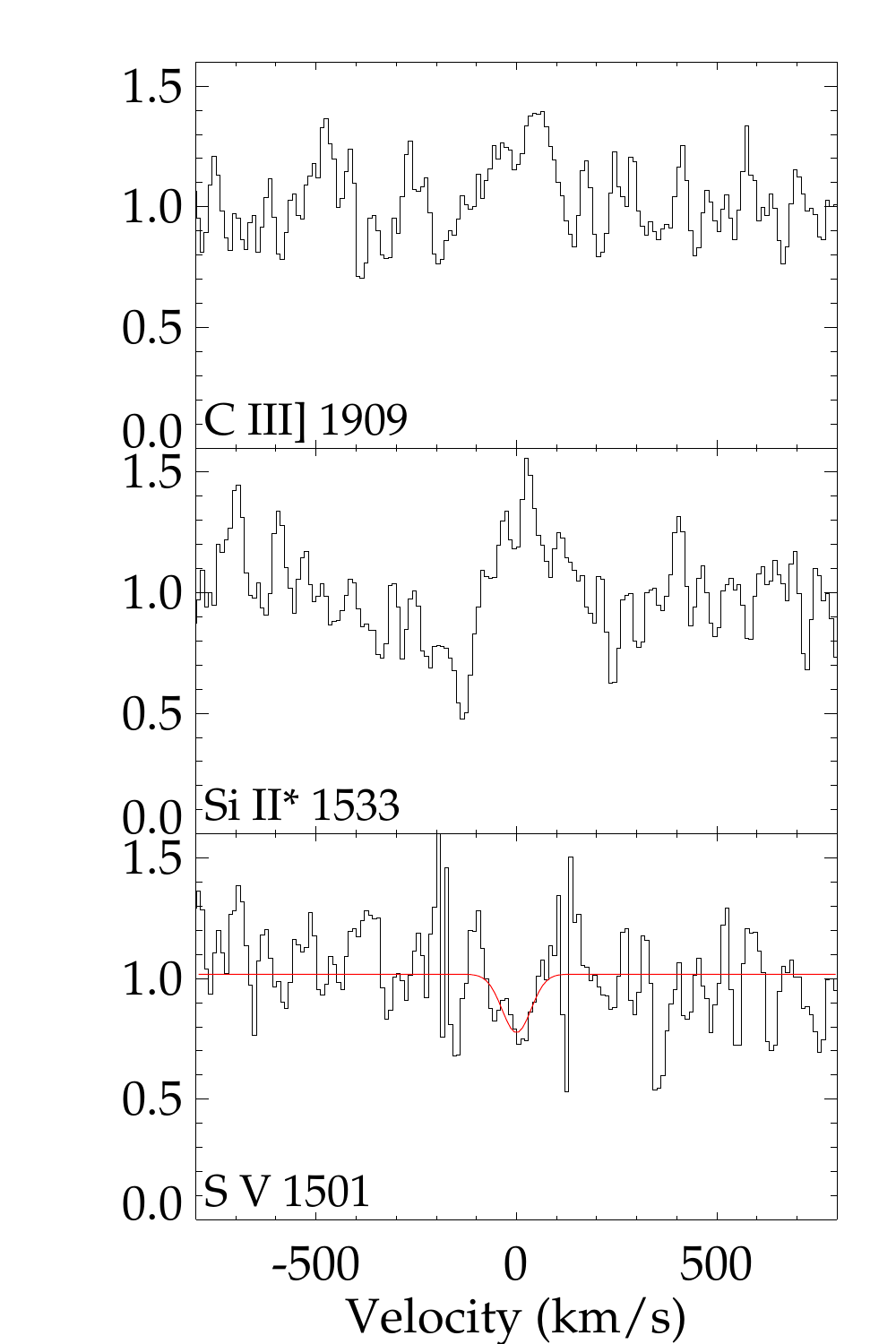}
}
\caption{
\label{fig:c38_zsys}
Features used to determine the systemic redshift of CSWA 38.
}
\end{figure}

In Figures~\ref{fig:c141_zsys} through \ref{fig:c38_zsys} we show a subset of the features used to derive systemic redshifts for each source. Line profiles are plotted as a function of velocity for the sake of comparison with interstellar absorption (e.g., Figure~\ref{fig:fcov_low}). For a subset of the features we also show best-fit Gaussian profiles as red lines.

\subsection{CSWA 141}

The redshift of CSWA 141 is well determined from several strong nebular emission lines. The most prominent of these is \Oii\ $\lambda\lambda$3727,9 which we fit with a double Gaussian function as described above. This yields the most precise redshift measurement in the sample due to the high SNR. The systemic feature with the second-highest SNR is the \Ciii\ $\lambda\lambda$1907,9 doublet. The redshift and velocity dispersion measured from \Ciii\ is in good agreement with [O {\sc ii}], with differences of $\Delta v_{sys} = 7 \pm 4$ \kms\ and $\Delta \sigma = -2 \pm 2$ \kms.

\subsection{CSWA 103}

The most significant systemic feature in the spectrum of CSWA 103 is the \Ciii\ $\lambda\lambda$1907,9 doublet. The redshift is corroborated by detection of fine-structure Fe {\sc ii}* emission lines with consistent line centroids ($< 20$ \kms offset in velocity). There is possible weak photospheric absorption from S {\sc v} although we are unable to measure a centroid for this feature.

\subsection{CSWA 19}

We adopt the redshift of CSWA 19 measured from \Ciii\ $\lambda\lambda$1907,9, which is consistent with the Fe {\sc ii}* $\lambda$2365 fine structure emission line as well as photospheric absorption. Fits to the individual photospheric lines Si {\sc iii} 1417 and C {\sc iii} 2297 give centroids which are offset by $-18$ and $+30$ \kms\, respectively.

\subsection{CSWA 40}

We measure the redshift of CSWA 40 from a combined fit to the photospheric absorption lines Si {\sc iii} $\lambda$1294, Si {\sc iii} $\lambda$1417, S {\sc v} $\lambda$1501, N {\sc iv}$\lambda$1718, and C {\sc iii} $\lambda$2297. We also identify possible Si {\sc ii}* emission at the same redshift but with a narrower line profile.

\subsection{CSWA 2}

We adopt the redshift of CSWA 2 measured from \Ciii\ $\lambda\lambda$1907,9, which is the only feature for which we are able to measure a reliable systemic redshift. We identify possible Fe {\sc ii} $\lambda$2396 emission with a velocity offset of $-48$ \kms.

\subsection{CSWA 128}

We adopt the redshift of CSWA 128 measured from O {\sc iii}] $\lambda\lambda$1661,6, which is the only feature for which we are able to measure a reliable systemic redshift although the significance of the fit is modest (7$\sigma$). We identify Si {\sc ii}* $\lambda$1533 emission at similar velocities. Photospheric absorption features are not significantly detected despite the good SNR of this spectrum.

\subsection{CSWA 164}

We measure the redshift of CSWA 164 from a combined fit to the photospheric absorption lines Si {\sc iii} $\lambda$1294, Si {\sc iii} $\lambda$1417, S {\sc v} $\lambda$1501, N {\sc iv}$\lambda$1718, and C {\sc iii} $\lambda$2297. \Ciii\ $\lambda\lambda$1907,9 is also seen at low SNR with a best-fit centroid offset by $30\pm15$ \kms.

\subsection{CSWA 39}

We adopt the redshift of CSWA 39 measured from \Ciii\ $\lambda\lambda$1907,9 due to its relatively high SNR. The photospheric absorption lines are in good agreement, and we also weakly detect O {\sc iii}] $\lambda\lambda$ 1661,6 and Si {\sc ii}* emission with velocity offsets of $\sim 15$ \kms.

\subsection{CSWA 38}

We measure the redshift of CSWA 38 from a combined fit to the photospheric absorption lines Si {\sc iii} $\lambda$1294, Si {\sc iii} $\lambda$1417, S {\sc v} $\lambda$1501, and N {\sc iv}$\lambda$1718. C {\sc iii} $\lambda$2297 is excluded from the fit due to low SNR of the spectrum at that wavelength. We also identify Si {\sc ii}* $\lambda$1533 and possible \Ciii\ $\lambda\lambda$1907,9 emission at similar redshifts.

\end{appendix}

\begin{deluxetable*}{lcccccccc}
\tablecolumns{8}
\tablewidth{0pt}
\tablecaption{Observing log\label{tab:arcs}}

\tablehead{
\colhead{ID} & \colhead{Dates observed} & \colhead{$z$\tablenotemark{a}} & \colhead{RA} & \colhead{Dec} & \colhead{PA} & \colhead{t$_{\mathrm exp}$ (ks)} & \colhead{SNR\tablenotemark{b}} \\
}
\medskip
\startdata
CSWA 141  &  8 Nov 2012    &  1.425194$\pm$0.000003  &  08:46:47.53  &  +04:46:09.3  &  190  &  5.2  &  4.5  \\
CSWA 103  &  8-10 Nov 2012 &  1.95978$\pm$0.00008    &  01:45:04.38  &  -04:55:50.8  &  115  &  25  &  11.5 \\
CSWA 19   &  9-10 Nov 2012 &  2.03237$\pm$0.00011    &  09:00:02.80  &  +22:34:07.1  &  86   &  20  &  18.3 \\
CSWA 40   &  5-6 Mar 2013  &  2.18938$\pm$0.00007    &  09:52:40.29  &  +34:34:39.2  &  70   &  16.2  &  5.4  \\
CSWA 2    &  5 Mar 2013    &  2.19677$\pm$0.00007    &  10:38:41.88  &  +48:49:22.4  &  17   &  7.2  &  3.0  \\
CSWA 128  &  8-10 Nov 2012 &  2.22505$\pm$0.00003    &  19:58:35.44  &  +59:50:52.2  &  60   &  16.3  &  15.4 \\
CSWA 164  &  8-10 Nov 2012 &  2.51172$\pm$0.00007    &  02:32:49.93  &  -03:23:25.8  &  158  &  28  &  15.3 \\
CSWA 39   &  5-6 Mar 2013  &  2.76223$\pm$0.00008    &  15:27:45.16  &  +06:52:19.6  &  105  &  18  &  10.7 \\
CSWA 38   &  6 Mar 2013    &  2.92556$\pm$0.00007    &  12:26:51.48  &  +21:52:17.9  &  130  &  10.8  &  9.8  \\
\enddata
\tablenotetext{a}{Adopted systemic redshift and formal 1$\sigma$ uncertainty}
\tablenotetext{b}{Continuum signal-to-noise ratio per 1.0 \AA\ resolution element at $\lambda=6200$ \AA}
\end{deluxetable*}

\begin{deluxetable*}{lll}
\tablecolumns{8}
\tablewidth{0pt}
\tablecaption{Systemic redshift features\label{tab:zsys}}

\tablehead{
\colhead{ID} & \colhead{feature} & \colhead{$z$} \\
}
\medskip
\startdata
CSWA 141  &  [O {\sc ii}] $\lambda\lambda$3727,9  &  1.425194$\pm$0.000003\tablenotemark{a} \\
\smallskip
          &  C {\sc iii}] $\lambda\lambda$1907,9  &  1.425248$\pm$0.000033  \\

CSWA 103  &  C {\sc iii}] $\lambda\lambda$1907,9  &  1.95978$\pm$0.00008\tablenotemark{a} \\
          &  Fe {\sc ii}* $\lambda$2365  &  1.95957$\pm$0.00008 \\
\smallskip
          &  Fe {\sc ii}* $\lambda$2612  &  1.95987$\pm$0.00004 \\

CSWA 19   &  C {\sc iii}] $\lambda\lambda$1907,9  &  2.03237$\pm$0.00011\tablenotemark{a} \\
          &  Fe {\sc ii}* $\lambda$2365  &  2.03217$\pm$0.00009 \\
          &  Si {\sc iii} $\lambda$1417  &  2.03219$\pm$0.00009 \\
\smallskip
          &  C {\sc iii} $\lambda$2297   &  2.03268$\pm$0.00008 \\

CSWA 40   &  stellar absorption\tablenotemark{b}  &  2.18938$\pm$0.00007\tablenotemark{a} \\
\smallskip
          &  Si {\sc ii}* $\lambda$1533  &  2.18926$\pm$0.00007 \\

CSWA 2    &  C {\sc iii}] $\lambda\lambda$1907,9  &  2.19677$\pm$0.00007\tablenotemark{a} \\
\smallskip
          &  Fe {\sc ii}* $\lambda$2396  &  2.19625$\pm$0.00014 \\

\smallskip
CSWA 128  &  O {\sc iii}] $\lambda\lambda$1661,6  &  2.22505$\pm$0.00003\tablenotemark{a} \\

CSWA 164  &  stellar absorption\tablenotemark{b}  &  2.51172$\pm$0.00007\tablenotemark{a} \\
          &  C {\sc iii}] $\lambda\lambda$1907,9  &  2.51207$\pm$0.00016 \\
\smallskip
          &  Si {\sc ii}* $\lambda$1533  &  2.51222$\pm$0.00006 \\

CSWA 39   &  C {\sc iii}] $\lambda\lambda$1907,9  &  2.76223$\pm$0.00008\tablenotemark{a} \\
          &  O {\sc iii}] $\lambda\lambda$1661,6  &  2.76204$\pm$0.00007 \\
          &  Si {\sc ii}* $\lambda$1533  &  2.76244$\pm$0.00006 \\
\smallskip
          &  stellar absorption\tablenotemark{c}  &  2.76238$\pm$0.00007 \\

CSWA 38   &  stellar absorption\tablenotemark{c}  &  2.92556$\pm$0.00007\tablenotemark{a} \\
          &  C {\sc iii}] $\lambda\lambda$1907,9  &  2.92588$\pm$0.00019 \\
          &  Si {\sc ii}* $\lambda$1533  &  2.92599$\pm$0.00007 \\

\enddata
\tablenotetext{a}{Adopted systemic redshift}
\tablenotetext{b}{Simultaneous fit to stellar photospheric absorption features Si {\sc iii} $\lambda$1294, Si {\sc iii} $\lambda$1417, S {\sc v} $\lambda$1501, N {\sc iv}$\lambda$1718, and C {\sc iii} $\lambda$2297.}
\tablenotetext{c}{Simultaneous fit to Si {\sc iii} $\lambda$1294, Si {\sc iii} $\lambda$1417, S {\sc v} $\lambda$1501, and N {\sc iv}$\lambda$1718.}
\end{deluxetable*}

\begin{deluxetable*}{lcccccccccc}
\tablecolumns{10}
\tablewidth{0pt}
\tablecaption{Column densities of individual transitions\label{tab:transitions}}

\tablehead{
\colhead{Transition} & \colhead{CSWA 141} & \colhead{CSWA 103} & \colhead{CSWA 19} & \colhead{CSWA 40} & \colhead{CSWA 2} & \colhead{CSWA 128} & \colhead{CSWA 164} & \colhead{CSWA 39} & \colhead{CSWA 38} \\
\colhead{$\Delta$v [\kms]} & \colhead{-400 $-$ -150} & \colhead{-500 $-$ 0} & \colhead{-250 $-$ 0} & \colhead{-250 $-$ 50} & \colhead{-200 $-$ 300} & \colhead{-300 $-$ 100} & \colhead{-400 $-$ 150} & \colhead{-300 $-$ 0} & \colhead{-400 $-$ 0} \\
}

\medskip
\startdata

H I 1216\tablenotemark{a} &  --  &  --  &  --  &  --  &  --  & 21.22$_{-0.25}^{+0.23}$ & 20.28$_{-0.35}^{+0.26}$ & 20.13$_{-0.60}^{+0.71}$ & 20.57$_{-0.17}^{+0.16}$  \\

O I 1302\tablenotemark{b} &              -- &             -- &             -- & $>$           20 & $>$           51 & $>$           30 & $>$           28 & $>$           10 & $>$           34 \\

O I 1355\tablenotemark{c} &              -- &             -- & $<$        99988 & $<$       123841 &             -- & $<$       331014 & $<$        98680 & $<$        77083 & $<$        83327 \\

Mg II 2803\tablenotemark{b} &  $>$ 0.87 &  --  &  --  &  --  &  --  &  --  &  --  &  --  &  --  \\

Al II 1670 &  --  &  --  & 0.12$_{-0.02}^{+0.03}$ &  --  &  --  &  --  & 0.15$_{-0.02}^{+0.05}$ &  --  &  --   \\

Al II 1670\tablenotemark{b} &  --  & $>$ 0.69 &  --  & $>$ 0.44 & $>$ 1.10 & $>$ 0.65 &  --  & $>$ 0.22 & $>$ 0.75 \\

Si II 1808 & 17.06$_{-28.72}^{+46.39}$ & 79.13$_{-20.14}^{+44.67}$ & -3.56$_{-7.26}^{+14.15}$ & 58.79$_{-20.36}^{+45.31}$ & 9.02$_{-40.43}^{+86.45}$ & 52.85$_{-12.36}^{+19.20}$ & 1.81$_{-11.02}^{+17.94}$ & 22.30$_{-12.34}^{+33.76}$ & 40.62$_{-15.57}^{+26.29}$  \\

Cr II 2056 & 1.30$_{-0.58}^{+-NaN}$ & 0.28$_{-0.28}^{+0.47}$ & -0.00$_{-0.11}^{+0.19}$ &  --  &  --  & 0.16$_{-0.20}^{+0.30}$ & -0.09$_{-0.17}^{+0.24}$ &  --  & 0.85$_{-0.38}^{+0.63}$  \\

Cr II 2066 & 2.84$_{-1.39}^{+-NaN}$ & 0.79$_{-0.55}^{+1.20}$ & -0.28$_{-0.23}^{+0.46}$ &  --  &  --  & 0.05$_{-0.37}^{+0.53}$ &  --  & -0.60$_{-0.27}^{+0.60}$ & 0.75$_{-0.62}^{+1.12}$  \\

Fe II 1608 &  --  & 13.53$_{-1.69}^{+1.89}$ & 0.46$_{-0.35}^{+0.94}$ & 8.42$_{-1.70}^{+4.34}$ & 13.66$_{-2.73}^{+5.91}$ & 12.34$_{-1.41}^{+2.57}$ & -0.12$_{-0.43}^{+0.73}$ & 3.17$_{-0.96}^{+1.18}$ & 4.18$_{-0.91}^{+2.47}$  \\

Fe II 2249 &  --  & 27.60$_{-13.16}^{+22.99}$ &  --  & 38.11$_{-14.54}^{+34.75}$ & 39.21$_{-36.86}^{+114.88}$ & -11.78$_{-10.23}^{+15.78}$ &  --  &  --  &     \\

Ni II 1317 &  --  &  --  &  --  &  --  &  --  &  --  & 1.74$_{-0.71}^{+1.14}$ & 0.51$_{-0.60}^{+1.31}$ & 1.79$_{-0.54}^{+0.84}$  \\

Ni II 1370 &  --  &  --  & -0.28$_{-0.40}^{+0.93}$ & 0.07$_{-0.93}^{+4.39}$ &  --  & 2.56$_{-1.06}^{+2.83}$ & -0.67$_{-0.39}^{+0.68}$ & 0.10$_{-0.42}^{+0.85}$ & -0.11$_{-0.63}^{+1.34}$  \\

Ni II 1741 & -2.00$_{-1.72}^{+4.06}$ & 2.49$_{-0.87}^{+1.55}$ & 0.52$_{-0.37}^{+0.67}$ & 1.30$_{-0.93}^{+2.37}$ & 0.32$_{-1.73}^{+4.34}$ & 2.58$_{-0.54}^{+0.77}$ & 2.16$_{-0.64}^{+1.30}$ & 0.76$_{-0.62}^{+1.05}$ & 2.15$_{-0.97}^{+2.07}$  \\

Cu II 1358 &  --  &  --  & 0.07$_{-0.13}^{+0.28}$ & -0.24$_{-0.38}^{+1.21}$ &  --  & 0.14$_{-0.25}^{+0.65}$ & -0.04$_{-0.12}^{+0.19}$ &  --  & -0.00$_{-0.17}^{+0.34}$  \\

\bigskip

Zn II 2026 & 0.07$_{-0.09}^{+0.08}$ & 0.21$_{-0.08}^{+0.13}$ & -0.02$_{-0.02}^{+0.04}$ & 0.54$_{-0.14}^{+0.28}$ & 0.93$_{-0.23}^{+0.56}$ & 0.22$_{-0.07}^{+0.10}$ & 0.12$_{-0.04}^{+0.06}$ &  --  & -0.06$_{-0.06}^{+0.15}$  \\


Mg I 2852 & -0.01$_{-0.02}^{+0.05}$  &  --  &  --  &  --  &  --  &  --  &  --  &  --  &  --  \\

Si I 1693 &  --  & 0.41$_{-0.28}^{+0.48}$ & 0.37$_{-0.12}^{+0.23}$ & -0.17$_{-0.27}^{+0.90}$ & -0.82$_{-0.64}^{+1.32}$ & 0.10$_{-0.17}^{+0.27}$ & 0.08$_{-0.18}^{+0.41}$ &  --  & 0.18$_{-0.21}^{+0.34}$  \\

Si I 2515 & 0.52$_{-0.21}^{+0.32}$ & 0.25$_{-0.14}^{+0.20}$ &  --  &  --  & 0.42$_{-0.32}^{+0.78}$ &  --  &  --  &  --  &  --   \\

Fe I 2167 & -0.02$_{-0.21}^{+0.42}$ & 0.13$_{-0.18}^{+0.29}$ &  --  &  --  &  --  & -0.11$_{-0.09}^{+0.12}$ &  --  & -0.18$_{-0.14}^{+0.19}$ &  --   \\

Fe I 2484 & -0.06$_{-0.04}^{+0.10}$ &  --  &  --  &  --  &  --  &  --  &  --  &  --  &  --   \\

\bigskip

Fe I 2523 & -0.05$_{-0.12}^{+0.33}$ & 0.04$_{-0.06}^{+0.09}$ &  --  &  --  &  --  &  --  &  --  &  --  &  --   \\

Al III 1854 &  --  & 1.10$_{-0.12}^{+0.21}$ & 0.58$_{-0.08}^{+0.12}$ & 0.33$_{-0.11}^{+0.24}$ & 1.23$_{-0.25}^{+0.91}$ & 0.93$_{-0.11}^{+0.29}$ & 0.34$_{-0.06}^{+0.12}$ & 0.33$_{-0.06}^{+0.24}$ &  --   \\

Al III 1862 &  --  & 1.36$_{-0.18}^{+0.37}$ & 0.81$_{-0.13}^{+0.20}$ & 1.29$_{-0.31}^{+\infty}$\tablenotemark{d} & 2.27$_{-0.49}^{+\infty}$\tablenotemark{d} & 1.33$_{-0.16}^{+0.38}$ & 0.51$_{-0.13}^{+0.17}$ & 0.54$_{-0.12}^{+0.13}$ & 1.52$_{-0.25}^{+0.62}$  \\

\enddata
\tablenotetext{}{All column densities are in units of $10^{14}$ cm$^{-2}$ and are integrated over the velocity range $\Delta$v, except where noted otherwise. Missing values indicate cases where we are unable to obtain a reliable measurement due to spectral coverage, sky line residuals, low throughput, or other problems.}
\tablenotetext{a}{H I column densities are given as $\log (N \times f_{c})$ from Voigt profile fitting, with $N$ in units of cm$^{-2}$.}
\tablenotetext{b}{Lower limit corresponding to optical depth $\tau > 2$.}
\tablenotetext{c}{2$\sigma$ upper limit.}
\tablenotetext{d}{No formal upper bound.}

\end{deluxetable*}

\begin{deluxetable*}{lcccccccccc}
\tablecolumns{10}
\tablewidth{0pt}
\tablecaption{Column densities from multiple transitions\label{tab:transitions_mult}}

\tablehead{
\colhead{Ion} & \colhead{CSWA 141} & \colhead{CSWA 103} & \colhead{CSWA 19} & \colhead{CSWA 40} & \colhead{CSWA 2} & \colhead{CSWA 128} & \colhead{CSWA 164} & \colhead{CSWA 39} & \colhead{CSWA 38} \\
\colhead{$\Delta$v [\kms]} & \colhead{-400 $-$ -150} & \colhead{-500 $-$ 0} & \colhead{-250 $-$ 0} & \colhead{-250 $-$ 50} & \colhead{-200 $-$ 300} & \colhead{-300 $-$ 100} & \colhead{-400 $-$ 150} & \colhead{-300 $-$ 0} & \colhead{-400 $-$ 0} \\
}

\medskip
\startdata

Si II\tablenotemark{a} & 17.06$_{-28.72}^{+46.39}$ & 79.13$_{-20.14}^{+44.67}$ & 6.34$_{-1.24}^{+10.17}$ & 56.89$_{-11.85}^{+45.84}$ & 59.07$_{-10.50}^{+58.36}$ & 50.99$_{-8.15}^{+17.21}$ & 10.66$_{-2.09}^{+8.73}$ & 22.55$_{-8.01}^{+26.52}$ & 33.94$_{-6.51}^{+21.45}$  \\
\medskip
Transitions  &  4            &  4          &  3,4      &  2,3,4        &  1,2,3,4    &  1,2,3,4      &  1,2,3,4    &  2,3,4        &  1,2,3,4    \\

Fe II\tablenotemark{b} & 1.49$_{-0.22}^{+0.66}$ & 14.70$_{-1.17}^{+3.71}$ & 1.07$_{-0.20}^{+0.43}$ & 6.29$_{-0.93}^{+3.02}$ & 22.76$_{-3.17}^{+10.40}$ & 8.65$_{-1.12}^{+5.40}$ & 1.50$_{-0.24}^{+0.40}$ & 1.92$_{-0.90}^{+2.96}$ & 3.50$_{-0.41}^{+1.37}$  \\
\medskip
Transitions  &  3,4,5,6,7,8  &  1,2,3,5,6  &  1,2,4    &  1,2,3,4,5,6  &  1,2,4      &  1,3,6        &  1,2,4      &  1,2,4,5      &  1,4,6      \\

Ni II\tablenotemark{c} & -- & 0.94$_{-0.53}^{+1.11}$ & 0.19$_{-0.22}^{+0.35}$ & 1.02$_{-0.54}^{+0.97}$ & 0.87$_{-0.79}^{+3.15}$ & 2.84$_{-0.39}^{+0.46}$ & 0.42$_{-0.30}^{+0.45}$ & 0.10$_{-0.20}^{+0.53}$ & 0.93$_{-0.36}^{+0.44}$  \\
\medskip
Transitions  &               &  4,5        &  2,3,4,5  &  1,2,3,4,5,6  &  1,2,3,4,5  &  1,2,3,4,5,6  &  1,2,3,4,5  &  1,2,3,4,5,6  &  1,2,3,4,5  \\

Al III\tablenotemark{d} & -- & 1.60$_{-0.48}^{+0.68}$ & 0.92$_{-0.28}^{+0.47}$ & -- & -- & 1.77$_{-0.43}^{+0.84}$ & 0.61$_{-0.25}^{+0.50}$ & 0.58$_{-0.19}^{+0.55}$ & --  \\

\enddata
\tablenotetext{}{All column densities are in units of $10^{14}$ cm$^{-2}$ and are integrated over the velocity range $\Delta$v.}

\tablenotetext{a}{\Siiia\ transitions correspond to $\lambda$1260$\rightarrow$1, $\lambda$1304$\rightarrow$2, $\lambda$1526$\rightarrow$3, $\lambda$1808$\rightarrow$4.}

\tablenotetext{b}{\Feiia\ transitions correspond to $\lambda$1608$\rightarrow$1, $\lambda$2249$\rightarrow$2, $\lambda$2260$\rightarrow$3, $\lambda$2344$\rightarrow$4, $\lambda$2374$\rightarrow$5, $\lambda$2382$\rightarrow$6, $\lambda$2586$\rightarrow$7, $\lambda$2600$\rightarrow$8.}

\tablenotetext{c}{\Niiia\ transitions correspond to $\lambda$1317$\rightarrow$1, $\lambda$1370$\rightarrow$2, $\lambda$1703$\rightarrow$3, $\lambda$1709$\rightarrow$4, $\lambda$1741$\rightarrow$5, $\lambda$1751$\rightarrow$6.}

\tablenotetext{d}{\Aliiia\ values are based on the $\lambda\lambda$1854,1862 doublet.}

\end{deluxetable*}

\begin{deluxetable*}{lcccccccccc}
\tablecolumns{10}
\tablewidth{0pt}
\tablecaption{Gas kinematics \label{tab:kinematics}}

\tablehead{
\colhead{ } & \colhead{CSWA 103} & \colhead{CSWA 19} & \colhead{CSWA 40} & \colhead{CSWA 2} & \colhead{CSWA 128} & \colhead{CSWA 164} & \colhead{CSWA 39} & \colhead{CSWA 38} \\
}

\medskip
\startdata

Weighted by $f_c$ \\

low ion $\bar{v}$ (\kms)  &  -259 $\pm$ 9  &  -159 $\pm$ 17  &  -141 $\pm$ 16  &  -19 $\pm$ 27  &  -161 $\pm$ 10  &  -108 $\pm$ 
10  &  -117 $\pm$ 20  &  -183 $\pm$ 9 \\

\medskip
low ion $\sigma_v$ (\kms)  &  165 $\pm$ 6  &  162 $\pm$ 8  &  179 $\pm$ 7  &  202 $\pm$ 14  &  170 $\pm$ 5  &  220 $\pm$ 4  &  
172 $\pm$ 9  &  173 $\pm$ 4 \\ \hline

Weighted by $N_{tot}$ \\

Low ion $\bar{v}$ (\kms)  &  -212 $\pm$ 21  &  -143 $\pm$ 32  &  -156 $\pm$ 45  &  -32 $\pm$ 54  &  -133 $\pm$ 25  &  -42 $\pm$ 27  &  -115 $\pm$ 42  &  -260 $\pm$ 37 \\

Low ion $\sigma_v$ (\kms)  &  149 $\pm$ 5  &  134 $\pm$ 10  &  96 $\pm$ 17  &  93 $\pm$ 39  &  117 $\pm$ 10  &  137 $\pm$ 13  &  32 $\pm$ 17  &  105 $\pm$ 16 \\
\medskip
Transitions\tablenotemark{a}  &  1,2,4  &  2,3,4  &  1,2  &  1,2  &  1,2,5  &  3,6  &  1,2  &  1,2,3  \\

\Aliiia \, $\bar{v}$ (\kms)  &  -209 $\pm$ 33  &  -106 $\pm$ 20  &  -224 $\pm$ 116  &  52 $\pm$ 73  &  -113 $\pm$ 27  &  -117 $\pm$ 59  &  -160 $\pm$ 37  &  -171 $\pm$ 55 \\

\Aliiia \, $\sigma_v$ (\kms)  &  133 $\pm$ 8  &  128 $\pm$ 8  &  113 $\pm$ 30  &  124 $\pm$ 36  &  144 $\pm$ 10  &  170 $\pm$ 22  &  89 $\pm$ 22  &  134 $\pm$ 16 \\

\enddata

\tablenotetext{a}{Low ion transitions correspond to 
\Siiia$\lambda$1808$\rightarrow$1, 
\Feiia$\lambda$1608$\rightarrow$2, 
\Feiia$\lambda$2344$\rightarrow$3, 
\Feiia$\lambda$2374$\rightarrow$4, 
\Niiia$\lambda$1741$\rightarrow$5, 
\Aliia$\lambda$1670$\rightarrow$6.
}

\end{deluxetable*}

\end{document}